\documentclass[11pt]{article}
\usepackage[hmargin=1in,vmargin=1in]{geometry}
\usepackage{hchang}

\usepackage{comment}
\usepackage[hyperref]{ntheorem}
\usepackage{cleveref}
\usepackage{soul} 
\usepackage{thm-restate}

\theoremseparator{.}
\theorembodyfont{\slshape}
\newtheorem{theorem}{Theorem}[section]
\newtheorem{lemma}[theorem]{Lemma}
\newtheorem{claim}[theorem]{Claim}
\newtheorem{observation}[theorem]{Observation}
\newtheorem{problem}[theorem]{Problem}

\newtheorem{question}[theorem]{Question}
\theorembodyfont{\upshape}
\newtheorem{remark}[theorem]{Remark}
\newtheorem{definition}[theorem]{Definition}

\def\eps{\e}

\def\bdry{\partial\!}
\newcommand{\bd}{\partial\!}
\newcommand{\ldd}{\mathcal{L}}
\newcommand{\patt}{\mathbf{p}}
\newcommand{\Patt}{\mathbf{P}}

\def\reals{\mathbb{R}}

\newcommand{\dist}{\mathrm{dist}}
\newcommand{\ecc}{\mathrm{ecc}}
\newcommand{\spath}{\lambda}
\newcommand{\Cell}{\mathit{Cell}}

\newcommand{\cN}{\mathcal{N}}
\newcommand{\cB}{\mathcal{B}}
\newcommand{\GB}{\mathcal{GB}}
\newcommand{\cS}{\mathcal{S}}
\newcommand{\cT}{\mathcal{T}}
\newcommand{\cR}{\mathcal{R}}
\newcommand{\cD}{\mathcal{D}}
\newcommand{\Rep}{\mbox{\sf Rep}}
\newcommand{\RB}{\cS}
\newcommand{\cL}{\mathcal{L}}
\newcommand{\cO}{\mathcal{O}}

\newcommand{\OO}{\widetilde{O}}
\newcommand{\OOO}{O^*}

\def\note#1{}

\title{Truly Subquadratic Time Algorithms for Diameter and Related Problems in Graphs of Bounded VC-dimension}

\author{Timothy M. Chan%
    \thanks{Siebel School of Computing and Data Science,
    University of Illinois at Urbana-Champaign. Email: tmc@illinois.edu.  Supported by NSF grant CCF-2224271.}
    \and
    Hsien-Chih Chang%
    \thanks{Department of Computer Science, Dartmouth College. Email: hsien-chih.chang@dartmouth.edu.}
    \and
    Jie Gao%
    \thanks{Department of Computer Science, Rutgers University. Email: jg1555@rutgers.edu. Gao would like to acknowledge NSF support through CNS-2515159, IIS-2229876, DMS-2220271,  DMS-2311064, CCF-2208663, CCF-2118953.}
    \and
    S\'andor Kisfaludi-Bak%
    \thanks{Department of Computer Science, Aalto University, Finland. Email: 
    sandor.kisfaludi-bak@aalto.fi. Supported by the Research Council of
    Finland, Grant 363444.}
    \and
    Hung Le%
    \thanks{Manning CICS, UMass Amherst. Email: hungle@cs.umass.edu. Supported by NSF grants CCF-2517033 and CCF-2121952, NSF CAREER Award CCF-2237288, and a Google Faculty Research Award.}  
    \and 
    Da Wei Zheng%
    \thanks{Institute of Science and Technology Austria, Klosterneuburg, Austria. Work in this paper was done while at the Siebel School of Computing and Data Science, University of Illinois at Urbana-Champaign.}
}

\begin{document}

\maketitle
\thispagestyle{empty}

\begin{abstract}
We give the first truly subquadratic time algorithm, with $\OOO(n^{2-1/18})$ running time, for computing the diameter of an $n$-vertex unit-disk graph, resolving a central open problem in the literature. Our result is obtained as an instance of a general framework, applicable to different graph families and distance problems.  Surprisingly, our framework completely bypasses sublinear separators (or $r$-divisions) which were used in all previous algorithms. 
Instead, we use \emph{low-diameter decompositions} in their most elementary form.  We also exploit \emph{bounded VC-dimension} of set systems associated with the input graph, as well as new ideas on \emph{geometric data structures}.
Among the numerous applications of the general framework, we obtain:
\begin{enumerate}
    \item An $\OO(mn^{1-1/(2d)})$ time algorithm for computing the diameter of $m$-edge sparse unweighted graphs with constant VC-dimension $d$. 
    The previously known algorithms by Ducoffe, Habib, and Viennot \emph{[SODA 2019]} and  Duraj, Konieczny, and Pot\c{e}pa \emph{[ESA 2024]}  are truly subquadratic only when the diameter is a small polynomial. 
    Our result thus generalizes truly subquadratic time algorithms known for planar and minor-free graphs (in fact, it slightly improves the previous time bound for minor-free graphs).
    \item An $\OO(n^{2-1/12})$ time algorithm for computing the diameter of intersection graphs of axis-aligned squares with arbitrary size.  
    The best-known algorithm by Duraj, Konieczny, and Pot\c{e}pa \emph{[ESA 2024]} only works for unit squares and is only truly subquadratic in the low-diameter regime.  
    \item The first algorithms with truly subquadratic complexity for other distance-related problems, including all-vertex eccentricities, Wiener index, and exact distance oracles. In particular, we obtain the first exact distance oracle with truly subquadratic space and $\OO(1)$ query time for any sparse graph with bounded VC-dimension, again generalizing previous results for planar and minor-free graphs.
\end{enumerate}
\end{abstract}

\newpage
\thispagestyle{empty}
\tableofcontents
\thispagestyle{empty}

\newpage
\setcounter{page}{1}

\newpage
\section{Introduction} 
\label{sec:intro}

A simple algorithm for computing the diameter of an unweighted $n$-vertex graph is to run a BFS from every vertex of the graph. 
For sparse graphs or intersection graphs of various classes of geometric objects (such as unit disks), BFS can be implemented in $\OO(n)$ time, leading to an algorithm to compute the graph diameter in  $\OO(n^2)$ time%
\footnote{Throughout this paper, $\OO(\cdot)$ notation hides polylogarithmic factors, and $\OOO(\cdot)$ hides $n^{o(1)}$ factors.}. 
Can we beat this na\"ive quadratic-time algorithm? 
More precisely, can we compute the diameter in \emph{truly subquadratic time} $O(n^{2-\eps})$ for some fixed constant $\eps > 0$ for these graphs? This simple question has motivated the development of a broad range of techniques that have driven algorithmic research for decades. 

For general sparse graphs, even distinguishing the diameter between 2 and 3 in truly subquadratic time is impossible, assuming the Strong Exponential Time Hypothesis (SETH)~\cite{Roditty2013-zr}. (For dense undirected graphs, one can exploit the matrix multiplication subroutine to compute the diameter in $O(n^\omega)$ time~\cite{Seidel1995-hn}, where $\omega < 2.371339$ is the matrix multiplication exponent~\cite{Alman25asymmetry}.)
Given the negative result, it is natural to consider more structured classes of sparse graphs, such as planar and minor-free graphs. For planar graphs, Cabello~\cite{Cabello2018-gz} designed the first truly subquadratic algorithm for the diameter problem by introducing a new technique based on abstract Voronoi diagrams. This technique heavily exploits planarity and hence fails for minor-free graphs. Then Ducoffe, Habib, and Viennot~\cite{ducoffe2022diameter} devised a new technique based on VC-dimension to compute the diameter of minor-free graphs in truly subquadratic time. Both the Voronoi diagram and VC-dimension techniques are major milestones in algorithm design for planar and minor-free graphs, opening the door for solving other distance-related problems in truly subquadratic complexity (time or space), such as designing compact (exact) distance oracles and computing eccentricities or Wiener index in planar and minor-free graphs~\cite{Cabello2018-gz,Gawrychowski2018-zy,Li2019-li,ducoffe2022diameter,le2023vc,KZ25}.

For geometric intersection graphs of objects in the plane, designing a truly subquadratic time algorithm for the diameter problem has been much more challenging.  
A \EMPH{geometric intersection graph} is a graph whose vertices are associated with objects in the plane, and edges correspond to object pairs that intersect.\footnote{ 
We represent an intersection graph by the objects themselves, so the input size is $O(n)$ even if the graph could be dense.
}
Unit-disk graphs---the intersection graphs of unit disks---are among the most basic and well-studied graphs in the geometric setting; alternatively, this is equivalent to constructing an unweighted graph based on a set of points in a metric space by connecting pairs of points whose distance is below some fixed threshold. 

While truly subquadratic algorithms have been ruled out for intersection graphs of unit segments, unit equilateral triangles, or unit balls (in 3D) under standard fine-grained complexity assumptions~\cite{Bringmann2022-me}, the lower bound techniques for these objects fail for unit disks. 
Therefore, computing diameter for unit-disk graphs in truly subquadratic time has become a central open problem raised by many authors~\cite{Chan2016-sy,Bringmann2022-me,duraj2023better,CGL24}. 
Such an algorithm points to a larger landscape where truly subquadratic results for basic geometric intersection graphs are possible.
We note that even distinguishing the diameter between 2 and 3 in truly subquadratic time for unit-disk graphs remains open. 

\begin{question}
\label{quest:uit-disk}    
Can one compute the diameter of unit-disk graphs in truly-subquadratic time?
\end{question}

Unlike planar graphs which are sparse, unit-disk graphs (and intersection graphs in general) can be dense: they can contain cliques of arbitrary size. 
Even computing the BFS tree in $\OO(n)$ time becomes non-trivial~\cite{Cabello2015-vo}. 
Recently, Chang, Gao, and Le~\cite{CGL24} ported the VC-dimension technique for computing diameter of minor-free graphs to unit-disk graphs; 
similar to planar graphs, on a unit-disk graph, the radius-$r$ balls for all integer values $r$ also have bounded VC-dimension.
A one-sentence summary of their technique is that they treated a (possibly large) clique as a single vertex, and designed a clique-based separator hierarchy~\cite{BKMT23}.
As a result, they obtained a subquadratic ($\OO(n^{2-1/18})$-time) algorithm that could only compute an approximation of the diameter with an additive error at most $1$ in unit-disk graphs. 
While the additive error is very small, their algorithm falls short of distinguishing between diameters 2 and 3. 
This suggests that computing the diameter \emph{exactly} for unit-disk graphs requires a very different approach. 
(There are many examples in the general graph literature where allowing a small constant additive approximation can make the problem significantly easier to solve;
for example, see \cite{Aingworth1999-qc}.)
For exact algorithms,
Duraj, Konieczny, and Pot\c{e}pa~\cite{duraj2023better} adapted the technique by Ducoffe, Habib, and Viennot~\cite{ducoffe2022diameter}, which is also based on VC-dimension and a \emph{stabbing path data structure}, to the intersection graph of \EMPH{unit squares}.\footnote{
All squares are axis-aligned in this paper.
} However, their technique only works when the true diameter is small  $D = O(n^{1/4-\eps})$~\cite{duraj2023better} and more importantly,  their stabbing path data structure does not work for unit disks, or even (non-unit) square graphs, as they heavily exploit the nice geometry of unit squares.  (In fact, they explicitly asked, even when the diameter is a constant, if the diameter of a unit-disk graph can be computed in truly subquadratic time.)

\subsection{Main Results on Diameter}

In this paper, we give the first truly subquadratic algorithm for computing the diameter in unit-disk graphs, resolving \Cref{quest:uit-disk} affirmatively. 
Moreover, our framework has many other applications and  yields
the first truly subquadratic algorithms
for the intersection graph of axis-aligned (arbitrarily sized) squares, as well as
arbitrary sparse graphs with bounded VC-dimension.

\begin{theorem} Let $G$ be a graph on $n$ vertices. We can compute the diameter of $G$ by Las Vegas randomized algorithms in:
\begin{itemize}
    \item $\OOO(n^{2-1/{18}})$ time if $G$ is the intersection graph of unit disks, and
    \item $\OO(n^{2-1/{12}})$ time if $G$ is the intersection graph of axis-aligned squares.  For unit-square graphs, the running time is $\OOO(n^{2-1/{8}})$.
    \item $\OO(mn^{1-1/(2d)})$ time if $G$ has $m$ edges and 
    VC-dimension $d$.  For the special case of $K_h$-minor-free graphs for a fixed $h$, the running time becomes $\OO(n^{2-1/(2h-2)})$.
\end{itemize}
\end{theorem}

See \Cref{table:diam-intro} for the summary of our results on the diameter problem in comparison with previous work.  (Incidentally, our result even slightly improves previous time bounds in the special case of $K_h$-minor-free graphs.  The fact that the exponent of our algorithm for unit disks is the same as in Chang, Gao, and Le's $+1$-approximation algorithm~\cite{CGL24} is a complete coincidence---the algorithms are very different.)

\subsection{Technical Overview}

All previous subquadratic diameter algorithms for planar and minor-free graphs for arbitrary diameters~\cite{Cabello2018-gz,Gawrychowski2018-zy,le2023vc,ducoffe2022diameter,CGL24} use
sublinear separators (or $r$-divisions), which are not available for geometric intersection graphs that could be dense.
A key highlight of our framework is that we completely bypass sublinear separators!
Instead, we use \EMPH{low-diameter decompositions (LDD)}.
LDDs have been used in recent breakthrough results, such as negative-weight shortest paths~\cite{BNW22} and $(2-\e)$-approximation for vertex cover on string graphs~\cite{lps+-1avcs-2024}
(see the references in~\cite{BNW22} for more background).
We stress that we only need the most elementary, non-probabilistic form of LDDs (dating back to \cite{Awerbuch85}), which are constructible simply by a number of ``truncated'' BFSes, and do not require expanders or flows.  In some ways, they are even simpler than planar-graph separators or $r$-divisions.

In addition to LDD, our framework incorporates many new ideas about
the usage of \EMPH{bounded VC-dimension} as well as the design of \EMPH{geometric data structures}.
We will describe all three components of our framework in a little more detail below.

\begin{table}
\small\centering
\renewcommand{\arraystretch}{1.5}
\begin{tabular}{r|cl:cl}
graph class & \multicolumn{2}{c}{best previous} & new 
\\
\hline
planar & $\OO(n^{5/3})$ & \cite{Cabello2018-gz,Gawrychowski2018-zy} &  \\
$K_{h}$-minor-free & $\OO(n^{2-1/(3h-1)})$ & \cite{ducoffe2022diameter,le2023vc} & $\OO(n^{2-1/(2h-2)})$ \\
VC-dim.-bounded & $\OO(\min\{ Dmn^{1-1/d},\ mn\})$ & \cite{ducoffe2022diameter,duraj2023better} & $\OO(mn^{1-1/(2d)})$ \\
\hdashline
unit-square & $\OO(\min\{ Dn^{7/4},\ n^2\})$ & \cite{duraj2023better} & $\OOO(n^{2-1/8})$ \\
square & $\OO(n^2)$ & \cite{Chan2017-oa} & $\OO(n^{2-1/12})$ \\
unit-disk & $O(n^2\sqrt{\frac{\log\log n}{\log n}})$ & \cite{Chan2016-sy} & $\OOO(n^{2-1/18})$ 
\end{tabular}

\caption{Time bounds of exact diameter algorithms for different classes of unweighted graphs.  
Here, $n$ is the number of vertices, $m$ is the number of edges, $D$ is the diameter, and $d$ is the (generalized distance) VC-dimension. Squares are axis-aligned.}
\label{table:diam-intro}
\end{table}

\paragraph{Component 1: Low-diameter decomposition.} 
For a given parameter $\Delta > 0$, a \EMPH{low-diameter decomposition (LDD)} decomposes the input graph into \emph{pieces}  of diameter at most $\Delta$ such that the total number of boundary vertices of all the pieces is $\OO(n/\Delta)$. 
(It is helpful to imagine choosing $\Delta = n^{\delta}$ for some small constant $\delta$, and hence the number of boundary vertices is truly sublinear.) 
The ability to control the total number of boundary vertices is reminiscent of $r$-division~\cite{Frederickson1987} used for diameter computation in planar~\cite{Cabello2018-gz,Gawrychowski2018-zy} and minor-free graphs~\cite{Le2022-cl}, but an important difference is that a piece in an LDD could have up to $\Omega(n)$ vertices, while in an $r$-division, every piece has truly sublinear size (for a typical choice of $r$). 
LDDs can be computed in $\OO(m)$ time for general graphs and $\OO(n)$ time for many classes of intersection graphs, as we will show (in \Cref{sec:low-diameter}).

\paragraph{Component 2: Bounded VC-dimension and stabbing paths.} 
Since any sparse graph has a good low-diameter decomposition, an LDD itself is not sufficient for constructing truly subquadratic algorithms due to the aforementioned conditional lower bound based on SETH~\cite{Roditty2013-zr}. 
A recent line of work on the diameter problem has hinted at bounded VC-dimension as an overarching property: 
planar graphs (more generally, 
minor-free graphs)~\cite{Chepoi2007,Bousquet2015,Li2019-li,ducoffe2022diameter,le2023vc} and 
intersection graphs of pseudo-disks~\cite{AbuAffash2021,duraj2023better,CGL24} (in particular, disks and squares) have bounded VC-dimension. 
Thus, we also assume that the input graph has a bounded VC-dimension. 

Given a set system  $(U, \mathcal{F})$ with a ground set $U$ and a family $\mathcal{F}$ of subsets of $U$, its \EMPH{VC-dimension} is the cardinality of the largest $S\subseteq U$ such that $S$ is \EMPH{shattered} by $\mathcal{F}$---for every $S'\subseteq S$, there is some $X\in \mathcal{F}$ such that $X\cap S = S'$. 
Given a graph $G$, there are several different ways to form a set system of bounded VC-dimension; see \Cref{sec:prelim}.  The simplest one is the set system of neighborhood balls $(V_G, \{N^r[v]\}_{r \geq 0})$: we say that a graph $G$ has \EMPH{VC-dimension}%
\footnote{A more precise terminology is \EMPH{distance VC-dimension} at most $d$; see \Cref{sec:prelim} for clarification.} 
at most $d$ if its system of neighborhood balls has VC-dimension at most $d$. ($N^r[v]$ is the set of all vertices that are at a distance at most $r$ from $v$, including $v$ itself.) 
It was known that planar graphs have VC-dimension at most $4$; $K_h$-minor-free graphs have VC-dimension at most $h-1$; and intersection graphs of pseudo-disks have VC-dimension at most $4$~\cite{CGL24}. 

There are two main ways that VC-dimension was used in the diameter computation: 
(1) \EMPH{stabbing path}: 
constructing a path that stabs each neighborhood ball $N^r[v]$ a sublinear number of times (in the worst case or on average), and 
(2) \EMPH{distance compression:} showing that there are few different distance vectors to a fixed set of important vertices (i.e., the boundary of a piece in an $r$-division). 
The first approach has been very successful in the \emph{low-diameter regime}: computing the diameter in time $\OO(Dn^{2-\eps_d})$ where $\eps_d$ is a constant depending on the VC-dimension $d$~\cite{ducoffe2022diameter,duraj2023better}. 
The second approach works for the arbitrary-diameter regime, but either requires sublinear separators~\cite{le2023vc} or allows distance approximation~\cite{CGL24}. 
We overcome the limitation and inherent obstacles from both approaches and devise a method in the presence of low-diameter decomposition to compute stabbing paths even when the graph diameter is large.
(In certain applications, we also manage to perform distance compression exactly without the presence of separators.)

The basic idea of the stabbing path approach is to order the vertices from $1$ to $n$, 
in such a way that each neighborhood ball $N^r[v]$ of radius $r$ can be represented as a union of $\OO(n^{1-1/d})$ many intervals on the stabbing path.%
\footnote{For a simpler exposition, we assume the worst-case bound on $\OO(n^{1-1/d})$ on the number of intervals representing $N^r[v]$. In the detailed implementation, we work with an \EMPH{amortized bound} which is faster to compute.}
The existence of a spanning path with $O(n^{1-1/d})$ stabbing (or ``crossing'') number was first shown in a seminal paper by Chazelle and Welzl~\cite{cw-qorss-1989}, and had found numerous applications in computational geometry, for example, in geometric range searching.
Constructing a good stabbing path may seem to require knowledge of the entire set system of balls $N^r[v]$ in the first place (which we do not have, since our problem is to compute all $N^r[v]$!).  Fortunately, it turns out that by known random sampling techniques\footnote{
This is where all our algorithms use Las Vegas randomization.
}, we only need to evaluate a small subset of balls to compute a good stabbing path; for example, in the unit disk or square case, the construction time is $\OO(n^{1+1/d})$ (more generally, the construction time is $\OO(n\rho)$ for stabbing number $\OO(n/\rho +\rho^{d-1})$ for a trade-off parameter $\rho$).

\begin{figure}[h!]
\centering
\includegraphics[scale=1]{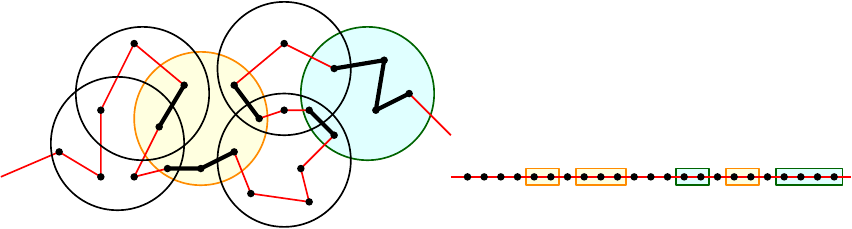}
\caption{Stabbing path and interval representation of disks.  The yellow disk is represented by three yellow intervals, and the green disk is represented by two intervals. The intervals representing different disks could overlap.}
\end{figure}

Let's say $v$ is one vertex in a diametral pair, whose shortest path distance realizes the diameter $D$.
Given the interval representation, we can check in time linear to the number of intervals ($\OO(n^{1-1/d})$) whether the union of all the intervals (and hence $N^r[v]$) covers $[1:n]$. 
If the answer is yes, then $r$ is at least the diameter $D$.   
By iterating through every vertex~$v$ as a potential endpoint of a diametral path, we can check if $r$ is greater than the true diameter $D$ in $\OO(n^{2-1/d})$ time. 
To compute the interval representations of $N^r[v]$ for all vertices in $V$, the pioneering work of Ducoffe, Habib, and Viennot~\cite{ducoffe2022diameter} introduced a \EMPH{ball growing} process: 
For each $r$, one computes the interval representations of $\set{N^r[v] : v\in V}$ from the interval representations of  $\set{ N^{r-1}[v] : v\in V}$ via the identity $N^r[v] = \bigcup_{u\in N[v]} \hat N^{r-1}[u]$.
(For the base case $r=0$, $N^{0}[v] = \set{v}$.)
This approach leads to running time $\smash{\OO(Dmn^{1-1/d})}$ for sparse graphs~\cite{duraj2023better}. 
For geometric intersection graphs, we cannot afford to access the neighbors of every vertex as it would result in $\Omega(n^2)$ time, and hence, different ideas are needed to avoid explicitly accessing neighbors. 
For the intersection graphs of unit squares, Duraj, Konieczny, and Pot\c{e}pa~\cite{duraj2023better}
devised a certain ``neighbor-set data structure''
to achieve total running time $\OO(Dn^{2-1/4})$.  
The factor of $D$ in the running time seems inherent to this approach, and $D$ could be as big as $\Omega(n)$.
Furthermore, their data structure does not work for arbitrary squares or unit disks.

To handle possibly large $D$, our new approach is to \emph{combine with low-diameter decomposition}.  
First, by computing BFS trees from the boundary vertices of the LDD with parameter $\Delta$, we could compute an estimated value $\tau\in [D-\Delta: D]$. 
Note that there are only a truly sublinear number of boundary vertices, and hence, the total running time of this step remains truly subquadratic. 
(In the case of geometric intersection graphs, we can implement BFS in $\OO(n)$ time using known techniques such as bichromatic intersections~\cite{Chan2016-sy,Klost2023-xs,CGL24}. 
See \Cref{subsec:LDD-geometric} for details.)
If we have $\set{N^{\tau}[v] : v\in V}$, then we only need to grow balls for another $\Delta$ iterations (here $\Delta \ll D$). 
However, we do not have access to  $\set{N^{\tau}[v] : v\in V}$ explicitly. 
Our key idea here is to define a \emph{modified} neighborhood ball $\hat{N}^r[v]$ in a way that we can initialize $\hat{N}^{\tau}[v] = \varnothing$ to kick-start the ball growing process, and at the same time the information computed is sufficient to answer the question about diameter $D$. 
Therefore, the precise definition of $\hat{N}^r[v]$ is somewhat tricky, tying directly to the pieces in the LDD; see \Cref{sec:framework} for the details.  
For sparse graphs (with bounded VC-dimension), we can afford to access the neighbors of every vertex explicitly. 
Hence, we could simply grow the modified balls $\set{\hat{N}^r[v]}$ in $O(\Delta)$ rounds in  $\OO(\Delta \cdot mn^{1-1/d})$ time. 
By choosing $\Delta = n^{1/2d}$ (to balance with the $\OO(mn/\Delta)$ running time of BFS computation), this leads to a relatively simple diameter algorithm running in $\OO(mn^{1-1/(2d)})$ time, which is truly subquadratic for the arbitrary-diameter regime.

\paragraph{Component 3: Geometric data structures.}  
For geometric intersection graphs that are not sparse, we cannot afford to access the neighbors directly even with the modified neighborhoods, so more ideas are needed. 
The data structure subproblem for the ball growing step we need to solve is the following: assume the interval representations of $\hat N^{r-1}[v]$ for every vertex $v$ are precomputed and stored;
given a query object $s$, compute the union of intervals in $\hat N^{r-1}[v]$ for all objects $v$ that intersect~$s$.
We can reduce this problem to the following:

\begin{restatable}[Interval Cover]{problem}{IntervalCover}
\label{def:DS-1} 
Given a set of $N$ objects $\mathcal{O}$ and each object $o\in \mathcal{O}$ is  associated with an interval $I_o\subseteq [1:n]$. Design a data structure to answer the following query:
\begin{itemize}
    \item \textsc{Covers?}$(q,I)$:  Given a query object $q$ and a query interval  $I\subseteq [1:n]$, decide whether the union of intervals associated with the objects intersecting\footnote{Here we mean the objects intersect, not their associated intervals.} $q$ in  $\mathcal{O}$ covers the whole $I$.  
\end{itemize}
\end{restatable}

\Cref{def:DS-1} can be viewed as a generalization of \emph{range searching}~\cite{AgarwalE99}:
given a query object $q$, find the objects intersecting $q$. 
In the computational geometry literature,
\emph{colored} variants of range searching have been studied~\cite{gjs-frgis-1995,krsv-ecorc-2008,GuptaSURVEY,chn-frcrs-2020}.  The above problem is an even more challenging variant, where each object is equipped with not a color but an interval.  This interesting generalization has not been considered before, to the best of our knowledge. 
(There have been some prior works on \emph{time-windowed} geometric data structures \cite{BannisterDGST14,BokalCE15,ChanHP19}, but typically queries are associated with a time interval but not the input objects; even more crucially, the queries in those works are mainly about whether a property is true for \emph{some} time value in a query interval $I$, rather than for \emph{all} time values in $I$.)
One reason the problem is more challenging than standard range searching is that it is not \emph{decomposable} (if the input set is divided into two subsets, knowing the answers of a query for the subsets does not necessarily help with the overall answer).

We note that en route to their unit square result, Duraj, Konieczny, and Pot\c{e}pa~\cite{duraj2023better} also formulated
a non-standard geometric ``neighbor-set data structure'' problem, but their formulation appears more complicated, as they (and  Ducoffe, Habib, and Viennot~\cite{ducoffe2022diameter} earlier) worked with symmetric differences of neighborhood sets.  Our approach using interval representations is in some sense ``dual'' to these previous approaches, and is more natural, leading to a geometric data structure problem that is simpler to state.

For unit squares, we give a solution to \Cref{def:DS-1}
with $N^{1+o(1)}$ preprocessing time and $N^{o(1)}$ query time.
Our data structure is deterministic, in contrast to Duraj et al.'s, which uses hashing techniques and inherently requires Monte Carlo randomization.
For arbitrary squares, Duraj et al.'s data structure approach does not work at all.  Although we are not able to obtain $N^{o(1)}$ query time for \Cref{def:DS-1} either, we propose a simple method which divides the range $[1:n]$ into blocks of size $b$, and builds a
data structure for \emph{rainbow colored intersection searching} (a version of colored range searching) for each block.  (See \Cref{subsec:DS-reduction} for details.)
This yields $\OO(N\cdot b)$ preprocessing and $\OO(L/b)$ query time, where $L$ is the length of the query interval.  This trade-off turns out to be sufficient to obtain a truly subquadratic algorithm in the end, for an appropriate choice of the parameter $b$.

For unit disks, \Cref{def:DS-1} is related to the well-known \emph{Hopcroft problem}\footnote{The Hopcroft problem tests, for a given system of points and lines in the Euclidean plane, whether any point lies on any line. The total number of points and lines is assumed to be $n$.}, and hence a query time $o(N^{1/3})$ appears unlikely~\cite{Erickson1996}; however, to obtain a truly subquadratic time for diameter, we need $O(N^\delta)$ query time for tiny $\delta>0$ (since the total number of input and query intervals is $\Omega(n^{2-1/4})$ or worse in our application). 
We circumvent this issue entirely by partitioning the given set of unit disks into constantly many \emph{modulo classes} (i.e., we
partition the plane into cells of constant side-length (say $1/2$), and take modulo classes of the index pairs of the cells). This way, if we take one cell $\square$, the collection of disks from the same modulo class intersecting $\square$ forms a \emph{pseudoline arrangement}. 
When input disks are restricted to one modulo class,
we are able to solve \Cref{def:DS-1} with $N^{1+o(1)}$ preprocessing time and $N^{o(1)}$ query time.\footnote{We do not break the Hopcroft problem's lower bound as we only solve the data structure problem for one modulo class.}
These data structure results may be of independent interest to computational geometers. 
They do not follow directly from existing techniques.  Instead, we propose a clever recursion, repeatedly and alternately taking \emph{lower envelopes and upper envelopes} of pseudo-segments~\cite{AGARWAL20001,Pettie_DSseq_15}.  (Experts in geometric data structure may find this part interesting, and are encouraged to read Section~\ref{SS:ds-unitdisk} for the details.)

\paragraph{Additional complications for unit disks.}
The fact that we have efficient data structures for unit disks only when restricted to a fixed modulo class creates a number of extra technical challenges:

\begin{itemize}
    \item Because we can only take union over balls from a fixed modulo class, the intermediate sets are no longer neighborhood balls, i.e., we need to work with a new set system.
    Fortunately, we can still prove that the (dual) VC-dimension of the new set system is at most 4, but \emph{only when the balls have the same radius $r$}.
    \item This condition in turn forces us to change the stabbing path---and all of its associated interval representations---every time we increment $r$.
    Fortunately, we show that the interval representations can be updated efficiently using random sampling techniques (with slightly worse amortized stabbing number $\OO(n/\rho + \rho^d)$).  
    \item At intermediate steps, we may now need to work with balls from two or three set systems across different types.  Fortunately, the combined set systems still have bounded VC-dimension (at most~8).
    \item The extra overhead in switching stabbing paths is too costly since each stabbing path computation costs $O(n\rho)$ time, and we have to compute for $\OO(n/\Delta)$ pieces and $O(\Delta)$ rounds.
    To achieve overall subquadratic time, we only work with pieces larger than a certain threshold $A$; for small pieces, we need to switch to a different method (based on distance compression),
    which achieves running time 
    $\OO(n\cdot |\bdry P| + |P|\cdot ( |P| + (|\bdry P| \Delta)^d))$ for each piece $P$ of size at most $A$ with boundary $\bdry P$.
\end{itemize}
All these details are explained in \Cref{sec:unit-disks}, but to illustrate the intricacies of the overall algorithm to the curious readers, the time bound for diameter for unit-disk graphs has the following form, where the sums are over all pieces $P$ of the LDD (which satisfies $\sum_P |P| = O(n)$ and $\sum_P |\bdry P| = \OO(n/\Delta)$):
\begin{align*} 
& \OOO\left( \Delta\cdot  n\rho + 
\sum_{P:\, |P|>A} \Paren{\Big. |\bd P| \cdot n + \Delta \cdot (n + |P|\cdot (n/\rho + \rho^8)) } +  \sum_{P:\, |P|\le A} \Paren{\Big. n \cdot |\bdry P| + |P| \cdot (|P| + (|\bdry P| \Delta)^4 )  } \right).
\end{align*}
Balancing cost by setting parameters $\Delta=n^{1/18}$ and $\rho=A=\Delta^2$ then yields $\OOO(n^{2-1/18})$. 
(Other variants of the algorithm for different graph classes and other related problems will have different expressions and different settings of parameters.)

\subsection{Other Distance-related Problems} 

Our framework for computing diameter naturally opens up the possibility of solving other distance-related problems. 
Here, we focus on three well-studied problems:
all-vertex eccentricities, exact distance oracles, and Wiener index.

\begin{table}
\small\centering
\renewcommand{\arraystretch}{1.5}
\begin{tabular}{r|cl:cl}
graph class & \multicolumn{2}{c}{best previous} & new
\\
\hline
planar & $\OOO(n^{3/2})$, $\OOO(n)$  & \cite{Charalampopoulos2023} &   \\
$K_{h}$-minor-free & $\OO(n^{2-1/(3h-1)})$ & \cite{le2023vc} &  
\\
VC-dim.-bounded & $O(mn),\ O(n^2)$ & folklore & $\OO(mn^{1-1/(4d+1)})$ \\
\hdashline
unit-square & $\OO(n^2)$ & \cite{Chan2017-oa} & $\OOO(n^{2-1/16})$ \\
square & $\OO(n^2)$ & \cite{Chan2017-oa} & $\OO(n^{2-1/20})$ \\
unit-disk & $O(n^2\sqrt{\frac{\log\log n}{\log n}})$ & \cite{Chan2016-sy} & $\OOO(n^{2-1/20})$ 
\end{tabular}
\caption{Construction time and space bounds of exact distance oracles for different classes of unweighted undirected graphs, with $\OO(1)$ query time.  We write out both construction time and
space bounds only when they are different.}
\label{table:oracle-intro}
\end{table}

\paragraph{Eccentricities.}
To highlight the new challenges beyond diameter computation, let us begin with eccentricities.
The \EMPH{eccentricity} of a vertex $v$, denoted by \EMPH{$\ecc(v)$}, is the maximum distance from $v$ to any other vertex in $G$. 
Our goal is to compute $\ecc(v)$ for every $v\in V_G$ in truly subquadratic time. 
Observe that the diameter is the maximum eccentricity and hence, computing all eccentricities is often more difficult. 

For computing diameter, we kick-start the ball growing process with radius $\tau\in [D-\Delta: D]$ and therefore we only need to grow in $O(\Delta)$ interactions. 
The key challenge in computing eccentricities is that $\ecc(v)$ of some vertex $v$ could be as small as $D/2$, and hence any ball growing process has to cover radii in the entire range $[D/2: D]$, which can be as large is $\Omega(n)$.
Interestingly, our framework for the diameter problem also points us to a way to resolve this issue. 
Specifically, we grow the modified neighborhood ball $\hat{N}^r[v]$ only for vertices in the same piece $P$ of the low-diameter decomposition. 
The observation is that for any two vertices $u$ and $v$ in $P$, $|\ecc(u) - \ecc(v)| = O(\Delta)$. 
Hence, to restrict to computing eccentricities of vertices in $P$, it suffices to grow modified neighborhood balls in $O(\Delta)$ steps. 
(For different pieces, the ranges of radii could be vastly different.) 
As the range of radii is piece-specific, the stabbing path data structure also has to be piece-specific instead of being ``global'' as in the case of computing graph diameter. 
Our results are summarized in the following theorem.

\begin{theorem} Let $G$ be a graph on $n$ vertices. We can compute all-vertex eccentricities of $G$ by Las Vegas randomized algorithms in: 
\begin{itemize}
    \item $\OOO(n^{2-1/{20}})$ time if $G$ is the intersection graph of unit disks, and
    \item $\OO(n^{2-1/{12}})$ time if $G$ is the intersection graph of axis-aligned squares.  For unit-square graphs, the running time is $\OOO(n^{2-1/{8}})$.
    \item $\OO(mn^{1-1/(2d)})$ time if $G$ has $m$ edges and (generalized distance) VC-dimension $d$.
\end{itemize}
\end{theorem}

\paragraph{Exact distance oracle.}
An \EMPH{exact distance oracle} is a data structure that, when given a pair of vertices, returns the shortest path distance of the vertices quickly. 
Our goal is to construct an oracle with a truly subquadratic \emph{space}. 
For geometric intersection graphs, known oracles with a truly subquadratic space can answer a distance query approximately within an additive error of $1$~\cite{ABT24,CGL24}; for sparse graphs, the query time is close to linear ($\Omega(n^{1-\eps_d})$) for some small constant $\eps_d = 1/2^{O(d)}$ depending on the VC-dimension $d$~\cite{ducoffe2022diameter}. 

For square and unit-disk graphs and sparse graphs with bounded VC-dimension, we provide an exact oracle with a truly subquadratic space and polylogarithmic query time.  
Furthermore, our oracle can be constructed in truly subquadratic time; therefore, our result can be interpreted as solving the all-pairs shortest-path problem in truly subquadratic time. 
(Of course any such algorithm has to output an implicit representation of the shortest distances since the explicit output size is $\Omega(n^2)$.)

Constructing an exact distance oracle is more difficult than computing all-vertex eccentricities: the queried distance range is $[0: n]$. In computing diameter and eccentricities, the modified ball $\hat{N}^r[s]$ is a subset of the true neighborhood ball $N^{r}[s]$ and we compute $\hat{N}^r[s]$ for all $r\in [\ecc(s)-O(\Delta): \ecc(s) + \Delta]$. 
However, if we query distance between $s$ and $t$ where $d_G(s,t) \ll \ecc(s)-O(\Delta)$, then knowing the true neighborhood ball $N^{r}[s]$ for $r\geq \ecc(s)-O(\Delta)$ (let alone its subset) does not tell us anything about $d_G(s,t)$. 
Our idea is to assign weights to vertices of $G$ appropriately and incorporate vertex weight in the definition of $\hat{N}^r[s]$, so that every vertex $t$ belongs to $\hat{N}^r[s]$ for some value of $r \in [-\Delta: \Delta]$; the radius could be negative, which is somewhat counterintuitive.  
As the range of the (weighted) radii is now $\Theta(\Delta)$, the ideas we develop for computing the diameter and eccentricities now can be applied here.  
As distances with vertex weights are closely connected to the notion of \emph{generalized} VC-dimension (formally defined in \Cref{sec:prelim}), we assume the input graph has a bounded generalized VC-dimension in the case of sparse graphs. 
All other graphs, such as geometric intersection graphs and minor-free graphs, have their generalized VC-dimension equal to the regular VC-dimension (of the neighborhood ball system). 

\medskip
All of these ideas lead to our exact distance oracles for various types of graphs.  See \Cref{table:oracle-intro} for a comparison of existing results and ours.

\begin{theorem}\label{thm:oracle-main} Let $G$ be a graph on $n$ vertices. We can compute an exact distance oracles for $G$ (by randomized Las Vegas algorithms)  with the following guarantees:
\begin{itemize}
    \item $\OOO(n^{2-1/20})$ construction time and size
     and $\OO(1)$ query time if $G$ is a unit-disk graph.
    \item  $\OO(n^{2-1/20})$ construction time and size
     and $\OO(1)$ query time if $G$ is a square graph.  For unit-square graphs, the construction time and size can be improved to $\OO(n^{2-1/{16}})$.
    \item $\OO(mn^{1-1/(4d+1)})$ construction time, $\OO(n^{2-1/(4d+1)})$ size, and $\OO(1)$ query time if $G$ has $m$ edges and (generalized distance) VC-dimension $d$.
\end{itemize}
\end{theorem}

Interestingly, if we ignore construction time, the above theorem implies the existence of subquadratic-size, $\OO(1)$-time distance oracles for all (not necessarily sparse) graphs with bounded (generalized distance) VC-dimension, in particular, all pseudo-disk graphs.

\paragraph{Wiener index.~} 
The Wiener index of a graph $G$ is the sum of the distances between all pairs of vertices. Computing the Wiener index has been studied~\cite{CK97,CK09,WulffNilsen2009}; truly subquadratic algorithms are only known for planar and minor-free graphs~\cite{Cabello2018-gz,Gawrychowski2018-zy,le2023vc}.  Here we provide the first such algorithms for graphs with bounded generalized VC-dimension and several geometric intersection graphs. Indeed, the algorithms for  Wiener index are simple corollaries of our algorithms for exact distance oracles in \Cref{thm:oracle-main} and therefore have the same running time guarantees.

\begin{theorem}\label{thm:Weiner-main} Let $G$ be a graph on $n$ vertices. We can compute the Weiner index of $G$ (by randomized Las Vegas algorithms) in: 
\begin{itemize}
    \item $\OOO(n^{2-1/20})$ time if $G$ is the intersection graph of unit disks.
    \item  $\OO(n^{2-1/20})$ time if $G$ is the intersection graph of axis-aligned squares.  For unit-square graphs, the running time is $\OO(n^{2-1/{16}})$.
    \item $\OO(mn^{1-1/(4d+1)})$ time if $G$ has $m$ edges and (generalized distance) VC-dimension $d$.
\end{itemize}
\end{theorem}

\section{Preliminaries}
\label{sec:prelim}
\subsection{Graphs and Low-diameter Decomposition}
\label{subsec:prelim-graph-ldd}

\paragraph*{Graph notation.} 
Let $G = (V_G, E_G)$ be an unweighted undirected graph with $n$ vertices and $m$ edges. For two vertices $u, v\in V$, let $d_G(u, v)$ denote the distance between $u$ and $v$ in $G$. Often we will omit the subscript and simply write $d(u,v)$ when the graph $G$ is clear.
The \EMPH{neighborhood} of a vertex $v\in V_G$ is the set of vertices that are distance at most $1$ to $v$, denoted by $\EMPH{$N[v]$} \coloneqq \set{u\in V_G: d(u,v) \le 1}$. 
The \EMPH{$k$-neighborhood ball} of a vertex $v\in V_G$ is the set of vertices with distance at most $k$ from $v$, denoted by $\EMPH{$N^k[v]$} \coloneqq \set{u\in V: d(u,v) \le k}$.
(Notice that $N[v] = N^1[v]$ and $N^k[v] = N[N^{k-1}[v]]$.)
Define the set of $k$-neighborhood balls as $\EMPH{$\cN^k_G$} \coloneqq \set{N^k[v]: v\in V}$, and the set of all neighborhoods balls as $\EMPH{$\cB_G$} \coloneqq \bigcup_k \cN^k_G$.

\paragraph*{Geometric intersection graphs.~} 
Consider a set $S$ of $n$ geometric objects in the plane. We define the \EMPH{geometric intersection graph} $G$ of $S$ as the graph obtained by creating a vertex for every geometric object, and connecting two geometric objects if they intersect. When $S$ consists of unit disks, i.e., disks of radius $1$, we refer to the geometric intersection graph $G$ as a \EMPH{unit-disk graph}. If $S$ consists of axis-aligned unit squares, we refer to the geometric intersection graph $G$ as a \EMPH{unit-square graph}. We will also consider when $S$ consists of axis-aligned squares (of arbitrary size). We refer to such graphs as \EMPH{square graphs}. In \Cref{subsec:LDD-geometric}, we describe a near-linear time algorithm for computing a BFS tree for square graphs, as stated below; the algorithm for unit-disk graphs is known~\cite{Klost2023-xs}.

\begin{lemma}\label{lm:SSP-geo-intersection} Let $G$ be the geometric intersection graph of squares or unit disks with $n$ vertices. We can compute a BFS tree from any given vertex of $G$ in $\OO(n)$ time.
\end{lemma}

\paragraph*{Low-diameter decomposition.}
Let $G$ be a graph with $n$ vertices and  $\Delta > 0$ be a diameter parameter. A  \EMPH{low-diameter decomposition} (LDD) of $G$ with parameter $\Delta$ is a decomposition of the vertex set $V$ into disjoint sets $V = V_1\cup \ldots \cup V_k$ and corresponding induced subgraphs $P_i \coloneqq G[V_i]$ called \EMPH{pieces}, such that:

\begin{itemize}
    \item \textbf{Low diameter:}  Piece $P_i$ is a single connected component of (strong) diameter\footnote{By strong diameter we mean that the shortest path between any two vertices in $P_i$ \emph{within} the subgraph $P_i$ is at most $\Delta$.} at most $\Delta$. 
    \item \textbf{Small boundary:}  Denote the boundary vertices of $P_i$ as \EMPH{$\bdry P_i$}, that is, the subset of vertices of $P_i$ that has an edge to a vertex in $V_G \setminus V_i$.  The decomposition satisfies $\sum_{i=1}^k |\bdry P_i| = \OO(n/\Delta)$.
    \item \textbf{No small pieces:} Each piece has size at least $\widetilde{\Omega}(\Delta)$.
\end{itemize}

We show in \Cref{sec:low-diameter} that such a decomposition always exists. 
Furthermore, in \Cref{subsec:LDD-sparse}, we show an efficient algorithm for computing this decomposition.
\begin{restatable}{theorem}{lddsparse}
\label{thm:ldd-sparse}
Let $G$ be a graph with $n$ vertices and $m$ edges. For any parameter $24\log n < \Delta \le n$, we can compute a low-diameter decomposition for $G$ in $O(m+n)$ time.
\end{restatable}

For unit-disk graphs and square graphs, we prove in \Cref{subsec:LDD-geometric} that the low-diameter decomposition is efficiently computable in near-linear time.
\begin{restatable}{theorem}{lddgeo}
\label{thm:ldd-geo}
Let $G$ be an intersection graph of $n$ unit disks or an intersection graph of $n$ axis-aligned squares. For any parameter $24\log n < \Delta \le n$, we can compute a low-diameter decomposition for $G$ in $\OO(n)$ time.
\end{restatable}

\subsection{VC-dimension}
A \EMPH{set system} is a pair $(X, \cS)$, consisting of a ground set $X$ and a collection of ranges that are subsets of $X$; in notation, $\cS \subseteq 2^{X}$. 
A subset $Y \subseteq X$ is said to be \EMPH{shattered} by $\cS$ if the collection $\{Y\cap S: S\in\cS\} = 2^{Y}$, that is, all possible subsets of $Y$ can be obtained by $\cS$.
The \EMPH{shatter function}, denoted by $\pi_{(X, \cS)}(k)$ is the largest number of sets that is created by the set system when restricted to $Y\subseteq X$ of size $k$. Formally it is:
\[ \pi_{(X,\cS)}(k) = \max_{\substack{Y\subseteq X \\ |Y| = k}}\left|\{Y\cap S: S\in \cS\}\right|.\]
The \EMPH{shatter dimension} of a set system is the smallest value $d$ such that $\pi_{(X,\cS)}(k) = O(k^d)$ for all $k$.
The \EMPH{VC-dimension} of a set system $(X,\cS)$ is the size of the largest subset of $Y\subseteq X$ that can be shattered by $\cS$.
The \EMPH{dual set system} of $(X, \cS)$ is the set system $(\cS^*, X^*)$, where the ground set $\cS^* = \{w_S:S\in \cS\}$ consists of elements indexed by $\cS$,
and each $s\in S$ induces a range $s^*=\{w_S\in \cS^*: S \ni s \}$ in $\cS^*$.
The \EMPH{dual VC-dimension} of a $(X, \cS)$ is the VC-dimension of the dual set system, and analogously the \EMPH{dual shatter dimension} is the shatter dimension of the dual set system.
We state some well-known results \cite{har2011geometric}.
\begin{lemma}\label{lem:vcfacts}
Let $(X, \cS)$ be a set system of VC-dimension $d$. The following is true:
\begin{enumerate}
    \item The dual set system $(\cS^*, X^*)$ has VC-dimension at most $2^{d+1}-1$.
    \item For $Y\subseteq X$, the set system $(Y, \cS)$ has VC-dimension at most $d$.
    \item (Sauer-Shelah Lemma~\cite{Shelah1972-ke, Sauer1972-ci}.)~ If $|X| \le n$ then $|\cS| \le O(n^d)$, so the shatter dimension of $(X, \cS)$ is at most $d$.
\end{enumerate}
\end{lemma}

\paragraph{VC-dimension in graphs.}
The \EMPH{$k$-distance VC-dimension} of a graph $G = (V_G, E_G)$ is the VC-dimension of the set system of $k$-neighborhood balls $(V_G, \cN^k_G)$. (Sometimes in the literature, e.g.,~\cite{ducoffe2022diameter}, the VC-dimension of $G$ is defined to be the $1$-distance VC-dimension.) The \EMPH{distance VC-dimension} of $G$ is the VC-dimension of the set system of balls $(V_G, \cB_G)$.  
Observe that the $k$-neighborhood set system $(V_G, \cN^k_G)$ is equivalent to its dual, so the dual VC-dimension is the same as the primal. This is not the case for the set system of arbitrary balls since the ground set and the set of ranges have different sizes. 

Karczmarz and Zheng \cite{KZ25} introduced\footnote{\cite{KZ25} consider what they call a \emph{multiball} set system where the ground set is $V_G\times M$ for a set of real weights $M\subseteq \R$.} a natural generalization: a set system $(U, \GB_G)$ whose ground set is $U = V_G\times \Z = \{(u, r):u\in V_G, r \in \Z\}$, and the ranges $\GB_G$ consists of \EMPH{generalized neighborhood balls} for $v\in V_G$ and $k\in \Z$ of the form:
\[ \tilde N^k[v] \coloneqq \Set{\big. (u, r)\in V_G\times\Z: d(u, v) \le r+k}. \]
Note that values of $r$ and $k$ are allowed to be negative.
We call the VC-dimension of $(U, \GB_G)$ the \EMPH{generalized distance VC-dimension} of a graph. It can be observed that this set system is equivalent to its dual. Furthermore, we can observe the following relationship between these VC-dimensions.

\begin{observation}
   $k$-distance VC-dimension of $G$ 
$\le$ 
distance VC-dimension of $G$ 
$\le$
generalized distance VC-dimension of $G$. 
\end{observation}

Throughout this paper, when we refer to graphs of bounded VC-dimension, we will be referring to families of graphs whose generalized distance VC-dimension of the graph is bounded by an absolute constant. 
Many of our results can also be adapted with more work to graphs that have bounded $k$-distance VC-dimension for all $k$, or graphs with bounded distance VC-dimension. We will focus on generalized distance VC-dimension as it holds for minor-free graphs and the geometric intersection graphs we care about, and also leads to the simplest exposition of our ideas.

\paragraph{Connection to distance encoding VC-dimension.}
Distance encodings were used by Li and Parter \cite{Li2019-li} to compute the diameter in a planar graph. This was later modified to a more general setting by Le and Wulff-Nilsen \cite{le2023vc}, whose definition we present below (restricted to unweighted graphs).

\begin{definition}
Let $G = (V_G, E_G)$ be an undirected unweighted graph.
Let $M \subseteq \Z$ be a set of integers. Let $S\subseteq V_G$ be an ordered set of $\ell$ vertices $S=\{s_0,s_1, \dots, s_{\ell-1}\}$. 
For every vertex $v\in V_G$ define the set:
\[ 
X_{S,M}(v) \coloneqq \Set{\big. (s_i, \delta): s_i\in S, \delta\in M,  d(v, s_i) - d(v, s_0) \le \delta}. 
\]
Let $X_{S, M} \coloneqq \Set{\big. X_{S,M}(v): v\in V}$ be the set of subsets of the ground set $S\times M$.
The \EMPH{distance encoding VC-dimension} of $G$ is the maximum VC-dimension of set systems of the form $(S\times M, X_{S, M})$ for all possible $S$ and $M$. 
\end{definition}

Observe that the set $X_{S, M}(v)$ is isomorphic to $\tilde N^{d(v,s_0)}[v] \cap (S\times M)$. 
Restricting the ground set of the set system $(U,\GB_G)$ to $(S\times M, \GB_G)$ does not increase the VC-dimension by \Cref{lem:vcfacts}, so we conclude the following observation.

\begin{observation}
Distance encoding VC-dimension of $G$ 
$\le$
generalized distance VC-dimension of $G$.
\end{observation}

\paragraph*{Graphs of bounded generalized distance VC-dimension.}
It was shown by Chepoi, Estellon, and Vaxes \cite{Chepoi2007} that planar graphs have distance VC-dimension at most $4$ by explicitly constructing a $K_5$ minor (by contradiction). This argument was extended by Bousquet and Thomass\'e \cite{Bousquet2015} to show that $K_h$-minor-free graphs have distance VC-dimension at most $h-1$.
Le and Wulff-Nilsen \cite{le2023vc} used a variation of this argument to show that $ K_h$-minor-free graphs have distance encoding VC-dimension at most $h-1$.
This argument was adapted by Karczmarz and Zheng \cite{KZ25} to show that $K_h$-minor-free graphs have generalized distance VC-dimension at most $h-1$ as well. 

\begin{theorem}[\cite{KZ25}]
Any $ K_h$-minor-free graph has generalized distance VC-dimension at most $h-1$.
\end{theorem}

For unit-disk graphs, it was shown by Abu-Affash \etal~\cite{AbuAffash2021} that the distance VC-dimension is $4$. 
Later, by Chang, Gao, and Le \cite{CGL24}, the intersection graph of pseudo-disks has distance VC-dimension and distance encoding VC-dimension of $4$ as well. The bound on distance VC-dimension was also independently shown by Duraj, Konieczny, and Pot\k{e}pa \cite{duraj2023better} for intersection graphs of fixed translates of geometric objects in the plane.
The proof in~\cite{CGL24} can be easily adapted to also bound the generalized distance VC-dimension.

\begin{theorem}
\label{thm:pd_vcdim}
Any geometric intersection graph of pseudo-disks in the plane has generalized distance VC-dimension at most $4$.
\end{theorem}

\subsection{Stabbing Path and Interval Representation}
\label{SS:stabbing-path}

Let $(X,\cS)$ be a set system with $|X| \le n$ and $|\cS| \le m$.
Let $\spath$ be an ordering of the elements of $X$.
Given a set $S\in\cS$, define the \EMPH{$\spath$-interval representation $\Rep_\spath(S)$} ({$\spath$-representation} for short) as the collection of maximal contiguous subsequences of $\spath$---called \EMPH{intervals}---whose union is $S$.
The size of the representation \EMPH{$|\Rep_\spath(S)|$} refers to the number of such intervals.
For a parameter $1\le \rho \le m$, a \EMPH{$\rho$-stabbing path $\spath$} of a set system $(X,\cS)$ of dual VC-dimension $d$ is an ordering of $X$ such that $\sum_{S\in\cS} |\Rep_\spath(S)| = \OO(mn/\rho +m\rho^{d-1})$.
Observe that if $n^{1/d}\le m$, this quantity is minimized when $\rho = n^{1/d}$ so $\sum_{S\in\cS} |\Rep_\spath(S)| = \OO(mn^{1-1/d})$. We will sometimes refer to an $n^{1/d}$-stabbing path $\spath$ simply as a \EMPH{stabbing path}.
We assume the existence of an \EMPH{element reporting oracle} that, given $S\in\cS$, can enumerate all elements of $S$ in \EMPH{$T_0(n)$} time, where $T_0(n) \ge n$.

\medskip
In \Cref{SS:reordering}, we show the following lemma to construct $\rho$-stabbing paths with high probability\footnote{In this paper we say an event $E$ happens \EMPH{with high probability} if $\Pr[E] \ge 1 - n^{-c}$ for some big enough constant $c$.}.

\begin{restatable}{lemma}{stabbingpathgeneral}
\label{lem:stabbing-path-general}
Let $(X,\cS)$ be a set system with $|X| \le n$ and $|\cS| \le m$ with dual shatter dimension at most $d$. For any parameter $1\le \rho \le m$, 
we can construct a stabbing path of $(X,\cS)$, that is, an ordering $\spath$ of $X$ such that 
$\sum_{S\in\cS}|\Rep_\spath(S)| = \OO(mn/\rho + m\rho^{d-1})$ in $\OO(T_0(n) \cdot \rho)$ time with high probability.     
\end{restatable}

\subsection{Geometric Data Structures}\label{subsec:DS-prelim}

We consider three different geometric data structures in decreasing difficulty that we will use in our algorithms, and the reduction from difficult problems to easier problems. We provide the details of the reductions in \Cref{subsec:DS-reduction}.

\paragraph{Interval searching.~} The interval searching problem directly captures the ball growing process for various objects in the frameworks  we study in \Cref{sec:framework} and \Cref{sec:framework-do}. 

\begin{problem}[Interval Searching]\label{prob:DS0} Let $\mathcal{O}_{IS}$ be a given set of objects, where each object $o\in \mathcal{O}_{IS}$ is associated with a set of intervals (of integer points) of $[1:n]$, denoted by \EMPH{$\mathcal{I}_o$}.  
Design a data structure that answers the following query: 
\begin{itemize}
    \item \textsc{IntervalSearch}$(q)$:  Given an object $q$, return (the interval representation of) all the integer points in $[1:n]$ associated with objects in $\mathcal{O}_{IS}$ that intersect $q$.
\end{itemize}
\end{problem}

For each query object $q$, let \EMPH{$\mathcal{I}_{out}(q)$} be the set of intervals representing all the integer points of the \EMPH{output}. Ideally, we want to construct a data structure for the interval searching problem that has near-linear preprocessing time and poly-logarithmic query time. However, this is difficult even when the objects are unit-disk graphs. 

In our context,  we will be querying interval searching for each object in $\mathcal{O}_{IS}$, and therefore, we will be solving the offline version of \Cref{prob:DS0}.  
Let $\EMPH{$N_{IS}$} \coloneqq \sum_{o\in \mathcal{O}_{IS}} \Paren{|\mathcal{I}(o)| + |\mathcal{I}_{out}(o)|}$ be the total number of input and output intervals.  
Let $\EMPH{$L_{IS}$} \coloneqq \sum_{o\in \mathcal{O}_{IS}}\sum_{I\in \mathcal{I}(o) \cup \mathcal{I}_{out}(o)}~|I|$ be the total length of the input and output intervals.  We want to construct a data structure $\mathcal{D}_{IS}$ for solving \Cref{prob:DS0} that has a small total run time as a function of $N_{IS}$ and $L_{IS}$. Here, the \EMPH{total run time} of the $\mathcal{D}_{IS}$ includes:  (1) the preprocessing time and (2) the total time to answer all the queries.%
\footnote{
Alternatively, the offline version of \Cref{prob:DS0} is equivalent to computing a Boolean matrix product $C=AB$, where $A$ is the adjacency matrix of an intersection graph, and $B$ and $C$ are Boolean matrices whose 1 entries can be covered by a small number of row intervals.  We will not adopt this viewpoint here. 
}

\paragraph{Interval cover.~} This is the data structure \Cref{def:DS-1}. Recall that $N$ is the number of input objects. Let $N_{IC}$ be the total number of input objects \emph{and} query objects, and $L_{IC}$ be the total length of the input and query intervals. Similar to the interval searching problem, we want to construct a data structure $\mathcal{D}_{IC}$  for solving \Cref{def:DS-1} with small total running time, as a function of $N_{IC}$ and $L_{IC}$. We will show (in \Cref{subsec:DS-reduction}) that if we can solve the interval cover problem efficiently, then we can solve the interval searching efficiently. 

\begin{restatable}{lemma}{CoverToSearching}
\label{lm:DS1-to-DS0}
If one can construct a data structure $\mathcal{D}_{IC}$ for solving \Cref{def:DS-1} with total run time $T(N_{IC},n,L_{IC})$ (for some polynomial function $T$),  then we can construct a data structure $\mathcal{D}_{IS}$  for solving \Cref{prob:DS0} in total run time $\OO(T(N_{IS},n,L_{IS}))$.
Furthermore, if $\mathcal{D}_{IC}$ has preprocessing time $P(N)$ and query time $Q(N)$, then $\mathcal{D}_{IS}$ has preprocessing time $\OO(P(\tilde{N}_{IS}))$ and query time  $\OO(Q(\tilde{N}_{IS})\cdot |\mathcal{I}_{out}(q)|)$ where $\tilde{N}_{IS}\coloneqq \sum_{o\in \mathcal{O}_{IS}}|\mathcal{I}(o)|$ is the total number of input intervals and $\mathcal{I}_{out}(q)$ is the set of output intervals from the interval search query of $q$ to $\mathcal{D}_{IS}$. 
\end{restatable}

\paragraph{Rainbow colored intersection search.~} 
On the surface, the next problem we present seems to be a strict special case of \Cref{def:DS-1} by requiring the interval to be a singleton.
However, we will show that a solution to this problem gives us solutions to the two other problems.

\begin{problem}[Rainbow Colored Intersection Searching]\label{def:DS-2} 
Given a set of objects $\mathcal{O}_{RC}$, each object $o\in \mathcal{O}_{RC}$ is associated with a color. Design a data structure to answer the following query:
\begin{itemize}
    \item \textsc{RainbowCover?}$(q)$: Given a query object $q$, decide whether all the colors appear in the set of objects intersecting $q$.
\end{itemize}
\end{problem}

In \Cref{subsec:DS-reduction}, we show how to use a data structure $\mathcal{D}_{RC}$ for solving \Cref{def:DS-2} to design a data structure $\mathcal{D}_{IC}$  for solving \Cref{def:DS-1}. 

\begin{restatable}{lemma}{RainbowToCover}
\label{lm:DS2-to-DS1}
If we can construct in $\OO(|\cO_{RC}|)$ time a data structure $\mathcal{D}_{RC}$ with $\Tilde{O}(1)$ query time for solving \Cref{def:DS-2}, then for any parameter $b \in [1,n]$, we can construct a data structure $\mathcal{D}_{IC}$ for solving \Cref{def:DS-1} that has total run  time $\OO(N_{IC} \cdot b + L_{IC}/b)$. 
\end{restatable}

This together with \Cref{lm:DS1-to-DS0} implies a solution to the interval searching problem, in particular, a data structure $\mathcal{D}_{IS}$ for solving \Cref{prob:DS0} that has total run time $\OO(N_{IS} \cdot b + L_{IS}/b)$.

\subsection{Handling of Small Pieces}
While the low-diameter decomposition guarantees that all pieces have size at least $\widetilde{\Omega}(\Delta)$, sometimes this guarantee is not enough, and we will switch to a different algorithm. 

\paragraph*{Diameter and eccentricities.~} 
For computing diameter and eccentricities, we use the notion of \emph{patterns}~\cite{Li2019-li}, and present the following lemma implicit in the work of Le and Wulff-Nilsen~\cite{le2023vc}. 

\begin{restatable}{lemma}{diamsmallpiece}
\label{lm:diameter-smallpiece}
    Let $G$ be a graph on $n$ vertices with distance encoding VC-dimension $d$.
    Let $P$ be a piece in $G$ with boundary $\bdry P$ and diameter $\Delta$. 
    If distances from $\bdry P$ to all vertices of $G$ are known, the eccentricity of all vertices in $P$ can be computed in $O\Paren{n \cdot |\bdry P| + (|P|+|\bdry P|^d\Delta^d)\cdot T(P)}$ where $T(P)$ is the time it takes to run boundary weighted BFS on $P$ with weights at most $\Delta$.
\end{restatable}

We add one small optimization to the result of Le and Wulff-Nilsen~\cite{le2023vc} using the notion of boundary weighted BFS, a BFS where boundary vertex distances are initialized. This boundary weighted BFS can be performed in time linear in the number of edges of the piece for sparse graphs, and in time near-linear in the number of vertices of $P$ for geometric intersection graphs. 
See \Cref{SS:diameter-smallpiece} for further details and a complete proof of \Cref{lm:diameter-smallpiece}.

\paragraph*{Distance oracles.}
Similarly, for distance oracles, we will use the following lemma, also implicit in the work of Le and Wulff-Nilsen~\cite{le2023vc}. 

\begin{restatable}[Section 4.3.1 of \cite{le2023vc}]{lemma}{lmdoexact}
\label{lm:do-sparse-exact-log}
Let $G = (V_G, E_G)$ be a graph with bounded distance VC-dimension $d$, and $P$ be an induced subgraph of $G$ with boundary $\bdry P$ and diameter $\Delta$. 
There exists a distance oracle that answers distances from any vertex $s\in P$ and any vertex $t\in V_G$ with $O(n \cdot |\bdry P| + |P|^d)$ space and $O(\log |\bdry P|)$ query time.

Furthermore, if $G$ also has bounded generalized distance VC-dimension $d$ and distances from $\bdry P$ to all vertices of $G$, the distance oracle can be computed in $O\Paren{n \cdot |\bdry P|+(|\bdry P|^d \Delta^d + |P|) \cdot T(P)}$ time, where $T(P)$ is the time it takes to run vertex weighted BFS on $P$ with weights at most $\Delta$.
\end{restatable}

For completeness, we provide the proof of \Cref{lm:do-sparse-exact-log} in  \Cref{app:oracle-smallsize}.

\section{Framework for Diameter and Eccentricities}
\label{sec:framework}

\label{subsec:framework-diameter}

In this section, we outline the algorithmic framework for computing the all-vertex eccentricity of different graph families in truly subquadratic time. 
(Recall that the \EMPH{eccentricity} of a vertex $u$ is defined to be $\EMPH{$\ecc(u)$} \coloneqq \max_{v\in V_G} d(u,v)$.)
Note that as diameter is the maximum eccentricity of any vertex in the graph, we can also compute diameter in truly subquadratic time.
Our framework can be tweaked for other problems, such as constructing distance oracles and Weiner index; we defer to \Cref{sec:framework-do}. 

Now we formally set up the framework, which
consists of a few high-level instructions, with the goal to compute for every vertex $u$, the $r$-neighborhood balls $N^r[u]$ iteratively for growing values of $r$.  
This is enough to answer the diameter problem exactly because a graph $G$ has diameter at most $D$ if and only if every radius-$D$ neighborhood ball contains all vertices in $G$.

Let $G$ be the input graph, given either explicitly using adjacency lists or implicitly as the intersection graph of objects.
Our framework has three steps.

\paragraph{Step~1: Low-diameter decomposition (LDD).}
Compute a low-diameter decomposition $\mathcal{L}$ of $G$ with a diameter parameter $\Delta > 0$. $\mathcal{L}$ has $\OO(n/\Delta)$ pieces, each of strong diameter at most $\Delta$. Furthermore, $\sum_{P\in \mathcal{L}} |\bdry P|=\OO(n/\Delta)$.
The vertices in $\bigcup_P \bdry P$ are called the \EMPH{boundary vertices}.

\paragraph{Step~2: Shortest-path computations.}
For each boundary vertex $v$ in $\bigcup_P \bdry P$, compute a breadth-first search tree in $G$ rooted at $v$.  We obtain $\ecc(v)$ as a byproduct. 
Define
\[
\EMPH{$\bdry\ecc$} \coloneqq \max_{\text{boundary vertex $v$}} \ecc(v).
\]

\paragraph{Step~3: Growing neighborhood balls.}
Consider one piece $P$ in the low-diameter decomposition $\cL$. 
Our goal is to compute \emph{some modified version of $N^r[s]$} for every vertex $s$ in $P$ and \emph{necessary values of $r$}. 
Fix an arbitrary vertex \EMPH{$s_P$} in~$\bdry P$, and define \EMPH{$\ecc_P$} as the corresponding eccentricity $\ecc(s_P)$.  
(Notice that $\ecc(s_P)$ is known after the shortest-path computation in Step~2 because $s_P$ is a boundary vertex of $P$.)
We set the \EMPH{modified neighborhood ball} for each vertex $s$ in $P$ to be 
\[
\EMPH{$\hat N^r[s]$} \coloneqq N^r[s] \cap R_P \text{, with }\, \EMPH{$R_P$} \coloneqq \Set{\big. t \in V_G : \dist(s_P,t) \ge \ecc_P - 2\Delta},
\] 
where $\Delta$ was defined to be the strong diameter bound of the pieces in the LDD
and $R_P$ is called the \EMPH{relevant region} for the eccentricity computation for $P$.
We will compute $\hat N^r[s]$ for every $s$ in $P \setminus \bdry P$ iteratively using the inductive formula 
\begin{align}
\label{Eq:inductive}
\hat N^r[s] = \bigcup_{v\in N[s]} \hat N^{r-1}[v].
\end{align}

We emphasize that while the notation seems to suggest otherwise, the definition of modified neighborhood balls $\hat N^r[s]$ \emph{depends on the piece $P$}.
Define the set of \EMPH{relevant balls} to be
\[
\EMPH{$\RB_P$} \coloneqq \Set{\big. \hat N^r[v] : v\in P\/, r\in [\ecc_P - 3\Delta, \ecc_P + \Delta] }.
\]
Let $\EMPH{$\RB$} \coloneqq \bigcup_{P\in \ldd} \RB_P$. 
The \EMPH{ball growing process} consists the following substeps:
\begin{enumerate}
\item[3.1.]
For every $s\in \bdry P$ and every $r\in [\ecc_P - 3\Delta, \ecc_P + \Delta]$, compute 
modified balls $\hat N^r[s]$ using Step~2.
\item[3.2.]
As a base case, we initialize $\hat{N}^r[s] = \varnothing$ for every $s\in P\setminus \bdry P$ when $r = \ecc_P - 3\Delta - 1$. 
\item[3.3.]
For other values of $r \in [\ecc_P-3\Delta, \ecc_P+ \Delta]$, compute  $\hat N^r[s]$ using the inductive formula~(\ref{Eq:inductive}).
\end{enumerate}
Then $\ecc(s)$ is the smallest value $r$ such that $\hat N^r[s]$ is the whole relevant region $R_p$. 
Therefore, we can compute $\ecc(s)$ from $\Set{\hat{N}^r[s] : r\in [\ecc_P-3\Delta-1, \ecc_P+\Delta] }$.

Note that we will assume that the diameter of the graph is at least $4\Delta$, otherwise the entire graph $G$ is a low-diameter decomposition of parameter $4\Delta$, and we can simply apply Step~$3$ with the relevant neighborhood balls being all balls, the relevant region $R_P$ being $V_G$, and the modified neighborhood balls being normal neighborhood balls.  

\paragraph*{Correctness.~}
To show that our algorithmic framework is correct, we show that we correctly computed all (modified) neighborhood balls, assuming the ball growing process is correct.
First note that for $s\in \bdry P$, we have correctly computed the modified neighborhood balls in Step~3.1.
If $s \in P \setminus \bdry P$, given the pair $(s, t)$ realizing $\ecc(s)$, we can guarantee that the vertex $t$ must lie in the relevant region $R_P$:
Denote $t_P$ to be the vertex that has distance $\ecc_P$ to $s_P$, then because $\dist(s,s_P) \le \Delta$, we have
\begin{align*}
\dist(s_P, t) 
&\ge \dist(s, t) - \Delta 
\ge \dist(s, t_P) - \Delta 
\ge \dist(s_P, t_P) - 2\Delta,
\end{align*}
and thus $t$ can be found in $\hat N^r[s]$ if $\hat N^r[s]$ is a relevant ball in $\cS_P$.
Furthermore, again by triangle inequality, $\ecc(s)$ is at least $\ecc_P - \Delta$ and at most $\ecc_P + \Delta$.
Thus it is sufficient to initialize $r$ to be $\ecc_P-3\Delta-1$ (in which case $N^r[s] \cap R_P = \varnothing$), 
so the initialization in Step~3.2 is correct.
Thus, assuming the ball expansion step is correct, all modified neighborhood balls are computed correctly in Step~3.3.

\paragraph{VC-dimension of neighborhood balls and stabbing paths.}
We cannot afford to store the (modified) neighborhood balls $\hat N^r[v] \in \cS_P$ explicitly.
Instead, we will rely on a compact representation of a set system with bounded VC-dimension to store the neighborhood balls \emph{implicitly} in a data structure. 
Given the (modified) neighborhood ball system $(V_G, \cS_P)$, we are responsible for bounding the (dual) VC-dimension of $(V_G, \cS_P)$ to be a constant $d$.
Then we compute {stabbing path $\spath$} for $(V_G, \cS_P)$, such that the {interval representation $\Rep_\spath(\hat N^r[v])$} of set $\hat N^r[v]$ has sublinear size. 
(See \Cref{SS:stabbing-path} for definition.)

\paragraph*{Implementing the ball growing process.}
To implement the ball growing process, we will use a stabbing path $\spath$ for the modified neighborhood balls to ensure we can compactly store all such balls. The exact details on how we implement the process will depend on the type of graph we are dealing with.

In a sparse graph $G$, we will show how to implement the ball expansion data structure in $G$ directly by explicitly considering the neighbors $N[v]$ of each vertex $v$ in the graph $G$. 

In a geometric intersection graph $G$, we instead implement the ball growing process for a piece $P\in \ldd$ with a data structure for the interval searching problem defined in \Cref{prob:DS0}. 
Each vertex $v$ in $P$ is associated with a geometric object $o_v$. Let $\cO_P$ denote the set of these geometric objects.
Suppose we have computed compact interval representations $\Rep_\spath(\hat N^{r-1}[v])$ for every vertex $v$ in $P \setminus \bdry P$, so we can associate these intervals to $o_v$. Using a data structure $\cD_{IC}$ for \Cref{prob:DS0}, the union of intervals of objects in $\cO_P$ that intersect with $o_v$ is exactly $\hat N^r[v]$ by \Cref{Eq:inductive}. 
Thus, we can implement the ball growing process in a geometric intersection graph if we have an offline data structure for \Cref{prob:DS0}. The efficiency of the algorithm will depend on the number of intervals in the representation with respect to the stabbing path $\spath$.

\paragraph*{Organization.}
In the next four sections, we will apply our framework to devise algorithms for diameter and eccentricities for different graph classes: sparse graphs of bounded VC-dimension (\Cref{sec:ecc-sparse}),
arbitrary-square graphs (\Cref{sec:ecc-square}),
unit-square graphs (\Cref{sec:ecc-unit-square}),
and unit-disk graphs (\Cref{sec:unit-disks}).
Sections \ref{sec:ecc-sparse}, \ref{sec:ecc-square}--\ref{sec:ecc-unit-square}, and \ref{sec:unit-disks} can be read independently, depending on the interest of the reader.
The sparse graph case is perhaps the simplest, not requiring geometric data structures.  The unit-disk case is the most involved and requires overcoming a number of (interesting) technical challenges.

\section{Diameter/Eccentricities in Sparse Graphs of Bounded VC-dimension}\label{sec:ecc-sparse}
We begin by applying the framework in \Cref{subsec:framework-diameter} to sparse graphs of bounded VC-dimension.
In this setting, the low-diameter decomposition could be constructed in $O(m)$ time using \Cref{thm:ldd-sparse}. Computing the BFS tree for every boundary vertex takes $\OO(mn/\Delta)$ total time where $\Delta$ is the diameter parameter in the low-diameter decomposition. Thus we focus on the third step of performing ball expansion.

To begin, we construct a global ordering $\lambda$ on all the vertices for our stabbing path data structure. The following is analogous to Corollary 15 in~\cite{duraj2023better} that we tailor to our setting.   
\begin{lemma}\label{lm:graph-ordering-global} We can compute in $\OO(mn^{1/d})$ time an ordering $\lambda$ of the vertices in $V$ such that for the system $\RB = \bigcup_{P\in \ldd} \RB_P$ 
such that $\sum_{P\in\ldd}\sum_{s\in P} \sum_{r=\ecc_P-3\Delta}^{\ecc_P+\Delta} \deg(s)\cdot |\Rep_\lambda(N^r[s])| = \OO(\Delta mn^{1-1/d})$.
\end{lemma}
\begin{proof} Let $\check{\mathcal{S}}$ be the set obtained by taking each set  ${N}^r[s]$ in $\mathcal{S}$ and adding  $\deg(s)$ copies of ${N}^r[s]$ to $\check{\mathcal{S}}$. Observe that:
\begin{equation*}
    |\check{\mathcal{S}}| = \sum_{P\in\ldd}\sum_{s\in P} \sum_{r=\ecc_P-3\Delta}^{\ecc_P+\Delta} \deg(s) =  O(\Delta \cdot m)
\end{equation*}
We then apply \Cref{lem:stabbing-path-general} to $X = V(G)$ and $\check{\cS}$ with $\rho = n^{1/d}$. 
Since we can implement the element reporting oracle in $O(m)$ time via BFS, the result follows.
\end{proof}

Now for every relevant piece $P$ in the low-diameter decomposition $\ldd$, we will restrict our attention to only the relevant region $R_P$ for the eccentricity computation. To do so, we consider the ordering $\lambda_P$ of $R_P$ obtained from $\lambda$ by restricting to the vertices of $R_P$. Observe that doing so does not increase the size of the interval representation of any sets.

\begin{observation} \label{obs:subordering} Let $R$ be a subset of $V$.  Let $\lambda$ be an ordering of the vertices $V$, and $\lambda'$ be an ordering of $R$ obtained by restricting $\lambda$ to the vertices in $R$. Then for any set $S\subseteq V$, $|\Rep_{\lambda'}(S\cap R)| \le |\Rep_{\lambda}(S)|$.
\end{observation}
\begin{proof} For any interval $I\in \Rep_{\lambda}(S)$, $I\cap R$ is also an interval in $\lambda'$.   
\end{proof}

\paragraph*{Ball expansion data structure.} 
To implement the ball expansion data structure, we will store each neighborhood ball in interval form.
For a boundary vertex $s\in \bdry P$ , we can compute $\hat{N}^r[s]$ for all $r\in[\ecc_P-3\Delta, \ecc_P+\Delta]$ in $O(n)$ time using the BFS tree we have computed from step 2, and in addition represent these balls in interval form.

Next we describe how to perform the ball expansion operation.
For vertices $s\in P\setminus \bdry P$, each neighbor $v\in N[s]$ is also in $P$ and we have a compact interval representation for $\hat{N}^{r-1}[s]$. 
We can take the union of the set of intervals by doing a line sweep in time $\OO\left(\sum_{v\in N[s]} |\Rep_{\lambda_p}(N^{r-1}[v])|\right)$. 

Furthermore, it is easy to detect if $\hat{N}^r[s] = R_P$ if the interval representation is all of $\lambda_P$.

\paragraph*{Time analysis.}  The amount of time taken for computing the global ordering in \Cref{lm:graph-ordering-global} is $\OO(mn^{1/d})$.
The runtime for ball expansion of the boundary vertices is:
\begin{align}
\sum_{P\in \ldd}\sum_{s\in \bdry P} O(n) ~=~ \OO(n^2/ \Delta)
\end{align}

\noindent By \Cref{obs:subordering}, the ball expansion for a non-boundary vertex $s\in P$ and $s\notin \bdry P$ from radius $r-1$ to $r$ takes time 
\[
\OO\left(\sum_{v\in N[s]}|\Rep_{\lambda_P}(\hat{N}^{r-1}[v])|\right) ~\le~ \OO\left(\sum_{v\in N[s]}|\Rep_{\lambda}({N}^{r-1}[v])|\right). 
\]

\noindent The total time taken for all ball expansion steps for non-boundary vertices across all the pieces is at most:
\begin{align*}
\sum_{P\in \ldd} \sum_{s\in P}\sum_{r=\ecc_P-3\Delta}^{\ecc_P+\Delta} \!\OO\left(\sum_{v\in N[s]} \! |\Rep_{\lambda}({N}^{r-1}[v])|\right) 
&~=~ \OO\left(\sum_{P\in\ldd}\sum_{s\in P} \sum_{r=\ecc_P-3\Delta}^{\ecc_P+\Delta} \!\! \deg(s)\cdot |\Rep_\lambda(N^r[s])|\right) \\
&~=~ \OO(\Delta mn^{1-1/d}).
\end{align*}
The last equality follows from \Cref{lm:graph-ordering-global}.

Recall that the first two steps of the framework can be implemented in $\OO(mn/\Delta)$ time. 
The total runtime for all three parts is:
\[\OO(mn/\Delta + mn^{1/d} +  n^2/\Delta + \Delta mn^{1-1/d}) ~=~ \OO(mn/\Delta + \Delta mn^{1-1/d}).\]
Setting $\Delta = O(n^{1/(2d)})$ yields a that this algorithm runs in $\OO(mn^{1-1/(2d)})$ time.
\begin{theorem}
\label{thm:ecc-sparse}
The diameter problem in a sparse undirected graph $G$ with $n$ vertices and $m$ edges and general distance VC-dimension at most $d$ can be solved in $\OO(mn^{1-1/(2d)})$ time.
\end{theorem}

\begin{remark}
We can also obtain similar results (albeit with possibly worse exponents) for other VC-dimension bounds. If the distance VC-dimension is bounded by $d$ or even if the $k$-neighborhood VC-dimension is bounded by $d$ for all $k$, we can follow an approach similar to what we have for unit-disk graphs (see \Cref{sec:unit-disks}). The main difference is an extra step to reorder the vertices when we transition from $k-1$-neighborhoods to $k$-neighborhoods using \Cref{SS:reordering}.
\end{remark}

\section{Diameter/Eccentricities in Square Graphs}\label{sec:ecc-square}

Next, we apply the framework in \Cref{subsec:framework-diameter} to intersection graphs of squares.
In step 1, we apply \Cref{thm:ldd-geo} to obtain our low-diameter decomposition $\ldd$ of $G$ in $\OO(n)$ time into pieces of size $\log n \leq \Delta \leq n$, where $\Delta$ is a parameter we will choose later.
In step 2, we compute BFS trees from each $v\in \bigcup_P \bd P$ using \Cref{lm:SSP-geo-intersection}. The algorithm takes $\OO(n)$ time per vertex, so this step takes $\OO(n^2/\Delta)$ time. Note that we also explicitly store all distances from these vertices, which takes $\OO(n^2/\Delta)$ space.

Recall that when $s\in \bdry P$ then we can explicitly compute $\hat N^r[s]$ for all values of $r$ in $O(n)$ time using the distances computed in step 2 of our framework.

\paragraph{Stabbing path.~} We now compute a global stabbing path $\spath$. We use the following lemma.
\begin{lemma}
\label{lem:stabbing-fatpseudodisks}
Let $G$ be a graph with generalized distance VC-dimension $d$, and a single-source distance finding algorithm with running time $T(n)$.
Then the modified neighborhood ball system has a path $\spath$ such that we have
\[
\sum_{P\in \ldd}\sum_{s\in P}\sum_{r=\ecc_P-3\Delta}^{\ecc_P+\Delta} |\Rep_\spath(\hat N^r[s])| ~=~ \OO(\Delta\cdot  n^{2-1/d})
\]
with high probability, i.e., $\spath$ is a stabbing path of the modified $r$-balls.
Furthermore, $\spath$ can be computed with a randomized algorithm in $\OO(n^{1/d}T(n))$ time.
\end{lemma}

\begin{proof}
We apply \Cref{lem:stabbing-path-general} with $\rho = n^{1/d}$ for the set system
\[
\mathcal{S} \coloneqq \Set{\big. N^r[s] : \text{$P \in \ldd$, $s\in P\setminus \bd P$, $r\in [\ecc_P-3\Delta,\ecc_P+\Delta]$} }.
\]
Notice that the system has size at most $|\cS|=3\Delta n$, and we can use the BFS algorithm to report the squares in the modified ball. By \Cref{obs:subordering}, as $\hat N^r[s] = R_P\cap N^r[s]$, we obtain the bound
\[
\sum_{P\in \ldd}\sum_{s\in P}\sum_{r=\ecc_P-3\Delta}^{\ecc_P+\Delta} |\Rep_\spath(\hat N^r[s])| 
~\le~  
\sum_{P\in \ldd}\sum_{s\in P}\sum_{r=\ecc_P-3\Delta}^{\ecc_P+\Delta}  |\Rep_\spath(N^r[s])| 
= \OO(\Delta\cdot  n^{2-1/d})
\]
\aftermath
\end{proof}

In all of the intersection graphs studied in this paper, we have $T(n)=\OO(n)$ and $d=4$. 
This leads to a stabbing path $\spath$ that is computed in $\OO(n^{5/4})$ time and has the property that the total size of the representation is $\OO(\Delta\cdot n^{7/4})$. Given that there are $O(\Delta \cdot n)$ modified balls we consider, the \emph{amortized} interval count to represent a single modified ball is $O(n^{3/4})$.

\paragraph{Growing balls.~} To grow the modified neighborhood balls, we will design a data structure for solving the interval searching problem (\Cref{prob:DS0}) for squares, which we restate here: we are given a set of square $S$, where each square $s \in S$ is associated with a set of intervals (of integer points) of $[1:n]$, denoted by $\mathcal{I}_s$.  Design a data structure  that answers the following query: 
\begin{itemize}
    \item \textsc{IntervalSearch}$(q)$:  Given a square $q$, return (the interval representation of) all the integer points in $[1:n]$ associated with squares in $S$ that intersect $q$.
\end{itemize}

We will be querying the data structure once for each square $s\in S$. Therefore, we are interested in minimizing the total query time. Let $\mathcal{I}_{out}(s)$ be the set of output intervals for a query square $s$. 
Let $\EMPH{$N$} \coloneqq \sum_{s\in S} \Paren{|\mathcal{I}(s)| + |\mathcal{I}_{out}(s)|}$ be the total number of input and output intervals.  Let $\EMPH{$L$} \coloneqq  \sum_{s\in S}\sum_{I\in \mathcal{I}(s) \cup \mathcal{I}_{out}(s)}~|I|$ be the total length of the input and output intervals.

\begin{lemma}\label{lm:square-interval-searching} For any parameter $b\in [1:n]$, we can construct a data structure $\check{\cD}$ for solving the interval searching problem for squares such that the total time to (i) construct $\check{\cD}$  and (ii) answer $|S|$ queries, one for each square $s\in S$, is $\OO(N \cdot b + L/b)$. 
\end{lemma}
\begin{proof}
  By \Cref{lm:DS2-to-DS1}, it suffices to construct a data structure $\mathcal{D}_{RC}$ for the rainbow colored intersection searching for squares that has nearly linear preprocessing time and poly-logarithmic query time. We provide such a data structure in \Cref{subsec:ds-square}.  
\end{proof}

Next, we present a simpler (but slower) algorithm for computing all eccentricities. Then we show how to improve the running time. 

\paragraph*{First version.} To compute all eccentricities, for each piece $P$ we restrict $\spath$ to $R_P$ in $O(n)$ time, and denote the resulting ordering by $\spath_P$. Next, we set $r=\ecc_P-3\Delta$ and compute the balls $\{\hat N^r[s]\}$ for each $s$. In general, once the representations of $\hat N^{r-1}[s]$ are known, the data structure to set up for computing modified balls of radius $r$ will have $\sum_{s\in P} |\Rep_{\lambda_P}(\hat N^{r-1}[s])|$ intervals in it.

To compute the representations of $\{\hat N^{r}[s]\}_{s\in P}$, we setup the data structure $\check{\cD}$ that takes as input: (i) a set of squares corresponding to vertices of $P$ and (ii) the interval representation  $\{\Rep_{\lambda_P}(\hat N^{r-1}[s])\}_{s\in P}$ for radius $r-1$. Then we apply $|P|$ queries $\{\textsc{IntervalSearch}(s): s\in P\}$ to output the interval representations of $\{\hat N^{r}[s]\}_{s\in P}$. Observe that the total length of all the intervals is at most $2|P|\cdot|R_P| = O(|P|\cdot n)$. Thus, the total running time for each $r$ is:

\begin{equation*}
    \OO\Paren{b\cdot \sum_{s\in P} (|\Rep_{\lambda_P}(\hat N^{r-1}[s])|  + |\Rep_{\lambda_P}(\hat N^{r}[s])|)+ |P|n/b }
\end{equation*}

Therefore, the total running time of computing all-vertex eccentricities, including the running time of the first two steps in the framework, is:
\begin{equation}
\label{eq:square-simpe-rep}
\begin{split}
      &~ \OO(n^2/\Delta + n^{5/4})   + \sum_{P\in \cL}\sum_{r=\ecc_P-3\Delta}^{\ecc_P+\Delta} \OO\left(b\cdot \sum_{s\in P} (|\Rep_{\lambda_P}(\hat N^{r-1}[s])|  + |\Rep_{\lambda_P}(\hat N^{r}[s])|)+ |P|n/b\right) \\
~=~   &~ \OO(n^2/\Delta + n^{5/4}) + \OO(b) \left( \sum_{P\in \cL} \sum_{r=\ecc_P-3\Delta}^{\ecc_P+\Delta}|\Rep_\spath(\hat N^r[s])|\right) + \OO(n^2\Delta/b)\\
~=~   &~ \OO(n^2/\Delta + n^{5/4}) + \OO(b \Delta\cdot  n^{7/4})+  \OO(n^2\Delta/b) \quad (\text{by \Cref{lem:stabbing-fatpseudodisks} and }d=4)\\
~=~   &~ \OO(n^2/\Delta + b \Delta\cdot  n^{7/4} +  n^2\Delta/b) \\ ~=~   &~ \OO(n^{2-1/16}).  \quad \text{(for optimal choices of $b = \Delta^2$ and $\Delta = n^{1/16}$)}
\end{split}
\end{equation}

\paragraph{Improved version.} We improve the running time by reducing the $\OO(n^2\Delta/b)$ in \Cref{eq:square-simpe-rep}, which is the \emph{total length of the intervals}, to $\OO(n^2/b)$ by keeping track of the sets $\hat N^{r}[s]\setminus \hat N^{r-1}[s]$ instead of $\hat N^{r}[s]$.  Notice that the eccentricity of $s$ will be the largest $r$ where $\hat N^{r}[s]\setminus \hat N^{r-1}[s]$ is non-empty. 
Let $\EMPH{$\hat{N}^{=r}[s]$} \coloneqq \hat{N}^{r}[s]\setminus \hat{N}^{r-1}[s]$. Then $|\Rep_{\lambda_P}(\hat N^{=r}[s])| \leq|\Rep_{\lambda_P}(\hat N^{r}[s])| + |\Rep_{\lambda_P}(\hat N^{r-1}r[s])|$ and therefore, $\{\hat N^{=r}[s]\}_{s\in P, P\in \mathcal{L}}$ has a compact representation:
\begin{equation*}
 \sum_{P\in \cL} \sum_{r=\ecc_P-3\Delta}^{\ecc_P+\Delta} |\Rep_\spath(\hat N^{=r}[s])|  \leq  \sum_{P\in \cL} \sum_{r=\ecc_P-3\Delta}^{\ecc_P+\Delta}(|\Rep_{\lambda_P}(\hat N^{r-1}[s])|  + |\Rep_{\lambda_P}(\hat N^{r}[s])|) = \OO(\Delta\cdot  n^{7/4}).
\end{equation*}

Observe that 
\begin{equation*}
    \hat{N}^{=r}[s] = \Big( \bigcup_{v\in N[s]} \hat{N}^{=r-1}[v] \Big) 
    \setminus \Paren{\Big. \hat{N}^{=r-1}[s]\cup \hat{N}^{=r-2}[s]}.
\end{equation*}

Thus, we could apply the same growing ball process for $\hat{N}^r[s]$. More precisely, we compute the interval representation of $\bigcup_{v\in N[s]} \hat{N}^{=r-1}[v]$ by querying the interval searching data structure, and then remove elements from $\hat{N}^{=r-1}[s]\cup \hat{N}^{=r-2}[s]$ using the interval representations of $\hat{N}^{=r-1}[s]$ and $\hat{N}^{=r-2}[s]$ computed from the previous iterations. 
On the other hand, the intervals in this representation are disjoint, so we can bound the total length $L$ over all $3\Delta$ iterations as $L\leq O(|P|n)$ instead of $O(\Delta|P|n)$. 
Therefore, by applying the same calculation in \Cref{eq:square-simpe-rep}, the final running time is:
\[\OO(n^2/\Delta + n^{5/4}) + \OO(b)\cdot N + \OO(n^2/b )  ~=~ \OO\left(n^2/\Delta +  b\Delta n^{7/4} + n^2/b \right)~=~ \OO(n^{2-1/12})
\]
for $b=\Delta=n^{1/12}$.

\begin{theorem}
\label{thm:ecc-square}
Computing the diameter and all-vertex eccentricities of square graphs with $n$ vertices can be done in $\OO(n^{2-1/12})$ time.
\end{theorem}

\section{Diameter/Eccentricities in Unit-square Graphs}
\label{sec:ecc-unit-square}

For unit squares, we can obtain a slightly faster algorithm.
The first two steps and the stabbing path computation are the same as the algorithm for square graphs. The total running time of these steps is $\OO(n^2/\Delta + \Delta n^{5/4})$. The growing ball step is more efficient since we can develop a better geometric data structure for unit squares.

\paragraph*{Growing balls.} For unit squares, we design a more efficient data structure for the interval cover problem (\Cref{def:DS-1}), and as a result, we obtain a more efficient data structure for the interval searching problem.  Let $S$ be a given set of unit squares where each square $q\in S$ is associated with a set of intervals of $[1:n]$. Each query \textsc{IntervalSearch}$(q)$ returns the intervals of integer points in $[1:n]$ associated with unit squares intersecting $q$. Let $\EMPH{$\Tilde{N}$} \coloneqq \sum_{s\in S}|\mathcal{I}(s)|$ be the number if input intervals.

\begin{lemma}\label{lm:unitsq-interval-searching} 
We can construct in $\Tilde{N}^{1+o(1)}$ time a data structure $\check{\cD}$ for solving the interval searching problem for unit squares that can answer each query  \textsc{IntervalSearch}$(q)$ in $\Tilde{N}^{o(1)} \cdot |\mathcal{I}_{out}(q)|$ time.
\end{lemma}

\begin{proof} In \Cref{subsec:ds-square} (and more specifically \Cref{thm:ds_unitsqure}), we construct a data structure for the interval cover problem (\Cref{def:DS-1}) for unit square with $\Tilde{N}^{1+o(1)}$ preprocessing time and $\OO(1)$ query time.  Then by \Cref{lm:DS1-to-DS0}, we obtain the preprocessing time and query time as in the lemma. 
\end{proof}

\paragraph{The algorithm.~} 
The algorithm is essentially the same for the square graphs in the previous section: restricting the ordering $\lambda$ to $R_p$ to get $\lambda_P$, and growing balls $\{\hat{N}^r[s]\}_{s\in P}$ for $r = \ecc_P-3\Delta$ to $\ecc_P+\Delta$ by applying a query \textsc{IntervalSearch}$(s)$ for each unit square $s\in P$ to the interval searching data structure in \Cref{lm:unitsq-interval-searching} built for $\hat N^{r-1}[s]$. Let $\Tilde{N}_{r-1} \coloneqq  \sum_{s\in P} (|\Rep_{\lambda_P}(\hat N^{r-1}[s])|$ be the number of input intervals to the data structure; the total output size is $\Tilde{N}_n$. 
Since $\Tilde{N}_{r-1}^{o(1)} = n^{o(1)}$, the total running time to grow all the balls per piece is 
\[
n^{o(1)} \cdot \sum_{r=\ecc_P-3\Delta}^{\ecc_P+\Delta}|\Rep_\spath(\hat N^r[s])|.
\] 
The total running time to compute all eccentricities is:
\begin{equation*}
\begin{split}
      &~ \OO(n^2/\Delta + \Delta n^{5/4}) + n^{o(1)} \sum_{P\in \ldd}\sum_{r=\ecc_P-3\Delta}^{\ecc_P+\Delta}|\Rep_\spath(\hat N^r[s])| \\ 
~=~   &~ O^*(n^2/\Delta + \Delta n^{5/4} + \Delta n^{7/4}) \quad \text{(by \Cref{lem:stabbing-fatpseudodisks} and $d=4$)}\\ 
~=~   &~ O^*(n^{2-1/8}) \quad \text{(for $\Delta = n^{1/8}$.)}    
\end{split}
\end{equation*}

\begin{theorem}
\label{thm:ecc-unitsq}
Computing the diameter and all-vertex eccentricities in unit square graphs with $n$ vertices can be done in $O^*(n^{2-1/8})$ time.
\end{theorem}

\section{Diameter/Eccentricities in Unit-disk Graphs}
\label{sec:unit-disks}

\newcounter{stepcount}
\def\cL{\mathcal{L}}

We now describe how to adapt the framework in \Cref{subsec:framework-diameter} to the more complicated setting of computing diameter and eccentricities for unit-disks.
The computation of LDD and BFS remains unchanged because unit-disks are fat pseudo-disks; 
we follow Step~1 and Step~2 of the framework (\Cref{subsec:framework-diameter}).
For LDD we use \Cref{thm:lld-pseudo-disks};
for BFS we use \Cref{lm:SSP-geo-intersection}.
However, Step~3 requires drastic changes in order to implement the ball expansion step.

In this section we assume the unit-disk graphs are in \emph{center-disk intersection model}: Unlike a typical geometric intersection graph where we create an edge between two objects if they intersect, here we add an edge between two unit-disks if the center of one disk lies in the other disk.
It is straightforward to see that the two models are equivalent by doubling the radii of all unit disks.  
For the sake of simplicity, we will scale the disks so that the radius is still one unit.

\subsection{Restriction to Fixed Types}

\paragraph{Partition into modulo classes.}
We partition the plane into square \EMPH{cells}: every unit-square is divided into $2 \times 2$ many cells, each of side length $1/2$. 
Each cell is indexed by the coordinates of its bottom-left corner modulo $3$; notice that the coordinates are multiples of $1/2$ and thus there are $6$ modulo classes per coordinate.
We collect all cells of the same index $(i,j)$ into a set $\Cell_{i,j}$; in other words,
\[ 
\EMPH{$\Cell_{i,j}$} \coloneqq \Set{\big. \text{cell $\square$} : \text{square $\square$  is located at $(x,y)$ where $x \equiv i$ and $y \equiv j$ $(\bmod~3)$}}.
\] 
We then classify the set of unit-disks $\mathcal{D}$ based on the cell classes where the center of the unit-disk lies:
\[ 
\EMPH{$\mathcal{D}_{i,j}$} \coloneqq \Set{\big. D \in \mathcal{D} : \text{disk $D$ has its center located in some cell in $\Cell_{i,j}$}}.
\] 
Notice that $\set{\mathcal{D}_{i,j}}_{i,j}$ is a partition of $\mathcal{D}$.
We say a disk in $\mathcal{D}_{i,j}$ has \EMPH{type $(i,j)$}.
Denote the number of types to be \EMPH{$\sigma$}; there are exactly $\sigma = 36$ types. 

The disks intersecting a query disk $D_q$ whose center point $q$ lies inside a cell $\square$ come in two flavors: those disks that completely contain the cell $\square$, and those that partially intersect the cell.  
We call the cells where the centers of these intersecting disks belong \EMPH{relevant} to $\square$; among them, we call those cells with disks partially intersecting $\square$ \EMPH{perimetric}.
Observe that there are only constantly many cells relevant to any fixed cell $\square$,
because we set the side length of each cell to be $1/2$. 

\begin{figure}[ht!]
\centering
\includegraphics[scale=1,]{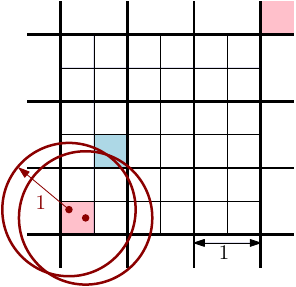}
\caption{The 36 cells formed by partitioning a $3\times 3$ square.  Cells of the same color are of the same type.  The disks in $\cD_{0,0}$ (in pink) intersects the blue cell as a pseudoline arrangement.}
\end{figure}

Fixed an arbitrary cell $\square$. 
A \EMPH{pseudoline arrangement} $\mathcal{L}$ inside $\square$ is a collection of boundary-to-boundary simple curves in $\square$, such that every pair of curves in $\mathcal{L}$ intersect each other at most once.
We now establish the main combinatorial property for disks of the same type: while two unit-disks intersect up to two times in the plane, if we focus on a single cell $\square$ and two disks whose centers are in some other cells of the same type, at most one intersection will appear in $\square$. 
(Indeed, based on the way we partition cells into modulo classes, at most one cell of each type is relevant to $\square$.)

\begin{lemma}
\label{L:pseudolines}
Let $\mathcal{D}$ be any set of unit-disks, partitioned into types as described above.  Given any cell $\square$ and a fixed type $(i,j)$, the boundary of the disks in $\cD_{i,j}$ intersects inside $\square$ as a pseudoline arrangement.
\end{lemma}

\begin{proof}
Assume for contradiction that there are two intersecting unit-disks $D$ and $D'$
with two intersection points inside $\square$ simultaneously.
As $\square$ has side-length $1/2$ and diameter at most $\sqrt{2}/2 \le 0.71$, the centers of $D$ and $D'$ must be at least $2\cdot\Paren{1^2 - (\sqrt{2}/4)^2}^{1/2} = \sqrt{14}/2 \ge 1.87$ units away.
Thus the two centers cannot be in the same cell (which has diameter at most $0.71$).
On the other hand, we reach a contradiction as any two distinct cells of the same type must be at least $2.5$ units away (because we index the cells by modulo $3$), while the centers of the intersecting disks $D$ and $D'$ can be at most $2$ units away.
\end{proof}

If we focus on one perimetric cell $\square'$ of $\square$ and rotate the plane so that $\square'$ lie about vertically above~$\square$ (the cells might not be parallel to the axis anymore), we can safely assume each pseudoline formed by the partial intersection by a disk in $\square'$ with $\square$ to be \EMPH{$x$-monotone}, that is, any vertical line intersects the pseudoline at most once.

\subsection{Implementation of the Neighborhood Growing Step}
\label{SS:ball-growing-unitdisk}

We first describe how to implement the ball growing process (Step~3.3) in the framework using the inductive formula~(\ref{Eq:inductive}),
which we recall here:
\begin{align}
\tag{\ref{Eq:inductive}}
\hat N^r[s] = \bigcup_{v\in N[s]} \hat N^{r-1}[v].
\end{align}

Fix a piece $P$ from the LDD, and some vertex $s$ in $P \setminus \bdry P$. 
Each modified neighborhood ball $\hat N^r[s]$ can be written as the union of a collection of $r$-balls with restrictions on the type of the second vertex in length-$r$ paths.  
More precisely, recall that a disk $D$ is of type $(i,j)$ if the center of the disk $D$ is located in some cell whose bottom-left corner has coordinates in the modulo class $(i,j)$.
We arbitrarily order the constantly many types and label them from $1$ to $\sigma$.
We say that a path in the intersection graph $G$ from a vertex $s$ to a vertex $v$ is a \EMPH{$\tau$-path} if the disk corresponding to the vertex following $s$ (the second vertex) in the path is of type $\tau \in [1:\sigma]$. 
For each vertex $s$ and an arbitrary subset $M \subseteq [1:\sigma]$, define
\begin{align}
    \EMPH{$\hat N^r_{M}[s]$} \coloneqq \Set{\big. v\in V_P : \text{there is a $\tau$-path from $s$ to $v$ of length at most $r$, for some $\tau \in M$} }.
\end{align}
We often use the $_{\le T}$ subscript to represent the subset $[1:T]$ below.
Notice that by restricting to $\tau$-paths from $s$ to $v$ in the definition, we can implement the inductive formula using
\[
\hat N^r_T[s] = \bigcup_{v\in N[s] \cap \Cell_{T}} \hat N^{r-1}_{\le \sigma}[v].
\]
We prove the following lemma in 
\Cref{sec:vcdim-type} (\Cref{lm:VC-type2-walk}). \note{primal-dual still not quite right.  This is the type-2 system, and the lemma statement say dual VC, but type-1 works for all $r$?}
\begin{lemma}
\label{lem:vcdim-type} 
For every $r$ and every $T \in [1:\sigma]$, both set systems 
\[
\Paren{V_G, \Set{\hat N^r_{T}[s] : s \in V_P}} \quad \text{and} \quad \Paren{V_G, \Set{\hat N^r_{\le T}[s] : s \in V_P}}
\]
have dual VC-dimension at most $4$. 
\end{lemma}

\noindent Notice that $\hat N^r[s] = \hat N^r_{\le \sigma}[s]$ and
we can compute $\hat N^r_{\le T}[s]$ iteratively by 
\begin{align}
\label{Eq:induct-on-types}
\hat N^r_{\le T}[s] 
\,=\, 
\hat N^r_{\le T-1}[s] \, \cup \, \hat N^r_{T}[s]
\,=\, 
\hat N^r_{\le T-1}[s] \, \cup \, \bigcup_{v \in N[s]\cap \Cell_{T}} \hat N^{r-1}_{\le \sigma}[v].
\end{align}
Now the strategy should be clear: 
We will compute $\hat N^r_{T}[s]$ from the previously stored $\hat N^{r-1}_{\le \sigma}[v]$ for every 1-neighbor $v$ of $s$, then take the union with $\hat N^r_{\le T-1}[s]$.
The first operation is done with the help of the geometric data structure;  to do so, one has to first switch the interval representation for the relevant neighborhood balls to be with respect to a unifying stabbing path over some \emph{combined} set system that has a bounded VC-dimension.  
After $\hat N^r_{T}[s]$ is computed, we switch the interval representation back then proceed to compute $\hat N^r_{T}[s] = \hat N^r_{\le T-1}[s]  \cup \hat N^r_{T}[s]$, this time switching the interval representations to be with respect to another unifying stabbing path over some \emph{auxiliary} set system.
Both set systems must have $O(1)$ (dual) VC-dimension.
This is the main technical hurdle which we will explain next.

\paragraph{Geometric data structure.}
Fix a cell $\square$ containing $s$ and a perimetric cell $\square'$ of $\square$ which lies vertically above $\square$ (after a rotation).
By \Cref{L:pseudolines}, the collection of disks whose center lies in $\square'$ intersects $\square$ as a collection of $N$ pseudolines $\cL$.
Assume each pseudoline $\ell_v$ in $\cL$ has an associated interval $I_v$ in some given stabbing path $\lambda$. 
Our next goal is to describe how to build a stabbing path data structure $\mathcal{D}_\spath$ that answers the \textsc{Covers?}$(s,I)$ query:  whether the union of intervals $I_v$ for every object~$v$ intersecting $s$ covers the whole~$I$. (This data structure would then be used to solve the interval searching problem, which gives us the interval representation of the modified neighborhood of $s$.)
A disk $D_v$ intersecting a query disk $D_s$ with center $s$ in $\square$ corresponds to a pseudoline $\ell_v$ that lies below the center $s$ of the query disk. 
Therefore it is equivalent if we can support the following query:
\begin{itemize}
\item \textsc{Covers?}$(s,I)$:
Given a query point $s$ and an interval $I$, test whether $\displaystyle \bigcup_{\substack{\ell_v \in \cL \\ \text{$\ell_v$ below $s$}}} \! I_v$\, contains~$I$.
\end{itemize}

\begin{lemma}
\label{L:ds-pseudolines}
Fix a radius $r$.
Let $\spath^\flat$ be a stabbing path defined for the union of set systems 
\[
\Set{\hat N^{r-1}_{\le \sigma}[v] : v \in V_P} \cup \Set{\hat N^r_T[s] : s \in V_P}.
\]
A stabbing path data structure $\mathcal{D}^r_{T}$ (with respect to $\spath^\flat$) can be constructed in $n^{1+o(1)}$ time, and support each \textsc{Covers?}$(s,I)$ query in $n^{o(1)}$ time for any $s$. \note{Missing the stabbing path $\spath_\flat$ in the description of DS, and the query should have cell constraints about $s$ and pseudolines.}
\end{lemma}

A proof of \Cref{L:ds-pseudolines} can be found in \Cref{SS:ds-unitdisk}. By the same argument in \Cref{lm:unitsq-interval-searching}, we can augment $\mathcal{D}^r_{T}$ to construct the interval representation of $\hat{N}^r_T[s]$ with respect to $\spath^r_T$ for every $s$ by calling an interval search query \textsc{IntervalSearch}$(s)$ to $\mathcal{D}^r_{T}$. 
The readers might have noticed that the stabbing order maintained by the data structure $\mathcal{D}^r_{T}$ is not the same as the stabbing order we would like to represent the neighborhoods $\hat{N}^r_T[s]$ in.
This discrepancy leads to the need for tool that allows the switch between different interval representations.

\paragraph{Switching between interval representations.}

We gain the ability to switch between different interval representations by computing a special kind of stabbing paths that ``respect'' some common $\rho$-sampling of the set system $(X,\cS)$.  This requires us to compute stabbing paths not using \Cref{lem:stabbing-path-general}, but something more sophisticated.  
Ultimately we will be able to shrink from $\cS$ to a subcollection $\cS'$ of $\cS$, and vice versa.
First we set up the terminologies.  

We are given a set system $(X,\cS)$ with at most $n = |X|$ elements and $m = |\cS|$ sets with dual shatter dimension of $(X,\cS)$ is $d$.
We fix a \EMPH{unique $\rho$-sampling $\cR$} of $\cS$, where each set in $\cS$ chosen with probability $\rho/m$.  (Later on we will restrict $\cR$ to subcollection $\cS'$ of $\cS$ and obtain $\cR'$; we can still think of $\cR'$ as obtained from $\cS'$ by sampling each element with probability $\rho/m$, even though we do not explicitly sample from $\cS'$. Notice that the parameter $m$ does \emph{not} change even if $\cS'$ gets smaller.)

Let $\spath$ be an ordering of $X$.
We say that a set $S$ \EMPH{crosses} a pair $(x,y)$ if $x\in S$ and $y\not\in S$, or vice versa.  The number of consecutive pairs in $\spath$ crossed by $S$ is at most twice the size $|\Rep_\spath(S)|$.
For any collection $\cR$, define the \EMPH{equivalence relation $\equiv_\cR$} over $X$, where $x\equiv_\cR y$ if and only if 
no set in $\cR$ crosses $(x,y)$. 
(In other words, 
$\{S\in\cR: x\in S\}=\{S\in\cR: y\in S\}$.) 
Then $\equiv_\cR$ has $O(|\cR|^d)$ equivalence classes since the dual shatter dimension is at most $d$.

Let $\cS'$ be an arbitrary subcollection of $\cS$.
Denote the restriction of the unique $\rho$-sampling $\cR$ of $\cS$ in $\cS'$ as \EMPH{$\cR'$}; in notation, $\cR' \coloneqq \cS' \cap \cR$. 
Notice that $\cR'$ is also a $\rho$-sampling.
Given any set system $(X,\cS)$,
a stabbing path $\spath$ of $(X,\cS)$ is \EMPH{$\cR'$-respecting} if 
each equivalence class of $\equiv_{\cR'}$ appears contiguously in $\spath$ for the restriction $\cR'$.  (The equivalence classes of $\equiv_{\cR'}$ are defined with respect to the restriction $\cR'$, not $\cR$.)

We compute specialized stabbing paths using the following lemma.

\begin{restatable}{lemma}{respect}
\label{lem:respecting-stabbing-path}
Assume the existence of an element reporting oracle that, given $S \in \cS$, can enumerate all elements of $S$ in $T_0(n)$ time.
Consider a fixed $\rho$-sampling $\cR$ of $\cS$.
We can compute the equivalence classes of $\equiv_\cR$ and construct an $\cR$-respecting stabbing path $\spath$ of $(X,\cS)$ such that
$\sum_{S\in\cS} |\Rep_\spath(S)| = \OO(mn/\rho + m\rho^{d-1})$ in $\OO(T_0(n) \cdot \rho)$ time with high probability.
In other words, one can compute a sampled $\rho$-stabbing path $\spath$ of $(X,\cS)$ and the equivalence classes of $\equiv_\cR$ as byproducts.
\end{restatable}

For the case of unit-disks, we have the element reporting oracle with query time $T_0(n) = \OO(n)$ by computing a BFS tree using \Cref{lm:SSP-geo-intersection}.
Once both stabbing paths (and their corresponding equivalence classes) were computed, respecting a common $\rho$-sampling (and its restriction), we can convert one interval representation to the other efficiently.

\begin{restatable}{lemma}{shrinkexpand}
\label{lem:shrink-expand}
\emph{[Conversion of interval representations.]}\ 
Let $(X,\cS)$ be a set system with $|X|\le n$ and $|\cS| \le m$.
Let $\cS'$ be a subcollection of $\cS$ and $\cT$ be a subcollection of $\cS'$.
Let $\cR$ be the unique $\rho$-sampling of $\cS$, and $\cR'$ be its restriction in $\cS'$. 
We are given an $\cR$-respecting stabbing path $\spath$ of $(X,\cS)$, and an $\cR'$-respecting ordering $\spath'$ of $(X,\cS')$ (along with the equivalence classes of $\equiv_\cR$ and $\equiv_{\cR'}$).
\begin{enumerate}
\item[(1)] \emph{[Shrinking from $\cS$ to $\cS'$.]}\ 
Given $\Rep_\spath(S)$ for all $S\in\cT$, we can compute $\Rep_{\spath'}(S)$ for all $S\in\cT$ in $\OO(mn/\rho + m\rho^d)$ total time with high probability.
\item[(2)] \emph{[Expanding from $\cS'$ to $\cS$.]}\ 
Given $\Rep_{\spath'}(S)$ for all $S\in\cT$, we can compute $\Rep_{\spath}(S)$ for all $S\in\cT$ in $\OO(mn/\rho + m\rho^d)$ total time with high probability.
\end{enumerate}
\end{restatable}

The proof of the two lemmas can be found in \Cref{SS:reordering}.

\paragraph{Neighborhood growing algorithm.}
Assuming we are equipped with the geometric data structure (\Cref{L:ds-pseudolines}) and the ability to switch between interval representations with respect to different stabbing paths (\Cref{lem:shrink-expand}), we can now formally describe the algorithm.

Fix a piece $P$.  For simplicity of the proof, we use \EMPH{$\cS^r$} to denote the collection $\Set{\hat N^{r}[v]: v\in V_P}$ and \EMPH{$\cS^r_{M}$} to denote $\Set{\hat N^{r}_{M}[v]: v\in V_P}$ for any subset $M \subseteq [1:\sigma]$.  
(Recall that the modified balls are defined by intersecting with the relevant region $R_P$, and thus are dependent on $P$.)
Similarly, we define \EMPH{$\spath^{r}$} be a stabbing path for the $(V_G,\cS^r)$, and 
\EMPH{$\spath^r_{M}$} be a stabbing path for $(V_G,\cS^r_M)$ for any subset $M \subseteq [1:\sigma]$.
Since all the set systems we need here are with respect to the same ground set $V_G$, we will slightly abuse the notation and use the shorthand $\cS$ to denote the set system $(V_G, \cS)$,
and use $\cS_1 \cup \cS_2$ to denote the union of the two set systems $(V_G,\cS_1 \cup \cS_2)$.

The algorithm has an \emph{outer-loop} and an \emph{inner-loop}.  
The \EMPH{outer-loop} has $4\Delta+1$ rounds,
iterating over every relevant radii $r \in [\ecc_P - 3\Delta : \ecc_P + \Delta]$; 
at the start of round $r$, we maintain the following \EMPH{invariants} that we have computed, 
\begin{enumerate}
\item[(1)] an $\cR^{r-1}_{\le \sigma}$-respecting $\rho$-stabbing path $\spath^{r-1}_{\le \sigma}$ for the set system $\cS^{r-1}_{\le \sigma}$ and its $\rho$-sampling $\cR^{r-1}_{\le \sigma}$; and
\item[(2)] $\spath^{r-1}_{\le \sigma}$-representation of the modified neighborhood balls $\hat N^{r-1}_{\le \sigma}[v]$ for every vertex $v$ in $P$.
\end{enumerate}

For each round with radius $r$, 
our algorithm now performs an \EMPH{inner-loop} by
repeating the following steps for $\sigma$ \EMPH{iterations}, where $T$ ranges from $1$ to $\sigma$; in iteration $T$ we take into account type-$T$ shortest paths using Eq.~(\ref{Eq:induct-on-types}), until we include all $\sigma$ types and thus finish computing the $\spath^r$-representation of $\cS^r$.
At the start of iteration $T$, we maintain the following \EMPH{invariants} that we have computed, for each type $T$,
\begin{enumerate}
\item[(i)] an $\cR^r_{\le T-1}$-respecting $\rho$-stabbing path $\spath^r_{\le T-1}$ for the set system $\cS^r_{\le T-1}$ and its $\rho$-sampling $\cR^{r}_{\le T-1}$;
\item[(ii)] $\spath^r_{\le T-1}$-representation of the modified neighborhood balls $\hat N^{r}_{\le T-1}[s]$ for every vertex $s$ in $P \setminus \bd P$.
\end{enumerate}

\noindent At every iteration $T$, we perform the following steps in order.  
For the base case when $T=1$, objects with the $_{\le T-1}$ subscript are considered to be null, which we omit from the algorithm.  
\begin{enumerate}\itemsep=0pt
\item 
Consider the \EMPH{combined set system} $\cS^{r-1}_{\le \sigma} \cup \cS^{r}_{T}$.  
Compute a $\rho$-sampling \EMPH{$\cR^r_T$} of $\cS^r_{T}$, and take union with the $\rho$-sampling $\cR^{r-1}_{\le \sigma}$ of $\cS^{r-1}_{\le \sigma}$ from invariant (1) to form a $2\rho$-sampling \EMPH{$\cR^\flat$} of $\cS^{r-1}_{\le \sigma} \cup \cS^{r}_{T}$.

\item
Compute an $\cR^\flat$-respecting $2\rho$-stabbing path \EMPH{${\smash{\spath^\flat}}$} along with the equivalence classes of $\equiv_{\spath^\flat}$ for the combined set system using
Lemma~\ref{lem:respecting-stabbing-path}.

\item 
Convert the $\spath^{r-1}_{\le \sigma}$-representation of $\hat N^{r-1}_{\le \sigma}[v]$ for every $v$ in $P$ given by invariant (2) into $\spath^\flat$-representation, using Lemma~\ref{lem:shrink-expand}(2) and the fact that $\cS^{r-1}_{\le \sigma}$ is a subcollection of the combined set~system. 

\item
Compute the geometric data structure \EMPH{$\mathcal{D}^{r}_{T}$} with respect to the $\rho$-stabbing path $\spath^\flat$ using Lemma~\ref{L:ds-pseudolines}. 

\item
Compute the $\spath^\flat$-representation of 
\[
\hat N^{r}_{T}[s] \,=\, \bigcup_{v \in N[s] \cap \Cell_{T}} \hat N^{r-1}_{\le \sigma}[v]
\]
for every vertex $s$ in $P \setminus \bdry P$ with the help of geometric data structure $\mathcal{D}^{r}_{T}$.

\item 
Convert the $\spath^\flat$-representation of $\hat N^{r}_{T}[s]$ for every $s$ in $P \setminus \bdry P$ back into $\spath^r_T$-representation, using Lemma~\ref{lem:shrink-expand}(1).

\setcounter{stepcount}{\value{enumi}}
\end{enumerate}

\noindent (At this point, we have successfully computed the spanning path $\lambda_T^r$ for $\cS_T^r$ and its interval representation with respect to $\lambda^r_T$.  We now proceed to take union with $\cS^r_{\le T-1}$, currently in $\lambda^r_{\le T-1}$-representation.)

\begin{enumerate}
\setcounter{enumi}{\value{stepcount}}
\item 
Define the \EMPH{auxiliary set system}:
\begin{align}
\label{eq:auxiliary-system}
\cS^r_{\le T-1} \cup \cS^r_{T} \cup \cS^r_{\le T} \,=\, 
\Paren{\big. V_G, \Set{\hat N^{r}_{\le T-1}[v]: v\in V_P} \,\cup\, \Set{\hat N^{r}_{T}[v]: v\in V_P} \,\cup\, \Set{\hat N^{r}_{\le T}[v]: v\in V_P} }.
\end{align}
Compute a $\rho$-sampling \EMPH{$\cR^r_{\le T}$} of $\cS^r_{\le T}$, and take union with the $\rho$-sampling $\cR^{r}_{\le T-1}$ of $\cS^{r}_{\le T-1}$ from invariant (i) and 
$\rho$-sampling $\cR^{r}_{T}$ of $\cS^{r}_{T}$ computed in Step~1
to form a $3\rho$-sampling \EMPH{$\cR^\sharp$} of the auxiliary set system $\cS^r_{\le T-1} \cup \cS^r_{T} \cup \cS^r_{\le T}$.

\item 
Compute a $\cR^\sharp$-respecting $3\rho$-stabbing path \EMPH{${\smash{\spath^\sharp}}$} along with the equivalence classes of $\equiv_{\spath^\sharp}$ for the auxiliary set system, using Lemma~\ref{lem:respecting-stabbing-path}.

\item
Convert $\spath^r_{\le T-1}$-representation of $\hat N^{r}_{\le T-1}[s]$ for every $s$ in $P \setminus \bd P$ given by invariant (ii) to $\spath^{\sharp}$-representations, using Lemma~\ref{lem:shrink-expand}(2).

\item
Convert $\spath^r_{T}$-representation of $\hat N^{r}_{T}[s]$ for every $s$ in $P \setminus \bd P$ given by Step~6 to $\spath^{\sharp}$-representations, using Lemma~\ref{lem:shrink-expand}(2).

\item
Compute $\hat N^{r}_{\le T}[s]$ by taking the union of $\hat N^{r}_{\le T-1}[s]$ and $\hat N^{r}_{T}[s]$ as $\spath^\sharp$-representations for every $s$ in $P \setminus \bd P$.
The output $\hat N^{r}_{\le T}[s]$ is again in the auxiliary set system and thus have $\spath^\sharp$-representation.

\item 
Compute an $\cR^r_{\le T}$-respecting $\rho$-stabbing path \EMPH{$\spath^r_{\le T}$} for set system $\cS^r_{\le T}$, by restricting $\spath^\flat$ to $\cS^r_{\le T}$. 
Convert the $\spath^\sharp$-representation of $\hat N^{r}_{\le T}[s]$ for every $s$ in $P \setminus \bdry P$ into $\spath^r_{\le T}$-representation, using Lemma~\ref{lem:shrink-expand}(1) and the fact that $\cS^{r}_{\le T}$ is a subcollection of the auxiliary set system.
\end{enumerate}
Notice that Step~12 of the algorithm maintains invariants (i) and (ii). 
After $\sigma$ iterations, the inner-loop ends.  We perform one extra step:
\begin{enumerate}
\item[13.] 
Convert the $\spath^\flat$-representation of
$\hat N^{r-1}_{\le \sigma}[s]$ for every $s$ in $\bdry P$ computed in Step~3 into $\spath^r_{\le \sigma}$-representation, using Lemma~\ref{lem:shrink-expand}(1).
Insert elements in the difference $\hat N^r_{\le \sigma}[s] \setminus \hat N^{r-1}_{\le \sigma}[s]$ to create $\spath^r_{\le \sigma}$-representation of $\hat N^r_{\le \sigma}[s]$ for every $s$ in $\bdry P$.
\end{enumerate}
We then proceed to the next round of the outer-loop.
Notice that invariant (1) follows directly from invariant (i), and
invariant (2) follows from invariants (ii) together with Step~13 from the previous round.

\paragraph{Handling small pieces.}
The most time-consuming part of our algorithm is to compute the $\rho$-stabbing paths; each computation takes $\OO(n\rho)$ time. 
But we have to compute $O(1)$ many stabbing paths \emph{for each piece and each round of the outer-loop}; there are $\OO(n/\Delta)$ many pieces (remember the modified neighborhood balls were defined differently for each piece $P$), but also for $O(\Delta)$ many rounds.  Therefore the computation of stabbing paths alone already takes $\OO(n\rho \cdot (n/\Delta) \cdot \Delta) = O(n^2\rho)$ time. 

To handle the issue, we only apply the above algorithm to \emph{large} pieces whose size is above certain threshold $A \ge \Delta$.  
This way, the number of such pieces is at most $\OO(n/A)$ instead of $\OO(n/\Delta)$.  (We eventually set $A = \Delta^{O(1)}$.)
To compute diameter or eccentricities for small pieces, we use \Cref{lm:diameter-smallpiece}.

\subsection{Analysis for Eccentricities}

We make the following observations about the shatter dimension of unions of set systems.
\begin{observation}
\label{obs:dual-shatter-facts}
Let $X$ be a ground set and let $\cS_1$ and $\cS_2$ be two set systems on $X$. Let us denote the set of ranges obtained by taking unions of ranges from $\cS^1$ and $\cS^2$ by $\widehat \cS \coloneqq \{S_1\cup S_2 : S_1\in \cS_1, S_2 \in \cS_2\}$.
Suppose that the shatter dimension of $\cS_1$ is $d_1$ and the shatter dimension of $\cS_2$ is $d_2$. Then the shatter dimension of $\cS_1\cup \cS_2$ is at most $d_1 + d_2$, and the shatter dimension of $\cS_1\cup \cS_2\cup\widehat \cS$ is also at most $d_1 + d_2$.
\end{observation}
The observation shows that the combined set system $\cS^{r-1}_{T} \cup \cS^{r}_{T}$ has dual shatter dimension $8$, because both $\cS^{r-1}_{T}$ and $\cS^{r}_{T}$ individually has dual VC-dimension (and thus dual shatter dimension) at most 4 by \Cref{lem:vcdim-type}. 
It also shows that the dual shatter dimension of the auxiliary set system $\cS^r_{\le T-1} \cup \cS^r_{T} \cup \cS^r_{\le T}$ is also $8$ again as the individual dual VC-dimensions are at most $4$~\Cref{lem:vcdim-type}.
Thus we set $d=8$ for the time analysis that follows.

\begin{lemma}
\label{lem:ball-growing-unit-disk}
Fix a piece $P$, a radius $r$, and some arbitrary parameter $\rho$.  
We can maintain invariants~(1) and (2) between iterations $T-1$ and $T$, in time $\OOO(n\cdot \rho + |P| \cdot (n/\rho + \rho^8))$,
by computing (1) an $\cR^{r}$-respecting $\rho$-stabbing path $\spath^{r}_{T}$ for the set system $\cS^{r}_{T}$ for some $\rho$-sampling $\cR^{r}$; and
(2) $\spath^{r}_{T}$-representation of the modified neighborhood balls $\hat N^{r}_{T}[v]$ for every vertex $v$.
\end{lemma}

\begin{proof}
Steps~1 and 7 take $O(n)$ time to compute $\rho$-samplings.
Steps 2 and 8 take $\OO(n\cdot \rho)$ time to compute $O(\rho)$-stabbing paths.
Steps~3, 6 and 13 take $\OO(|P| \cdot(n/\rho + \rho^8))$ time to convert interval representations, because the combined set system $\cS^{r-1}_{\le \sigma} \cup \cS^{r}_{T}$ has size $2|P|$.
Step~4 takes $O(n^{1+o(1)})$ time to compute the geometric data structure $\cD^r_T$.
Step~5 takes $O(n \cdot n^{o(1)})$ time to compute the union using $\cD^r_T$.
Steps 9, 10, 12 take $\OO(|P| \cdot(n/\rho + \rho^8))$ time to convert interval representations,
because the auxiliary set system $\cS^{r}_{\le T-1} \cup \cS^{r}_{T} \cup \cS^r_{\le T}$ has size $3|P|$. 
Step~11 takes $O(n)$ time to compute the union of two sets with the same $\spath^\sharp$-representation.
Overall the neighborhood growing step can be implemented in $\OOO(n\cdot \rho + |P| \cdot (n/\rho + \rho^8))$ time per piece per radius.
\end{proof}

To analyze the total running time, we separate the pieces of LDD into large and small, based on whether the size of the piece is at least $A$ or not for some parameter $A$.
For large piece $P$ of size at least $A$, we run the ball growing algorithm in \Cref{SS:ball-growing-unitdisk}.
The outer-loop is executed for $O(\Delta)$ rounds, each taking $\OOO(n\cdot \rho + |P| \cdot (n/\rho + \rho^8))$ time by \Cref{lem:ball-growing-unit-disk}, followed by Step 13, which takes $\OO(\sum_{s \in \bd P} \Abs{\hat N^r_{\le \sigma}[s] \setminus \hat N^{r-1}_{\le \sigma}[s]} )$ time to compute $\hat N^r_{\le \sigma}[s]$ for every $s$ in $\bdry P$ using binary search in the $\spath^r_{\le \sigma}$-representation.
There are $O(n/A)$ large pieces.
Overall, for each large piece, this takes time 
\begin{align*}
&~ \OO\Paren{ \sum_r \sum_{s \in \bd P} \Abs{\hat N^r_{\le \sigma}[s] \setminus \hat N^{r-1}_{\le \sigma}[s]} } + O(\Delta) \cdot \OOO(n\cdot \rho + |P| \cdot (n/\rho + \rho^8))  \\
= &~ \OOO\Paren{ |\bd P| \cdot n + \Delta \Paren{n\cdot \rho + |P| \cdot (n/\rho + \rho^8) }}.
\end{align*}
By \Cref{lm:diameter-smallpiece}, for each small piece $P$ of size less than $A$, it takes $\OO( n \cdot |\bdry P| + |P| \cdot (|P| + (|\bdry P| \Delta)^4) )$ time to compute eccentricity of all vertices in $P$.
There are at most $\OO(n/\Delta)$ small pieces.

\medskip
The final running time of the all-vertex eccentricities algorithm for unit-disk graphs is:
\begin{align*} 
& \OOO\left( \sum_{P:\, |P|>A} \Paren{\Big. |\bd P| \cdot n + \Delta \cdot (n\cdot\rho + |P|\cdot (n/\rho + \rho^8)) } +  \sum_{P:\, |P|\le A} \Paren{\Big. n \cdot |\bdry P| + |P| \cdot (|P| + (|\bdry P| \Delta)^4) } \right) \\
~\le~ & \OOO\left( n^2/\Delta + \Paren{\frac{n}{A} \cdot \Delta \cdot n\rho} + \sum_{P:\, |P|>A} \Paren{|P|\cdot (n/\rho + \rho^8)} +  \sum_{P:\, |P|\le A}  (n \cdot |\bdry P| + |P|^2)  +  \sum_{P:\, |P|\le A} \Paren{ |\bdry P| \cdot A^4 \Delta^4 } \right) \\
~=~ & \OOO\Paren{ n^2/\Delta + \Delta n^2\rho/A + \Delta n^2/\rho + \Delta n\rho^8 + n^2/\Delta + nA^2/\Delta + \Delta^3 A^4 n }.
\end{align*}
Balancing cost by setting parameters $\Delta=n^{1/20}$, $\rho=\Delta^2$ and $A=\Delta^4$ then yields $\OOO(n^{2-1/20})$. 

\begin{theorem}
Computing eccentricities of an $n$-node unit-disk graph can be done in $\OOO(n^{2-1/20})$ time.
\end{theorem}

\subsection{Analysis for Diameter}

For the special case of computing the diameter of unit-disk graphs, we can get a slight improvement in running time by making the following observation:
\begin{quote}
In the analysis for the all-vertex eccentricities algorithm, computing $\rho$-stabbing paths in Step 2 and Step 8 using Lemma~\ref{lem:respecting-stabbing-path} takes $\OO(n \cdot \rho)$ time per piece per $r$, which is the bottleneck.
We can instead compute a \emph{global} $\rho$-stabbing path per type for both the combined set system and the auxiliary set system at the start of each iteration of the inner-loop, then restrict these stabbing paths to each piece $P$.\note{Emphasize that this is possible because for diameter, the estimated $r$ is the same across all pieces.}
\end{quote}
More specifically, at the start of iteration $T$:
\begin{enumerate}
\item[0.1.]
Consider the \EMPH{global combined set system} $\Paren{V_G, \Set{N^{r-1}_{\le \sigma}[v] : v \in V_G} \cup \Set{N^r_T[v] : v \in V_G}}$.
This set system differs from $\cS^{r-1}_{\le \sigma} \cup \cS^r_T$ in two places: the neighborhood balls are \emph{not} modified, and the ball centers range over all vertices in $G$, not just in $P$.

Compute a $\rho$-sampling \EMPH{$\check \cR^r_T$} of $\Paren{V_G, \Set{N^r_T[v] : v \in V_G}}$, then take union with the $\rho$-sampling of $\Paren{V_G, \Set{N^{r-1}_{\le \sigma}[v] : v \in V_G}}$ computed from the previous round $r-1$ to form a $\rho$-sampling \EMPH{$\check \cR^\flat$} of the global combined set system.

\item[0.2.]
Compute an $\check \cR^\flat$-respecting $\rho$-stabbing path \EMPH{${\check\spath^\flat}$} along with the equivalence classes of $\equiv_{\check\spath^\flat}$ for the global combined set system using
Lemma~\ref{lem:respecting-stabbing-path}.

\item[0.3.]
Consider the \EMPH{global auxiliary set system} 
\[
\Paren{V_G, \Set{N^{r}_{\le T-1}[v] : v \in V_G} \cup \Set{N^r_T[v] : v \in V_G} \cup \Set{N^r_{\le T}[v] : v \in V_G} }.
\]

Compute a $\rho$-sampling \EMPH{$\check \cR^r_{\le T}$} of $\Paren{V_G, \Set{N^r_{\le T}[v] : v \in V_G}}$, then take union with the $\rho$-sampling of $\Paren{V_G, \Set{N^{r}_{\le T-1}[v] : v \in V_G}}$ computed from the previous iteration $T-1$ 
and $\rho$-sampling of $\Paren{V_G, \Set{N^{r}_{T}[v] : v \in V_G}}$ from Step~0.1 to form a $\rho$-sampling \EMPH{$\check \cR^\sharp$} of the global auxiliary set system.

\item[0.4.]
Compute an $\check \cR^\sharp$-respecting $\rho$-stabbing path \EMPH{${\check\spath^\sharp}$} along with the equivalence classes of $\equiv_{\check\spath^\sharp}$ for the global auxiliary set system using
Lemma~\ref{lem:respecting-stabbing-path}.
\end{enumerate}

\noindent Then, for each piece $P$, we modify the following steps in iteration $T$:
\begin{enumerate}
\item[2.]
Restrict the $\check \cR^\flat$-respecting $\rho$-stabbing path $\check\spath^\flat$ to another stabbing path \EMPH{$\spath^\flat$} for the combined set system $\cS^{r-1}_{\le \sigma} \cup \cS^{r}_T$ for piece $P$.  
This is done by first removing every neighborhood ball $N^r_T[v]$ not centered in $P$, then taking intersection between $N^r_T[v]$ and the relevant region $R_P$ to form $\hat N^r_T[v]$.

\item[8.]
Restrict the $\check \cR^\sharp$-respecting $\rho$-stabbing path $\check\spath^\sharp$ to another stabbing path \EMPH{$\spath^\sharp$} for the auxiliary set system $\cS^{r-1}_{\le \sigma} \cup \cS^{r}_T$ for piece $P$.  
\end{enumerate}

\paragraph{Analysis.}
We only count for the new changes in the diameter case; for the remaining steps, see the time analysis for computing eccentricities.

In the new Step~2, the removal of balls not centered in $P$ does not increase the stabbing number of $\check\spath^\flat$.
Taking intersection with $R_P$ does not change the stabbing number, because this is equivalent to restricting the stabbing path range from $[1:n]$ to $R_P$ (in the same order), and what was one interval in $[1:n]$ remains one interval in $R_P$.
So $\spath^\flat$ is still a $\rho$-stabbing path.
The $\rho$-sampling \EMPH{$\cR^\flat$} can be obtained by restricting $\check \cR^\flat$ to $\cS^{r-1}_{\le \sigma} \cup \cS^r_{T}$. 
The removal of balls not centered in $P$ only decreases the number of sets in consideration and thus makes $\equiv_{\spath^\flat}$ coarser than $\equiv_{\check\spath^\flat}$.
Taking intersection with $R_P$ does not change the status of $\hat N^r_T[v]$ being chosen in the sample or not.
Thus $\spath^\flat$ remains $\cR^\flat$-respecting.
As a result, $\spath^\flat$ is an $\cR^\flat$-respecting $\rho$-stabbing path.

For the new Step~8, using similar reasoning, $\spath^\sharp$ is an $\cR^\sharp$-respecting $\rho$-stabbing path.

Steps~0.2 and 0.4 take $\OO(n\cdot \rho)$ time.
Steps 2 and 8 now take $O(n)$ time to carry out the restriction.  (Only the part about restricting $[1:n]$ to $R_P$ needs to be implemented, not the removal of balls centered outside $P$.)
Overall the neighborhood growing step can be implemented in $\OOO(n + |P| \cdot (n/\rho + \rho^8))$ time per piece, plus another $\OO(n\rho)$ time across all pieces.

\medskip
The final running time of the diameter algorithm for unit-disk graphs is:
\begin{align*} 
& \OOO\left( \Delta\cdot  n\rho + 
\sum_{P:\, |P|>A} \Paren{\Big. |\bd P| \cdot n + \Delta \cdot (n + |P|\cdot (n/\rho + \rho^8)) } +  \sum_{P:\, |P|\le A} \Paren{\Big. n \cdot |\bdry P| + |P| \cdot (|P| + (|\bdry P| \Delta)^4 )  } \right) \\
~=~ & \OOO\Paren{ n^2/\Delta + \Delta n\rho +  \Delta n^2/A + \Delta n^2/\rho + \Delta n\rho^8 + n^2/\Delta + nA^2/\Delta + \Delta^3 A^4 n }.
\end{align*}
Balancing cost by setting parameters $\Delta=n^{1/18}$ and $\rho=A=\Delta^2$ then yields $\OOO(n^{2-1/18})$. 

\begin{theorem}
Computing diameter of an $n$-node unit-disk graph can be done in $\OOO(n^{2-1/18})$ time.
\end{theorem}

\section{Framework for Distance Oracles (and Wiener Index)}
\label{sec:framework-do}

Our algorithm for computing the Wiener index is a simple extension of our algorithm for computing the exact distance oracle. Therefore, in the following, we focus exclusively on describing the framework for computing an exact distance oracle. Then we give more details on how to compute the Wiener index with the same running time.

For distance oracles, the first two steps are the same as in the framework for eccentricities described in \Cref{sec:framework}. Nonetheless, we present a full description of the framework since there are significant differences in step 3. 
The key difference is that instead of a specific set of relevant vertices $R_P$ for piece $P$, we need to consider all vertices $V_G$. 
The distances we need to consider will vary depending on the vertex $t\in V_G$, and therefore, we could not use the same definition of $\hat{N}^r[s]$ in the diameter computation for distance oracles.
However, we observe that since we have a good additive estimate $\hat d$ that is within $\pm \Delta$ of the true distance between $t\in V_G$ and a vertex $s\in P$, we only need to consider distances in an $O(\Delta)$ range around $\hat d$. Our idea is to add a weight to each vertex $t$ and use vertex weights to define $\hat{N}^r[s]$ (\Cref{eq:defn-nhat-sparse}).

In our oracle construction, it is important to distinguish between large and small pieces (determined by some size threshold) in the LDD $\mathcal{L}$.  For large pieces, we will use the interval representation. For small pieces, we use the oracle construction of \Cref{lm:do-sparse-exact-log}.

\paragraph*{Oracle construction.} Let $A$ be the parameter chosen later. For each piece $P$ in an LDD $\mathcal{L}$.

\begin{enumerate}
\item Compute a low-diameter decomposition $\mathcal{L}$ of $G$ with a diameter parameter $\Delta > 0$. 
\item For each vertex $v\in \bigcup_{P\in \ldd} \bdry P$ compute a breath-first search tree in $G$ rooted at $v$. 
\item For each piece $P \in \mathcal{L}$ where $|P| > A$, let $s_P$ be an arbitrary vertex of $\bdry P$. For every vertex $v\in V_G$, we compute and store a weight $w_P(v) = d_G(v, s_P)$. 
 Observe by the triangle inequality that for any vertex $s\in P$: 
 \begin{equation*}
     w_P(v) - \Delta \leq d_G(s,v) \leq  w_P(v) + \Delta
 \end{equation*}
Now we define an adjusted neighborhood ball as follows:
\begin{align}\label{eq:defn-nhat-sparse}
\hat N^{r}[s] := \{ v\in V : d_G(v, s) \le r+w_P(v)\} \qquad \forall r\in [-\Delta, \Delta]
\end{align}
Then we compute $\hat N^r[s]$ with the ball expansion data structure $\cD$ and store all intermediate balls in the following procedure:
\begin{enumerate}
\item[3.1] For every $s\in \bdry P$, we can explicitly compute the modified balls $\hat N^r[s]$ for all $r\in [-\Delta, \Delta]$  as well as compute and store a compact interval representation with respect to an ordering $\lambda$.
\item[3.2] As a base case, we initialize $\hat{N}^r[s] = \varnothing$ for every $s\in P$ when  $r = -\Delta-1$.  
\item[3.3] 
For other  values of $r \in  [-\Delta,\Delta]$, compute $\Rep_\lambda(\hat N^r[s])$ using the inductive formula 
\begin{align}
\hat N^r[s] = \bigcup_{v\in N[s]} \hat N^{r-1}[v]
\end{align}
by taking the union of the intervals.
\end{enumerate}
\item For each piece $P\in \ldd$ with $|P| < A$, construct the distance oracle of \Cref{lm:do-sparse-exact-log}.
\end{enumerate}

\paragraph*{Correctness of ball expansion initialization.}
Since $d_G(s_P,t) - \Delta \le d_G(s,t) \le d_G(s_P, t) + \Delta$ for every $t\in V$, $t\notin\hat N^{-\Delta-1}[s]$ and $t\in \hat N^{\Delta}[s]$. Thus, the initialization is correct, and we have correctly computed the desired modified neighborhood balls.

\paragraph*{Answering queries.}
Suppose we get a distance query between a vertex $s$ that is in a piece $P\in \ldd$, and any other vertex $t\in G$.
If $|P|< A$, we query the distance oracle for small pieces, and by \Cref{lm:do-sparse-exact-log} the query time is $O(\log n)$. Otherwise, for any $r\in [-\Delta, \Delta]$, we can detect if $t\in \hat N^r[s]$ by checking if $t$ lies in an interval of $\Rep_\lambda(\hat N^r[s])$ by binary search in $O(\log n)$ time\footnote{We can reduce this running time to $O(1)$ by using the fractional cascading technique; this would complicate the details.}. Thus, we can binary search for the first radius $r_t$ such that $t\in \hat N^{r_t}[s]$ and $t\notin \hat N^{r_t-1}[s]$. By the definition of $\hat N$ of \Cref{eq:defn-nhat-sparse}, we can conclude:
\[ d_G(s,t) = d_G(t, s_p) + r_t.\]
In either case, we spend $\OO(1)$ time.

\paragraph{Computing the Wiener index.} In the oracle construction, we compute and store $\hat{N}^r[s]$ for every $s$ in a large piece $P$.
For every vertex $t \in \hat{N}^r[s]\setminus \hat{N}^{r-1}[s]$, the exact distance from $s$ to $t$ is $d_G(t,s_p)+r$, and hence $\sum_{t\in \hat{N}^r[s]\setminus \hat{N}^{r-1}[s]} d_G(s,t)= \sum_{t\in \hat{N}^r[s]\setminus \hat{N}^{r-1}[s]} d_G(t,s_p)+|\hat{N}^r[s]\setminus \hat{N}^{r-1}[s]|\cdot r$. This allows us to compute $\sum_{v\in V}d_G(s,v)$ in the same running time as it takes to construct the interval representation of $\{\hat{N}^r[s]\}_{r=-\Delta}^{\Delta}$. 
For small pieces, Le and Wulff-Nilsen~\cite{le2023vc} provided an algorithm for computing $\sum_{v\in V}d_G(s,v)$ that has the same running time as the construction time for exact oracles 
of small pieces. Therefore, the time to compute the Wiener index is the same as the time to construct an exact distance oracle.

\paragraph*{Organization.}
In the next four sections, we will apply our framework to devise algorithms for exact distance oracles (and thus Wiener index) for different graph classes: sparse graphs of bounded VC-dimension (\Cref{sec:do-sparse}),
arbitrary-square graphs (\Cref{sec:do-square}),
unit-square graphs (\Cref{sec:do-unit-square}),
and unit-disk graphs (\Cref{sec:do-unit-disk}).
There will be similarities with earlier sections on diameter (Sections \ref{sec:ecc-sparse}--\ref{sec:unit-disks}).

\section{Distance Oracles for Sparse Graphs of Bounded VC-dimension}\label{sec:do-sparse}

We begin by considering sparse graphs of bounded VC-dimension.

\paragraph*{Stabbing path construction.~}
For a piece $P\in \ldd$, let $\vol(P) = \sum_{s\in P}\deg(s)$ be the total degree of vertices in $P$, i.e., the \EMPH{volume} of $P$. 
We will construct a stabbing path $\lambda_P$ for each piece $P\in \ldd$ satisfying 
$\OO(1)\cdot \sum_{r=-\Delta}^{\Delta}\sum_{s\in P}\deg(s)\cdot |\Rep_{\lambda_P}(\hat N^r[s])| = \OO(\Delta\vol(P)(n/\rho+\rho^{d-1}))$ for a parameter $\rho$ to be specified later 
using \Cref{lem:respecting-stabbing-path}.

\paragraph*{Construction time.~} 
Computing the low diameter decomposition and the boundary distances stored in step 2 takes $\OO(mn/\Delta)$ time.
For large pieces, the total construction time involves computing the ordering $\lambda$ in $\OO(m\rho)$ time (by \Cref{lem:respecting-stabbing-path}) and the ball expansion procedure which takes $O(\Delta \vol(P) \cdot (n/\rho+\rho^{d-1}))$ time per piece. Thus, the total running time is: 
\begin{equation*}
    \sum_{\substack{P\in \ldd \\|V_p| \ge A}} \OO(m\rho +\Delta \vol(P)\cdot(n/\rho+\rho^{d-1})) 
    ~=~ \OO(nm\rho/A+\Delta mn/\rho+\Delta m\rho^{d-1}).
\end{equation*}

For small pieces, we observe that we can compute a vertex weighted BFS on $P$ with weights at most $\Delta$ in time $\OO(\vol(P))$. 
Therefore,  in \Cref{lm:do-sparse-exact-log}, $T(P) = \OO(\vol(P))$, giving the total running time for all the small pieces:
\begin{equation*}
\begin{split}
   \sum_{\substack{P\in\ldd\\ |V_P|\le |A|}}  O(n|\bdry P|+(|\bdry P|^d \Delta^d + |P|) \cdot T(P)) 
   & ~=~ \OO(n^2/\Delta ) + \OO(A^d\Delta^d + A) \cdot \sum_{P\in\ldd} \vol(P) \\[-11pt]
   & ~=~ \OO(n^2/\Delta  + mA^d\Delta^d).
\end{split}
\end{equation*}
The total running time for the algorithm is:
\[ \OO(mn/\Delta + nm\rho/A+\Delta mn/\rho+\Delta m\rho^{d-1}+ mA^d\Delta^d) = O(mn^{1-1/(4d+1)})
\]
for $\Delta = n^{1/(4d+1)}$, $\rho=\Delta^2$, and $A=\Delta^3$.

\paragraph*{Space usage.~} The boundary distances that we store in step 2 take $\OO(n^2/\Delta)$ space.
For large pieces, in step 3, we use $\OO(\Delta\vol(P)(n/\rho+\rho^{d-1}))$ space to store compact representations of the neighborhood balls and $O(n)$ space to store distances from each vertex to $s_P$. We also use $O(n)$ space per boundary vertex to store $\hat N^r[s]$ for all $r\in [-\Delta, \Delta]$ in step 3(0) by storing $\{\hat N^{r}[s]\setminus  \hat N^{r-1}[s]\}$ for every $r$. Thus, the total space is:
\begin{align*}    
    \sum_{\substack{P\in\ldd\\ |V_P|\ge |A|}} \OO(n\cdot |\bdry P| +\Delta\vol(P)(n/\rho+\rho^{d-1})) 
    &~=~ \OO(n^2/\Delta + \Delta m (n/\rho+\rho^{d-1})) \\[-11pt] 
    &~=~ \OO(n^2/\Delta + mn/\Delta + m\Delta^{2d-1}) \quad (\text{since } \rho = \Delta^2).
\end{align*}
For each small piece, step 4 requires $O(n|\bdry P| + |V_P|^d)$ space by \Cref{lm:do-sparse-exact-log}. The total space required for all small pieces is 
\[ 
\sum_{\substack{P\in\ldd\\ |V_P|\le |A|}} O(n|\bdry P| + |V_P|^d)  ~=~  \OO(n^2/\Delta +nA^{d-1})= \OO(n^2/\Delta +n\Delta^{3d-3}) \quad \text{space as} \quad A = \Delta^3.\]
Therefore, the total space of our oracle is:
\begin{align*}
    \OO\left( n^2/\Delta + mn/\Delta + m\Delta^{2d-1} + n\Delta^{3d-3}\right)
  \ &~=~ \OO(mn^{1-1/(4d+1)})
\end{align*}
for $\Delta = n^{1/(4d+1)}$. 

\begin{theorem}
Given undirected graph $G$ with $n$ vertices and $m$ edges that has generalized distance VC-dimension at most $d$, we can construct in $\OO(mn^{1-1/(4d+1)})$ time an exact distance oracle of $\OO(mn^{1-1/(4d+1)})$ space and $\OO(1)$ query time.
\end{theorem}

\begin{remark}\label{rm:graph-space} We chose our parameters to minimize the construction time. We can trade off between space and query time. In the extreme, if construction time does not matter, we can apply the large piece solution to all pieces to obtain a distance oracle using $\OO(mn^{1-1/(2d)})$ space.
\end{remark}

\begin{remark}\label{rm:graph-other-vc}
     In this section, we assumed the graph has a bounded distance VC-dimension. The exponent can be slightly optimized when the time it takes to perform BFS in a piece $P$ is $O(|P|)$ instead of $O(\vol(P))$. This is the case for minor-free graphs, where the space can be improved to $\OO(mn^{1-1/(4d)})$. We can also obtain similar results (albeit with worse exponents) if we make other bounds on VC-dimension, such as the distance VC-dimension, and even if we only assume that the $k$-neighborhood VC-dimension is bounded by $d$ for all $k$.
\end{remark}

\section{Distance Oracles for Square Graphs}
\label{sec:do-square}
For square graphs,
we follow the construction of the oracle in \Cref{sec:framework-do}. We note that the VC-dimension $d = 4$ in this case. We only analyze the construction time since space is bounded by it. 

\paragraph*{Stabbing path construction.~} For a piece $P\in \ldd$. We will construct a stabbing path $\lambda_P$ for each piece $P\in \ldd$ satisfying:
\begin{equation}\label{eq:sq-oracle-path}
    \OO(1)\cdot \sum_{r=-\Delta}^{\Delta}\sum_{s\in P}|\Rep_{\lambda_P}(\hat N^r[s])| = \OO(\Delta|P|(n/\rho+\rho^{3}))
\end{equation}
for a parameter $\rho$ to be specified later 
using \Cref{lem:respecting-stabbing-path}. The running time is $\OO(\rho n)$ as we show in \Cref{sec:low-diameter} that we can find a BFS tree in square graphs in $\OO(n)$  time. 

Given the interval representation $\{\hat N^{r-1}[s]\}_{s\in P}$ for radius $r-1$, we compute the interval representation of $\{\hat N^{r}[s]\}_{s\in P}$  using the data structure $\check{\cD}$ for the interval search problem for squares (\Cref{lm:square-interval-searching}) in the eccentricities computation with the same setup: the input contains a set of squares corresponding to vertices of $P$ and the interval representation  $\{\Rep_{\lambda_P}(\hat N^{r-1}[s])\}_{s\in P}$ for radius $r-1$. The queries are $\{\textsc{IntervalSearch}(s): s\in P\}$ whose outputs are the interval representations of $\{\hat N^{r}[s]\}_{s\in P}$. The total time to grow balls for all radii, using the same efficient encoding as in the improved algorithm for computing eccentricities in \Cref{sec:ecc-square}, is:
\begin{equation}\label{eq:sq-oracle-grow}
    \begin{split}
    &~ \sum_{r=-\Delta}^{\Delta} \OO(b\cdot \sum_{s\in P} (|\Rep_{\lambda_P}(\hat N^{r-1}[s])|  + |\Rep_{\lambda_P}(\hat N^{r}[s])|)) + \OO(|P|n/b) \\ 
    ~=~ &~ \OO(b\Delta |P|(n/\rho + \rho^{3}) + |P|n/b) \quad (\text{by \Cref{eq:sq-oracle-path}}).      
    \end{split}
\end{equation}

\paragraph{Construction time.} 
For small pieces, we show (in \Cref{lm:bdr-BFS-square} in the appendix) that we can compute a vertex weighted BFS on $P$ with weights at most $\Delta$ in time $\OO(|P|)$. Therefore,  in \Cref{lm:do-sparse-exact-log}, $T(P) = \OO(|P|)$, giving the total running time for all the small pieces:
\begin{equation*}
\begin{split}
   \sum_{\substack{P\in\ldd\\ |V_P|\le |A|}}  \OO(n|\bdry P|+(|\bdry P|^4 \Delta^4 + |P|) \cdot |P|)& ~=~ \sum_{P\in\ldd}\OO(n|\bdry P|+A^{4} \Delta^4  \cdot |\bdry P|)\\[-6pt]
   &~=~ \OO(n^2/\Delta  + nA^4\Delta^3).
\end{split}
\end{equation*}

For large pieces, the running time to grow balls (\Cref{eq:sq-oracle-grow}) plus the running time of $\OO(n\rho)$ to compute $\lambda_P$ for each piece $P$ is:
\begin{equation*}
      \sum_{\substack{P\in\ldd\\ |V_P|\ge |A|}}  \OO(n\rho  + b\Delta |P|(n/\rho + \rho^{3}) + |P|n/b)  ~=~ \OO(n^2 \rho/A + b\Delta n(n/\rho + \rho^3) + n^2/\rho).
\end{equation*}
Therefore, the total running time to construct the oracle is:
\begin{equation*}
     \OO(n^2/\Delta  + nA^4\Delta^3 + n^2 \rho/A + b\Delta n(n/\rho + \rho^3) + n^2/\rho) ~=~  \OO(n^{2-1/20}) 
\end{equation*}
by setting $b=\Delta=n^{1/20}$, $\rho=\Delta^3$, $A=\Delta^4$.

\begin{theorem} Given a square graph with $n$ vertices, we can construct in $\OO(n^{2-1/20})$ time an exact distance oracle of $\OO(n^{2-1/20})$ space and $\OO(1)$ query time.
\end{theorem}

\section{Distance Oracles for Unit-square Graphs}\label{sec:do-unit-square}

For unit square graphs,
we follow the oracle construction for square graphs above. The only difference is that we use \Cref{lm:unitsq-interval-searching} for solving the interval searching problem for unit squares.  Therefore, the running time to construct all the intervals is within an $n^{o(1)}$ factor of the total number of intervals. Since the stabbing path $\lambda_P$ for each piece $P$ still satisfies \Cref{eq:sq-oracle-path}, the total running time to grow all the balls for each large piece is $O^*(\Delta|P|(n/\rho+\rho^{3}))$. Therefore, the construction time for large pieces becomes:
\begin{equation*}
    \begin{split}
    \sum_{\substack{P\in \ldd \\|V_p| \ge A}} O^*(n\rho +\Delta |P|(n/\rho+\rho^{3})) ~=~ O^*(n^2\rho/A+\Delta n^2/\rho+\Delta n\rho^{3}).
    \end{split}
\end{equation*}
The construction time for small pieces is the same: $\OO(n^2/\Delta  + nA^4\Delta^3)$. Thus, the total construction time of the oracle is:
\begin{equation*}
    \begin{split}
       O^*(n^2/\Delta  + nA^4\Delta^3 + n^2\rho/A+\Delta n^2/\rho+\Delta n\rho^{3}) ~=~ O^*(n^{2-1/16})
    \end{split}
\end{equation*}
for $\Delta = n^{1/16}, A = \Delta^3, \rho = \Delta^2$.

\begin{theorem} Given a unit square with $n$ vertices, we can construct in $\OO(n^{2-1/16})$ time an exact distance oracle of $\OO(n^{2-1/16})$ space and $\OO(1)$ query time.
\end{theorem}

\section{Distance Oracles for Unit-disk Graphs}
\label{sec:do-unit-disk}

For unit-disk graphs, we follow the same strategy in \Cref{sec:unit-disks}, adapted to the distance oracle framework \Cref{sec:framework-do}.
\begin{itemize}
\item
We partition the neighborhood balls into types, so that within any cell, balls of a fixed type intersect the cell as a pseudoline arrangement.
\item
We use the same geometric data structure (\Cref{SS:ds-unitdisk}) and interval representation switching technique (\Cref{SS:reordering}), to implement the ball growing step (Step~3.3) in the framework using the inductive formula~(\ref{Eq:inductive}).
\item
We switch to a different distance oracle construction using \Cref{lm:do-sparse-exact-log} when the piece has size at most $A$.
\end{itemize}
We only analyze the construction time since space is bounded by it. 

\paragraph*{Ball expansion step.~} 
We now need to deal with modified balls of a fixed radius $r$ for different types with vertex weights on their endpoints.
To be precise:
\[ 
\hat N^r_M[s] \coloneqq \{(v, w(v)) : v\in V \text{~where the $\tau$-walk from $s$ to $v$ is at most $r+w(v)$ for~} \tau\in M\}  
\]
We bound the dual VC-dimension of the set system $((v, w(v))_{v\in V}, \{N^r_M[s]\}_{s\in P})$ in \Cref{lem:VC-adjusted-ball}.

For each type $T$,
given the $\spath^{r-1}_T$-representation for every modified balls in the set system $\cS^{r-1}_T$, we compute the $\spath^r_T$-representation for every modified balls in the set system $\cS^{r}_T$, using the same interval representation switching strategy and the data structure $\cD^r_T$ for the interval cover problem for unit-disks, similar to \Cref{SS:ball-growing-unitdisk}.
The total time to grow balls for all radii is
\(
n\cdot\rho + |P|\cdot (n/\rho + \rho^8).
\)

\paragraph{Construction time.} 
For small pieces, we show (in \Cref{lm:bdr-BFS-unitdisk} in the appendix) that we can compute a vertex weighted BFS on $P$ with weights at most $\Delta$ in time $\OO(|P|)$. 
Therefore, in \Cref{lm:do-sparse-exact-log}, $T(P) = \OO(|P|)$, giving the total running time for each small piece to be
\(
n \cdot |\bdry P| + |P| \cdot (|P| + (|\bdry P| \Delta)^4).
\)

For large pieces, we grow the balls for $O(\Delta)$ rounds, each taking time $n\cdot\rho + |P|\cdot (n/\rho + \rho^8)$.
Therefore, the total running time to construct the oracle is:
\begin{align*} 
& \OOO\left( n^2/\Delta +  \sum_{P:\, |P|>A}  \Delta (n\cdot\rho + |P|\cdot (n/\rho + \rho^8)) +  \sum_{P:\, |P|\le A} \Paren{ n \cdot |\bdry P| + |P| \cdot (|P| + (|\bdry P| \Delta)^4) } \right) \\
~=~ & \OOO\Paren{ n^2/\Delta +  \Delta n^2\rho/A + \Delta n^2/\rho + \Delta n\rho^8 + n^2/\Delta + nA^2/\Delta + \Delta^3 A^4 n }.
\end{align*}
Balancing cost by setting parameters $\Delta=n^{1/20}$, $\rho=\Delta^2$ and $A=\Delta^4$ then yields $\OO(n^{2-1/20})$. 

\begin{theorem} 
Given a unit-disk graph with $n$ vertices, we can construct in $\OO(n^{2-1/20})$ time an exact distance oracle of $\OO(n^{2-1/20})$ space and $\OO(1)$ query time.
\end{theorem}

\section{Conclusion and Open Questions}

In this paper, we have presented the first truly subquadratic algorithms for diameter and
related problems for many classes of geometric intersection graphs.
Naturally, many open questions follow, for example, improving the exponents of the time bounds of
any of our algorithms.  More intriguingly:
\begin{itemize}

\item Is there a truly subquadratic algorithm for computing the diameter
of arbitrary disk graphs?  Our algorithm can be extended to the case when the number
of different radii is $n^{o(1)}$, but the general case appears more difficult.

\item Could we prove any conditional lower bound on the running time of the form $\Omega(n^{1+\delta})$ for computing the diameter of unit-disk graphs?
Bringmann \etal~\cite{Bringmann2022-me} proved a near-quadratic conditional lower bound for 3D unit-ball graphs under the orthogonal vector (OV) hypothesis.

If one considers more difficult problems than diameter, e.g., counting the number of pairs with shortest-pair\note{shortest-path?} distance at most $r$ (which can be solved by our algorithms in subquadratic time), an $\Omega(n^{4/3})$ conditional lower bound follows for unit-disk graphs if one believes certain offline range searching problems similar to Hopcroft's problem require $\Omega(n^{4/3})$ time (namely, counting the number of pairs of points with Euclidean distance at most 1 in $\R^2$).

\item Is there a near-linear-time algorithm for distinguishing between diameter 2 vs.\ 3 for unit-disk graphs? 
Bringmann \etal~\cite{Bringmann2022-me} proved a near-quadratic conditional lower bound for 12D unit-hypercube graphs under the hyperclique hypothesis, and obtained an $O(n\log n)$-time algorithm for unit-square graphs. 
\end{itemize}

\noindent There are a few specific open questions related to our algorithms.  For example: 
\begin{itemize}
\item Is the VC-dimension of the set system in \Cref{lem:vcdim-type} bounded when we do not
restrict to a fixed $r$ and $T$?  If so, this might simplify our algorithms for unit disks.
\item Could we solve the interval cover data structure problem (\Cref{def:DS-1}) for 
arbitrary squares with $N^{1+o(1)}$ preprocessing time and $N^{o(1)}$ query time?  If so,
this would improve the exponent for our algorithms for arbitrary squares.  This appears difficult.
\item Less importantly, on the interval cover problem data structure problem for unit disks from a fixed modulo class,
could the extra $2^{O(\sqrt{\log N\log\alpha(N)})}\le N^{o(1)}$ factors be reduced to polylogarithmic?  A related question is to
determine tight bounds on the combinatorial complexity of the ``generalized envelopes'' from
\Cref{SS:ds-unitdisk}.
\end{itemize}

Besides unit squares, Duraj, Konieczny, and Pot\c{e}pa~\cite{duraj2023better} also considered translates of a convex polygon with constant complexity.
It is not difficult to similarly extend our algorithms for unit/arbitrary squares to translates/homothets of 
other convex polygonal shapes with constant complexity
(and our algorithms for unit disks to translations of fat convex non-polygonal shapes with constant complexity).

\small
\bibliographystyle{alphaurl}
\bibliography{unitdisk-diam}

\appendix
\newpage
\normalsize

\section{Low-diameter Decompositions}
\label{sec:low-diameter}
In this section, we construct the low-diameter decomposition of sparse graphs and geometric intersection graphs. 
Recall that we use $N^{r}[v] = \{u: d_G(u,v)\leq r\}$ to denote the set of vertices in the neighborhood ball of radius $r$ centered at $v$.

We give an algorithm for computing a low-diameter decomposition as claimed in \Cref{subsec:prelim-graph-ldd}. 
Our low-diameter decomposition for graphs is perhaps most similar to the low-diameter decomposition in~\cite{APWS95};  we are not aware of any work stating the exact guarantees with our definition of LDD\@. We first present the general algorithm and the properties of the LDD. We will discuss the detailed implementation and running time for sparse graphs and geometric intersection graphs separately.

\paragraph*{Basic algorithm.} 
Let $\phi$ be a parameter with $0< \phi < 1$. We will choose $\phi \coloneqq 24\log n/\Delta$.
Start with the entire graph $G_1 = G$.
Pick an arbitrary vertex $u\in V$, perform a BFS to compute neighborhood balls centered at $v$: $N^1[v]$, $N^2[v]$, \dots, $N^\ell[v]$. Stop when $|N^{\ell}[v]|/|N^{\ell-2}[v]| \le 1+\phi$, and set $V_1 = N^{\ell-1}[v]$ as one piece in the decomposition. Note that this is guaranteed to eventually happen because when $N^{\ell-2}[v]$ is the entire connected component of $v$, then $N^{\ell}[v] = N^{\ell-1}[v] = N^{\ell-2}[v]$. We mark the vertices in $N^{\ell}[v]\setminus N^{\ell-2}[v]$ as boundary vertices\footnote{The vertices in $N^{\ell}[v]\setminus N^{\ell-1}[v]$ are boundary vertices of later pieces constructed in the process. The vertices in $N^{\ell-1}[v]\setminus N^{\ell-2}[v]$ are boundary vertices of $V_1$, although there may be other boundary vertices in $N^{\ell-2}(v)$ that we accounted for earlier in the process. }. Repeat this procedure on $G_2=G_1\setminus V_1$ to find $V_2$, then on $G_3 = G_2 \setminus V_2$, and so on.

\paragraph*{Low diameter property.} 
First we bound the strong diameter of one such ball $N^{\ell-1}[v]$ that we included in our low diameter decomposition. Observe that for all $2\le i< \ell$, the ball $N^i[v]$ of radius $i$ satisfies $|N^i[v]|> (1+\phi)|N^{i-2}[v]|$. 
Since the size of the largest ball is at most $n$, if $\ell$ is odd, we have that:
\[ n \ge |N^{\ell}(v)| \ge |N^{\ell-1}(v)| > (1+\phi)\cdot  |N^{\ell-3}(v)| > (1+\phi)^{(\ell-1)/2} \cdot |N^{0}(v)| = (1+\phi)^{(\ell-1)/2} \]
Taking logarithms on both sides, and using the fact that $\phi < 1$, we obtain:
\[ \log n > \frac{\ell-1}{2} \cdot \log (1+\phi) \ge \frac{\ell-1}{2}\cdot (\phi - \phi^2/2) > \frac{(\ell-1) \cdot \phi}{4} = \frac{6(\ell-1) \cdot \log n}{\Delta} \]
Rearranging the inequality yields $\ell \le \Delta/6$. The diameter is at most $2\ell\le \Delta/3$.

\paragraph*{Small boundary property.}
Let $N^{r_1}_{G_1}[v_1]$ be the ball of largest radii we compute in $G_1$, $N^{r_2}_{G_2}[v_2]$ the ball in $G_2$, $\dots$, $N^{r_k}_{G_k}[v_k]$ the ball in $G_k$.
Observe that when we choose $P_i = G[N^{r_{i}-1}_{G_i}[v_i]]$, the vertices of $N^{r_{i}-1}_{G_i}[v_i]\setminus N^{r_i-2}_{G_i}[v_i]$ are potentially boundary vertices $\bdry P_i$, and a vertex in $N^{r_{i}}_{G_i}[v_i]\setminus N^{r_i-1}_{G_i}[v_i]$ is a boundary vertex in $\bdry P_j$ for some piece $P_j$ with $j> i$. Assume we have a total of $k$ pieces in the LDD. 
Thus it can be seen that:
\begin{align*}
\sum_{i=1}^k |\bdry P_i| &\le \sum_{i=1}^{k} |N^{r_i}_{G_i}[v_i]\setminus N^{r_i-2}_{G_i}[v]|\\
&\le 
\sum_{i=1}^{k} \phi \cdot |N^{r_i-2}_{G_i}[v_i]| \tag{since $|N^{r_i}_{G_i}[v_i]|\le (1+\phi)|N^{r_i-2}_{G_i}[v_i]|$}\\
&\le 
\sum_{i=1}^{k} \phi \cdot |N^{r_i-1}_{G_i}[v_i]| \tag{since $N^{r_i-2}_{G_i}[v_i]\subseteq N^{r_i-1}_{G_i}[v_i]$}\\
&\le \phi \cdot n = 
24n\log n/\Delta.
\end{align*}

\paragraph{No small pieces.}
To ensure that no piece is small, we will do some post-processing of the pieces obtained from the basic algorithm. We use the following claim.
\begin{lemma}
Let $P_i$ be a piece found in the basic LDD algorithm found by taking the vertices $N^{r-1}[v]$ in $G_i$. Either $P_i$ has at least $\Omega(\Delta/\log n)$ vertices, or $P_i$ is an entire connected component of $G_i$.
\end{lemma}
\begin{proof}
Suppose that $|N^{r}(v)| > |N^{r-2}(v)|$. Then as $(1+\phi)\cdot |N^{r-2}(v)| \ge |N^{r}(v)| \ge |N^{r-2}(v)|+1$, we conclude that $|N^{r-2}(v)| \ge 1/\phi = O(\Delta/\log n)$, so $P$ has size at least $O(\Delta/\log n)$. Otherwise $|N^{r}(v)| = |N^{r-1}(v)|=|N^{r-2}(v)|$ and thus $P_i$ is an entire connected component of $G_i$.
\end{proof}
In our post-processing, we will merge $P_i$ with an arbitrary neighboring component. Observe that since $P_i$ is an entire connected component of $G_i$, no later piece $P_j$ with $j>i$ will merge into $P_i$. 
Now consider a piece $P_j$ with multiple pieces $P_{i_1}, P_{i_2}, \cdots, P_{i_t}$ merging into it in the post-processing step, $j<i_1, \cdots, j<i_t$. Since all pieces have diameter at most $\Delta/3$, the resulting merged $P_j$ has diameter at most $\Delta$.

\subsection{Sparse Graphs}\label{subsec:LDD-sparse}

Consider the standard BFS algorithm that computes $N^{r}[v]$ by adding all neighbors incident to $N^{r-1}[v]$ into a queue. For every vertex $v$, the basic algorithm will add all its neighbors into a queue at most once, so the basic algorithm can be implemented in $O(m+n)$ time. The post-processing step involving merging components can also be done in $O(m+n)$ time. 

\lddsparse*

\subsection{Geometric Intersection Graphs}\label{subsec:LDD-geometric}

Here we consider geometric intersection graphs of fat pseudo-disks of similar size and squares of varying sizes. A family of objects are called \EMPH{pseudo-disks} if each one is the interior of a simple closed Jordan curve and two objects are either disjoint, have one object fully inside the other, or properly intersect each other at two boundary points. Disks are by definition pseudo-disks. The geometric intersection graph of a family of pseudo-disks can be considered, combinatorially, as a set of vertices $V$ representing the pseudo disks and two vertices are connected if their corresponding pseudo-disks have non-empty intersection. For the algorithm below, we consider fat pseudo-disks that are of roughly the same size and have constant complexity.
Specifically, a fat pseudo-disk is sandwiched between two disks of the same center $p$ of radius $r$ and $R$ with two fixed constants $r, R$ and $r\leq R$ and the boundary can be described by a constant number of algebraic curves. We call this pseudo-disk centered at $p$ as $C_p$. 
The input to our algorithm consists of the description of a family of $n$ fat pseudo-disks with input size $O(n)$. We assume that one can compute in $O(1)$ time whether two pseudo-disks have an edge or not. The geometric intersection graph of such pseudo-disks can be dense (i.e., having edges of size $\Theta(n^2)$). We show that the low diameter decomposition can still be computed in near linear time, similar to the running time for sparse graphs (\Cref{subsec:LDD-sparse}). 

Recall that the basic idea is to perform BFS from a vertex $v$ to compute balls centered at $v$:  $N^0[v]=\{v\}$, $N^1[v]$, $N^2[v]$, \dots, $N^\ell[v]$, and stop when $|N^{\ell}[v]|/|N^{\ell-2}[v]| \le 1+\phi$. Let $V_1 = N^{\ell-1}[v]$. Then repeat this procedure on $G_2=G_1\setminus V_1$ to find $V_2$, then on $G_3 = G_2 \setminus V_2$, and so on. 

We have to be careful in implementing the basic idea: we do not want to spend $\OO(n)$ time per iteration as the number of iterations could be $\Omega(n)$.  This is achievable by not explicitly constructing all the edges, an idea that is generally adopted for computing a breadth-first search tree for geometric intersection graphs (for fat objects of similar sizes)~\cite{Efrat2001-hm,Cabello2015-vo,Chan2016-sy}. We use the same algorithm as in~\cite{CGL24} for pseudo-disks of similar sizes. 
The core step in the BFS is to find the vertices that are exactly $j$-hops away from $v$, denoted by $Y_j$ --  from the vertices that are exactly $j-1$ hops away from $v$, $Y_{j-1}$. Put a grid of size $r\sqrt{2}$. Two pseudo-disks with centers in the same grid cell are connected by an edge for sure. Thus, if a pseudo disk centered at $p$ in one cell appears in $Y_{j-1}$, all pseudo-disks centered in the same cell will be included in $Y_{j}$ if they are not yet covered in $B^{j-1}(v)$. In addition, the other vertices to be included in $Y_j$ will come from cells that have distance at most $2R$ away from cells touched by $Y_{j-1}$. Since $R^2/r^2$ is a constant, we only need to check for each cell touched by $Y_{j-1}$, at most a constant number of nearby cells. This step can be implemented by using an operation called the red-blue intersection problem, which finds all the blue pseudo-disks that intersect at least one red pseudo-disks, where all red pseudo-disks and blue pseudo-disks are separated by a horizontal (or a vertical) line.
We use the following lemma from~\cite{CGL24}.

\begin{lemma}[\cite{CGL24}]\label{lm:red-blue-gen} 
In time $O(n_b\log n_b +n_r\alpha(n_r)\log n_r+n_r 2^{\alpha(n_r)})$, we can solve the red-blue intersection problem of $n_r$ pseudo-disks and $n_b$ blue pseudo-disks. Here $\alpha(n)$ is the inverse Ackermann function.
\end{lemma}

With this Lemma we can conclude the following theorem.

\begin{theorem}
\label{thm:lld-pseudo-disks} Let $G$ be the intersection graphs of $n$ fat pseudo-disks of similar size.  For any parameter $24\log n< \Delta \le n$, we can compute a low-diameter decomposition for $G$ in $\OO(n)$ time.
\end{theorem}
\begin{proof}
    We argue that for the entire algorithm, a non-empty cell in the grid of size $r\sqrt{2}$ is only visited a constant number of times. First, if cell $c$ has a vertex $p\in V_i$ and $p$ is not on the boundary of this piece $V_i$, then all pseudo-disks centered in the cell will be included in $V_i$. After $V_i$ is removed, cell $c$ becomes empty and will not be visited again in later iterations. Therefore, a cell $c$ visited by $V_i$ is only visited again by pieces $V_j$ with $j>i$ if $c$ has only vertices that are at the boundary of $V_i$. 
    That says, the cell $c$ has a neighboring cell $c'$ (within distance $2R$ from $c$), such that $c'$ contains a vertex $p$ of $V_i$ and $p$ is not on the boundary of $V_i$. In this case all vertices in $c'$ are entirely in $V_i$ (or earlier pieces). Thus, after the $i$-th iteration, at least one of the neighboring cells of $c$ is wiped out.  Since $c$ has only a constant of such neighboring cells, $c$ is only visited a constant number of times. This finishes the argument.
\end{proof}

As a corollary, since unit squares and unit disks are fat pseudo-disks of similar size, we conclude that a low diameter decomposition can be computed for these classes of intersection graphs in $\OO(n)$ time.

\paragraph{Axis-aligned squares.~}
We will need a similar theorem for axis-aligned squares (which might not be of similar size.) 
A BFS on the intersection graph of axis-parallel squares can be done in time $O(n\log n)$~\cite{Klost2023-xs}, by using data structures developed in~\cite{Baumann2024-gm}.
Again we focus on how to find the objects of $j$-hops away from a starting vertex $v$ from the objects of $j-1$ hops away.
When the squares have different sizes, instead of a grid of a single size, one can use a hierarchical structure such as the (compressed) quadtree. Each square is associated with a quad whose size is comparable with its size. Further the compressed quadtree can be decomposed into $O(n)$ canonical paths  such that each root to leaf path can be represented by $O(\log n)$ disjoint canonical paths.
A canonical path has a smallest cell $\sigma$ and largest cell $\tau$, and is associated with a constant number of regions, classified as inner, middle and outer regions. The inner region is a disk centered at the smallest cell $\sigma$ of the canonical path. Further, each region $A$ is associated with two sets, the first type $S_1(A)$ contains a collection of objects centered inside $A$ that form a clique, and the second type $S_2(A)$ contains objects that intersect with at least one site in $S_1(A)$. A similar red-blue intersection problem can be solved in linear time for axis-parallel squares, assuming sorting along $x$ and $Y$ coordinates is performed already, as shown in~\cite{Klost2023-xs}. In summary, to implement a BFS step, for each region $A$ touched by the vertices in $Y_{i-1}$, include all objects that are in $S_1(A)$ and then perform red-blue intersection modules with $A$ against a constant number of other regions. Since each object stays in at most $O(\log n)$ sets of the first type
and at most $O(\log n)$ sets of the second type, the total running time carries an extra $O(\log n)$ factor.
We can use this algorithm for the low-diameter decomposition and obtain the following. 

\begin{theorem}\label{thm:lld-squares} Let $G$ be the intersection graphs of $n$ axis-aligned squares.   For any parameter $24\log n< \Delta \le n$, we can compute a low-diameter decomposition for $G$ in $\OO(n)$ time.
\end{theorem}

\begin{proof}
The same argument as in \Cref{thm:lld-pseudo-disks} applies here: for each region $A$, either the type one objects $S_1(A)$ are completely included in a piece $V_i$ and this region disappears; or, one of the (constantly many) nearby regions are completely included in $V_i$ and disappears. By a charging argument, each region is only touched a constant number of times. Thus the total running time is in the order of $\OO(n)$.
\end{proof}

\section{VC-dimension Lemma}
\label{sec:vcdim-type}

In this section, we prove a lemma bounding the VC-dimension of certain set systems (\Cref{lem:vcdim-type}) from \Cref{sec:unit-disks}, which is needed in our algorithms for unit-disk graphs.

Let $G$ be the geometric intersection graphs of unit-disk graphs. Let $M$ be a subset of vertices, called \emph{a type}. We say that a walk $W$ from a vertex $v$ to a vertex $u$ is a \EMPH{Type-1 $M$-walk} if the vertex preceding $u$ (the second to last vertex) in the walk is in $M$. We say that the walk is a \EMPH{Type-2 $M$-walk} if the vertex following $v$ (the second to first vertex) in the walk is in $M$.  

For a technical reason explained later, we assume that every vertex in $G$ has a self-loop attached to it.  For every vertex $v$, define:
\begin{equation}
    \begin{split}
        B^{(1)}_M(v,r)  &= \{u\in V| ~\text{there is a Type-1 $M$-walk from $v$ to $u$ of length exactly $r$}\}\\
    B^{(2)}_M(v,r)  &= \{u\in V| ~\text{there is a Type-2 $M$-walk from $v$ to $u$ of length exactly $r$}\}
    \end{split}
\end{equation}

The reason for attaching a self-loop to every vertex is that if $d_G(v,x)\leq r-1$ for some vertex $x$ in $M$, then $x \in  B^{(1)}_M(v,r)$ since we can make a Type-1 $M$-walk of length $r$ by traversing from $v$ to $x$ along the shortest path (of length at most $r-1$) and then along the self-loop to get a walk of length at most $r$. The second to last vertex of the walk is $x$ itself, which is in $M$. Furthermore,  if there is a Type-1 $M$-walk from $v$ to $u$ of length less than $r$, then there is a Type-1  $M$-walk from $v$ to $u$ of length exactly $r$ by traveling the self-loop attached to the vertex in $M$ preceding $u$. The same holds for Type-2 $M$-walk. The main result of this section is to show that the system of balls deriving from Type-1 $M$-walk has a bounded VC-dimension.

\begin{lemma}
\label{lm:VC-Type1-walk}  $(V, \{B^{(1)}_M(v,r) \}_{r\in \reals, v\in V})$ has VC-dimension at most $4$.  
\end{lemma}

Observe that Type-1 and Type-2 $M$-walks are dual to each other:  a  Type-1 $M$-walk from $v$ to $u$ of length $r$ is a  Type-2 $M$-walk from $u$ to $v$ of length $r$. Therefore, for \EMPH{a given $r$}, $(V,\{B^{(2)}_M(v,r) \}_{v\in V})$ is the dual set system of  $(V,\{B^{(1)}_M(v,r) \}_{v\in V})$, and therefore, has VC-dimension at most $2^{4} = 16$. By modifying the proof of \Cref{lm:VC-Type1-walk}, get an improved bound for balls from  Type-2 $M$-walk:

\begin{lemma}
\label{lm:VC-type2-walk} 
For any $r\in \mathbb{N}$, $(V, \{B^{(2)}_M(v,r) \}_{ v\in V})$ has VC-dimension at most $4$.  
\end{lemma}

The set system in \Cref{lm:VC-type2-walk} only includes balls of fixed radius. It is possible that the more general set system $(V, \{B^{(2)}_M(v,r) \}_{ r\in \reals,  v\in V})$, which includes all balls of all radii, has VC-dimension at most $4$. However, for a technical reason, our proof of \Cref{lm:VC-Type1-walk} does not extend to this general case. See \Cref{rm:Type2-extend} for more details.

For the distance oracle application, we will need to handle vertices with weights. So we define the following set system for weighted vertices. Suppose each vertex $u$ has a weight $w(u)$, and the ground set is $\{(u, w(u))\}_{u \in V}$. Recall for the distance oracle, we maintain the adjusted neighborhood ball as follows (see \Cref{sec:framework-do} \Cref{eq:defn-nhat-sparse}, copied below):
\begin{align*}\label{eq:defn-nhat-sparse}
\hat N^{r}[s] := \{ v\in V : d_G(v, s) \le r+w_P(v)\} \qquad \forall r\in [-\Delta, \Delta]
\end{align*}
Further, the path connecting $s$ (the center of the neighborhood ball) to $v$ has the second vertex (adjacent to $s$) of a special type. Thus, we consider a Type-2 walk from $s$ to $v$. Therefore the set system we work with will be $(\{(u, w(u))\}_{u \in V}, \{B^{(2)}_{M, w}(v,r) \}_{ v\in V})$ where 
\begin{equation}
    B^{(2)}_{M, w}(s,r)  = \{v\in V| ~\text{there is a Type-2 $M$-walk from $s$ to $v$ of length exactly $r+w(v)$}\}
\end{equation}
Take this set system as the primal system, we can define the dual system as follows. Specifically, $v\in B^{(2)}_{M, w}(s,r)$ if and only if $s\in B^{(1)}_{M, w}(v,r)$ where
\begin{equation}
    B^{(1)}_{M, w}(v,r)  = \{s\in V| ~\text{there is a Type-1 $M$-walk from $v$ to $s$ of length exactly $r+w(v)$}\}
\end{equation}
Notice that $B^{(1)}_{M, w}(v,r)=B^{(1)}_{M}(v,r+w(v))$.
Therefore, the VC-dimension bound we need is provided precisely by \Cref{lm:VC-Type1-walk} for Type-1 walks, which fortunately works for neighborhood balls of varying radii. With this we immediately have the following.

\begin{lemma}\label{lem:VC-adjusted-ball}
    For any $r\in \mathbb{N}$, $(\{(u, w(u))\}_{u \in V}, \{B^{(2)}_{M, w}(v,r) \}_{ v\in V})$ has dual VC-dimension at most $4$. 
\end{lemma}

\subsection{Type-1 $M$-Walks}

In this section, we prove \Cref{lm:VC-Type1-walk}. As all $M$-walks in this section are of Type-1, we will drop the prefix Type-1, and only refer to Type-1 $M$-walks as $M$-walk. We also call the last edge of an $M$-walk to $u$ as an \EMPH{$M$-edge}. 

\begin{proof}[Sketch Proof of \Cref{lm:VC-Type1-walk}]  The strategy is basically the same as \cite{CGL24}\footnote{We refer to \url{https://arxiv.org/pdf/2401.12881}.}. We only show the steps needed for adapting the proof here. Consider four vertices $a, b, c, d$ representing four disks $D_a, D_b, D_c, D_d$ and assume that there are two (Type 1) $M$-walks $P(b, a)$ \emph{from $b$ to $a$} and $P(c, d)$ \emph{from $c$ to $d$}. (The vertices preceding $a$ and $d$ in the walks are in $M$.) We define a \EMPH{local crossing pattern} to be four distinct vertices \EMPH{$a', b', c', d'$} with $a', b'$ on $P(a, b)$ (with $a'$ closer to $a$ than $b'$) and $c', d'$ on  $P(c, d)$ (with $c'$ closer to $c$ than $d'$) such that one of the four vertices $a', b', c', d'$ has edges to all the other three vertices; see \Cref{fig:intersection-path}. The central claim is the following; if the claim holds, then the rest of the argument is standard. 

\begin{claim}\label{clm:intersection} Either there is an $M$-walk $P'(c, a)$ whose hop length is at most $|P(c, d)|$ or there is an $M$-walk $P'(b, d)$ whose hop length is at most $|P(b, a)|$.
\end{claim}

\begin{figure}
\centering
\begin{tabular}{cccc}
\includegraphics[width=0.22\linewidth]{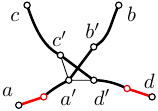}&
\includegraphics[width=0.2\linewidth]{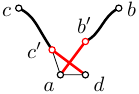}&
\includegraphics[width=0.2\linewidth]{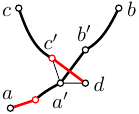}&
\includegraphics[width=0.2\linewidth]{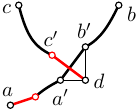}\\
(a) & (b) & (c) & (d)
\end{tabular}
\caption{If two  $M$-walk $P(b, a)$ and $P(c, d)$ intersect with a local crossing pattern $a', b', c', d'$, then there is an $M$-walk from $c$ to $a$ that are no longer than $|P(c, d)|$ or there is an $M$-walk from $b$ to $d$ that is no longer than $P(b, a)$. The vertices in $M$ are highlighted red. }
\label{fig:intersection-path}
\end{figure}

We consider a case study depending on whether the local crossing pattern involves an $M$-edge. 
In the first case when the local crossing pattern does not involve the last edge (from a vertex in M to the endpoint of the walk) of the two M-walks (see \Cref{fig:intersection-path} (a) for an example), the proof  follows exactly the same as that of~\cite{CGL24}.
The second case, which is also easy, is when the local crossing pattern involves two $M$-edges. In this case, both $c',b'\in M$. Either we have the edge $c'a'$ or the edge $b'd'$. In both cases the claim is true. \Cref{fig:intersection-path} (b) shows the case with edge $c'a$ present. In this case, we can find an $M$-walk  $P'(c, a)$ through $P(c, c')$ and then take edge $c'a$, which is not longer than path $P(c, d)$. If $b'd$ is present, then the path that follows $P(b, b')$ and then edge $b'd$ is an $M$-walk and not longer than $P(b,a)$.

The difficult case for the proof of the claim is when the local crossing pattern happens at an M-edge of one path with the non M-edge part of the other path. Without loss of generality, assume that the $M$-edge involved in a local crossing pattern is the edge $c'd$. See case (c) and (d) in \Cref{fig:intersection-path} for an example. 

We first consider the case when $c'a'$ and $a'd$ are present. We prove by contradiction. Consider an $M$-path from $b$ to $d$:
\[
    P'(b,d)  = P(b,b')\circ (b'a')\circ (a'c') \circ (c'd').
\]
Since the claim does not hold, the following holds:
\[  
    |P'(b,d) | > |P(b,a)|
    \Leftrightarrow |P(b,b')| + 3  > |P(b,b')| + 1 + |P(a,a')| 
    \Leftrightarrow   |P(a',a)|  <  2.
\]

If $|P(a', a)|=0$, then $a=a'$ and $b'$ must be a vertex in $M$. This is a contradiction, as the crossing occurs between two $M$-edges.

If $|P(a,a')| = 1$, then $a' \in M$. This means $a' a$ is an $M$-edge.
Now we define another walk $Q(b,d) = P(b,b')\circ (b'a')\circ (a'd)$. Since $a'\in M$, $Q(b,d)$ is an $M$-walk and, furthermore, $|Q(b,d)| = |P(b,a)|$. Thus,  $Q(b,d)$ is the $M$-walk that satisfies the claim.
Now, if $Q(b, d)$ is still longer than $P(b, a)$, by the same analysis we have $|P(a', a)|<1$. This leads to a contradiction.

The next case we consider is where the edges $a'd, b'd$ are present. See \Cref{fig:intersection-path} (d). Consider an $M$-path from $b$ to $d$:
\[
    P'(b,d)  = P(b,b')\circ (b'd)\circ (dc') \circ (c'd).
\]
Notice that this is an $M$-walk. If it is longer than $|P(b, a)|$, we have
\[   
    |P'(b,d) | > |P(b,a)|
    \Leftrightarrow |P(b,b')| + 3  > |P(b,b')| + 1 + |P(a,a')| 
    \Leftrightarrow   |P(a',a)|  <  2.
\]
For the same reason as explained earlier, $|P(a', a)|=0$ is not possible and $|P(a',a)|=1$ means that $a'\in M$ and we now find an $M$-walk $Q(b, d)$ by following $P(b, b')$ and then edges $b'a'$ and $a'd$. This path is one shorter than $P'(b, d)$ and this again gives a contradiction.

The other two cases, when either edges $c'b', c'a'$ or edges $c'b', b'd$ are present, are easy. Basically the edge $b'c'$ provides an $M$-walk from $b$ to $d$ which is not longer than $P(b, a)$.
\end{proof}

\begin{remark}\label{rm:Type2-extend} If we apply the same proof to Type-2 $M$-walks, \Cref{clm:intersection} remains true. However, what we need is a slightly different version:  Either there is a Type-2 $M$-walk $P'(a, c)$ whose hop length is at most $|P(a, b)|$ or there is a Type-2 $M$-walk $P'(d, b)$ whose hop length is at most $|P(d, c)|$.  The proof does not extend to show this version. On the other hand, if we fix a radius $r$, then everything goes through; see the next section.
\end{remark}

\subsection{Type-2 $M$-Walks}

$M$-walks in this section are referred to Type-2 $M$-walks.

\begin{proof}[Proof of \Cref{lm:VC-type2-walk}] We follow the same setup in the proof of \Cref{lm:VC-Type1-walk}. Assume that there are two (Type 2) $M$-walks $P(a, b)$ \emph{from $a$ to $b$} and $P(d, c)$ \emph{from $d$ to $c$}. (We switch the roles  of $a$ and $b$, and of $c$ and $d$, so that we can reuse \Cref{fig:intersection-path}.) The following claim implies the lemma:

\begin{claim}\label{clm:Type2} Either there is an $M$-walk $P'(a,c)$ whose hop length is at most $|P(a, b)|$ or there is an $M$-walk $P'(d, b)$ whose hop length is at most $|P(d, c)|$.
\end{claim}

Observe that $|P(a, b)|$ and $|P(d, c)|$ have length exactly $r$ each since they are from balls of radius exactly $r$. Therefore, $|P(a, b)| = |P(d, c)|$, and hence \Cref{clm:Type2} follows directly from \Cref{clm:intersection}.
\end{proof}

\section{Geometric Data Structures}

In this section, we describe how to solve
the interval searching problem (\Cref{prob:DS0}),
the main geometric data structure problem used by our diameter algorithms and distance oracles, for different types of geometric objects.
In \Cref{subsec:DS-reduction},
we first describe how to reduce the interval cover to the
rainbow colored intersection searching (\Cref{def:DS-2}) and then describe how to reduce 
the interval searching problem to the interval cover problem (\Cref{def:DS-1}), though with some loss of efficiency.
For squares, we solve the rainbow colored intersection searching problem in 
\Cref{subsec:ds-square}.
For unit disks of a fixed modulo class and for unit disks, we solve the interval cover problem directly (without going through rainbow colored searching), and thus more efficiently, in
\Cref{SS:ds-unitdisk} and \Cref{subsec:unitsq-DS} respectively, using an interesting recursive approach.

\subsection{Reductions Between Data Structure Problems}
\label{subsec:DS-reduction}

In this subsection, we provide the reductions between the data structure problems in \Cref{subsec:DS-prelim}, and in particular, proving \Cref{lm:DS1-to-DS0} and \Cref{lm:DS2-to-DS1}. 

\paragraph{Interval cover to rainbow colored intersection searching.} We reduce the interval cover problem (\Cref{def:DS-1}) to the rainbow colored intersection searching problem (\Cref{def:DS-2}).

\RainbowToCover*
\begin{proof}
Consider an instance of \Cref{def:DS-1}. Divide the range $[1, n]$ into $n/b$ blocks of length $b$, denoted by intervals $B_1, \ldots, B_{n/b}$, with $[1, n]=\bigcup_{k=1}^{n/b} B_k$.
Denote by $\mathcal{S}$ the set of all intervals associated with objects in  $\mathcal{O}$ and $\mathcal{S}(q)$ the set of intervals associated with objects intersecting $q$. 
Consider the query interval $I$. It intersects with a set of blocks $B_i, \dots, B_j$ such that $I$ overlaps with at most two of the blocks partially, namely, the two blocks at the end ($B_i$ or $B_j$), and fully contains all the middle chunks $B_{i+1}, \ldots, B_{j-1}$. To verify if the union of the intervals of $\mathcal{S}(q)$ covers the query interval $I$, we need to check for each of the blocks, $B_k$, $i\leq k\leq j$, if $B_k\cap I$ is covered by the union of the intervals of $\mathcal{S}(q)$, limited within block $B_k$. 
If for each $B_k$ the answer is true, we answer Yes. Otherwise, we answer No.
In the following we focus on answering the coverage query for a fixed block $B$ and check if the union of the intervals $\{I_s\cap B \mid I_s\in \mathcal{S}(q)\}$ covers $I\cap B$.

Now fix a block $B$. Take $I'=I\cap B$. Similarly, for each object $s$ we restrict the interval $I_s$ within $B$ and take $I'_s=I_s\cap B$ and take $\mathcal{S}'=\{I'_s\mid I_s\in \mathcal{S}\}$.
Each interval $I_s$ fully covers a set of middle chunks and only partially covers at most two extreme blocks at the end of $I_s$. Thus we can write $\mathcal{S}'=\mathcal{S}'_1 \cup \mathcal{S}'_2$, with the first category $\mathcal{S}'_1$ containing the intervals $I'_s=B$ (i.e., $I_s$ fully covers $B$) and the second category $\mathcal{S}'_2$ containing the intervals $I'_s\neq B$ (i.e., $I_s$ partially covers $B$).
We perform two queries for $I'$ against $\mathcal{S}'_1$ and $\mathcal{S}'_2$ respectively. 

For $\mathcal{S}'_1$, all the intervals are given the same color and we just check if at least one of them is associated with an object intersecting $q$. We solve this problem by issuing \textsc{RainbowCover?}$(q)$ against the objects whose intervals appear in $\mathcal{S}'_1$. If this rainbow query returns a positive answer, $I'$ is covered and we are done. Otherwise, we check for $I'$ against $\mathcal{S}'_2$.
This query is more complicated since the intervals $I'_s \in \mathcal{S}'_2$ only partially cover $B$. 
We give each of the elements in $B$ a unique color. There are at most $b$ colors. Also, for each object $s$ with $I'_s\in \mathcal{S}'_2$, we make a colored copy of the object $s$ for each element in $I'_s$; the color of the copy is equal to the color of the corresponding element. Now we discuss the case when $I'=B$ and when $I' \subset B$ separately. 
When $I'=B$, i.e., $B$ is an `internal' block, we build a rainbow colored query structure for all color/elements in $B$ and issue a query \textsc{RainbowCover?}$(q)$  to see if all colors show up. If the query returns no, we return negative to the interval cover query.
In the case when $I'$ is a boundary block ($I' \subset B$), we build the rainbow colored query structure for each color/element in $B$. To answer the query for $I'$, we 
issue \textsc{RainbowCover?}$(q)$ for each color in $I'$ against the corresponding data structure to see if this color appears among objects that intersect $q$. The total number of such queries is the number of elements in $I'$ and is at most $b$.
If all the rainbow queries return True -- that all colors in $I'$ appear -- then all elements in $I'$ are covered by the union of intervals in $\mathcal{S}'_2$ for those objects intersecting $q$. If any rainbow query returns a no, we return a negative answer for the interval cover query. 
 
To analyze the total running time, we need to account for the preprocessing time and the total query time for all the $n/b$ blocks. Recall that we solve the interval cover problem in an `off-line' version and assume all input intervals and query intervals are given. 
$N_{IC}$ is the total number of input objects and query objects, and $L_{IC}$ is the total length of the input and query intervals. 
We issue a total of $O(L_{IC}/b)$ rainbow queries in the first category since for each query $(q, I)$ we only consider the blocks that overlap with $I$. For the second category we issue a total of $O(L_{IC}/b)$ rainbow queries for the blocks that are internal to the interval queries and $O(N_{IC} b)$ rainbow queries for the boundary blocks. Thus the total query time is $\OO(N_{IC}b+L_{IC}/b)$. 

For the preprocessing time, we consider the time spent to prepare for the rainbow query in the first and second category separately. For the second category, we have a total of $2N_{IC}$ intervals since each interval $I_s$ of an input object $s$ only contributes at most two boundary intervals. Each interval generates at most $b$ colored objects so we have a total of $O(N_{IC}b)$ objects, over all the $n/b$ blocks. We build the rainbow colored query data structure for each block separately. The total preprocessing time for rainbow query in the second category is thus $\OO(N_{IC}b)$. For the rainbow query in the first category, we perform a linear scan of the blocks and only update the rainbow query data structure $\mathcal{D}_{RC}$ when needed -- an input object appears (starts to fully cover a new block) or disappears (stops covering the current block). Each input object only triggers two updates.  
In fact, for each update, we simply rebuild the rainbow query data structure from scratch. For each input interval $I_s$, the amortized run time attributed to $I_s$ in these preprocessing and rebuilding efforts is $\OO(|I_s|/b)$ and therefore the total running time remains $\OO(L_{IC}/b)$, where $L_{IC}$ is the total length of the input and query intervals.
Therefore, the total run time is bounded by $\OO(N_{IC}b + L_{IC}/b)$. This finishes the proof. 
\end{proof}

We also need a data structure that can answer interval avoidance queries. Specifically,

\begin{problem}
[Interval Avoidance Problem] 
\label{def:DS-avoidance} 
Given a set of $N$ objects $\mathcal{O}$ and each object $o\in \mathcal{O}$ is associated with an interval $I_o\subseteq [1:n]$. Design a data structure to answer the following query:
\begin{itemize}
    \item \textsc{Avoids?}$(q,I)$:  Given a query object $q$ and a query interval  $I\subseteq [1:n]$, decide whether the union of intervals associated with the objects intersecting\footnote{Here we mean the objects intersect, not their associated intervals.} $q$ in $\mathcal{O}$ is disjoint from the interval  $I$.  
\end{itemize}
\end{problem}

The interval avoidance problem is easier than the interval cover problem, as it is decomposible -- we can partition the input objects into two sets and check the query $(q, I)$ against each set for avoidance separately. 

\begin{lemma}\label{lm:DS2-to-DS-avoidance}
If we can construct in $\OO(|\cO_{RC}|)$ time a data structure $\mathcal{D}_{RC}$ with $\Tilde{O}(1)$ query time for solving \Cref{def:DS-2}, then we can construct
 a data structure $\mathcal{D}_{IA}$ for solving \Cref{def:DS-avoidance} that has a preprocesing time of $\OO(N_{IC})$ such that each interval avoidance query takes time $\OO(1)$. 
\end{lemma}
\begin{proof}
An interval $I_o$ intersects $I$ if either at least one endpoint of $I_o$ is inside $I$ or one endpoint of $I$ is inside $I_o$. Therefore, to answer the interval avoidance query, we run two types of queries. In the first type we verify if $I$ includes any endpoints of intervals whose associated objects intersect $q$. If yes, we immediately return no to the interval avoidance query. If not, we proceed to the second type of queries where we check if an endpoint of $I$ stabs any intervals whose associated objects intersect $q$. The first type is a range query, and the second type is an interval stabbing query. We explain the two operations separately.

For the range query, we take the set $\mathcal{S}$ of all intervals associated with objects in $\mathcal{O}$ and build a binary tree $\mathcal{T}$ on all the $2|\mathcal{S}|$ endpoints of the intervals. Further, for each node $v$ on the tree $\mathcal{T}$ we build a rainbow colored query structure on the objects in $\mathcal{O}$ whose associated intervals have at least one endpoint in the subtree of $v$. In particular, the data structure at the root of $\mathcal{T}$ includes all objects in $\mathcal{O}$. The total preprocessing time for these query data structures is $\Tilde{O}(N_{IC})$, since each object in $\mathcal{O}$ only appears in $O(\log N_{IC})$ of the rainbow colored query structures. Next we run a standard range query with $I$ on tree $\mathcal{T}$ to find a set $Q(I)$ of $O(\log |\mathcal{S}|)$ vertices of $\mathcal{T}$ such that each vertex $v\in Q(I)$ has the entire subtree fully inside $I$, but its parent does not meet this condition. We issue a query of $q$ on the rainbow colored structure at each vertex in $Q(I)$. If any query returns a positive answer (indicating intersection), then $I$ does not avoid the objects intersecting $q$. We issue at most $O(\log N_{IC})$ rainbow colored range queries with a total cost of $\Tilde{O}(1)$.

For the interval stabbing query, we build an interval tree on the intervals $\mathcal{S}$. Specifically, we have a binary tree $\mathcal{Y}$ where the root $r$ is associated with value $\ell(r)=\lfloor n/2\rfloor$ (the median of $[1, n]$) as well as a subset of intervals $S(r)$ -- all the intervals in $\mathcal{S}$ that are stabbed by  $\ell(r)$. Recursively, we build the left (right) subtree by using all the intervals to the left (right) of $\ell(r)$ respectively. Further, for each node $v$ in the interval tree, we build two binary trees, $\mathcal{Z}_1(v)$ on the left endpoints of the intervals in $S(v)$ (that are all smaller than or equal to $\ell(v)$), and $\mathcal{Z}_2(v)$ on the right endpoints of the intervals in $S(v)$ (that are all greater than or equal to $\ell(v)$). For each node $u$ on a tree $\mathcal{Z}_i(v)$, $i=1,2$, we build a rainbow colored query structure for all the intervals in the subtree of $u$. Again, these objects are given the same color.

The total preprocessing time for these query data structures is $\Tilde{O}(N_{IC})$, since each interval in $\mathcal{S}$ only appears in the set $S(v)$ of one vertex $v$ on tree $\mathcal{Y}$ and then at most $O(\log N_{IC})$ vertices in the secondary level trees $\mathcal{Z}_i(v)$. 

Next we take one endpoint $p$ of $I$ and issue a stabbing query on $\mathcal{Y}$. We first issue stabbing query against the root vertex $r$  of $\mathcal{Y}$ and depending on whether $p$ is less than or greater than $\ell(r)$, recursively query either the left subtree or the right subtree of $\mathcal{Y}$. We just explain how to query $p$ against a node $v$ of $\mathcal{Y}$. The total query cost is just an extra $\log$ factor more. Specifically, if $p\leq \ell(v)$, we issue a query to $\mathcal{Z}_1(v)$; if $p \geq \ell(v)$, we issue a query to $\mathcal{Z}_2(v)$. 
Suppose we query $p$ on $\mathcal{Z}_1(v)$.  The other case is symmetric. We take all the vertices $Z_1(p)$ of $\mathcal{Z}_1(v)$: $u \in Z_1(p)$ if all vertices in the subtree of $u$ are completely to the left of $p$ but $u$'s parent fails to meet this condition. $|Z_1(p)|=O(\log N_{IC})$. Now we query $q$ again the rainbow colored range query structure for all vertices in $Z_1(p)$. If any  of these queries return a Yes, the interval avoidance query is negative. 
In total the query cost adds a total factor of $O(\log^2 N_{IC})$ on top of the cost of a single rainbow colored range query.

In summary, we only add extra poly-logarithmic factors on top of the rainbow colored range query structure. Thus, we can implement the interval avoidance query with a preprocesing time of $\OO(N_{IC})$ such that each interval avoidance query takes time $\OO(1)$.
\end{proof}

\paragraph{Interval searching to interval cover.} We reduce the interval searching problem (\Cref{prob:DS0}) to the interval cover problem (\Cref{def:DS-1}) with polylogarithmic loss.

\CoverToSearching*

\begin{proof}
We take an instance of \Cref{prob:DS0}. For each object $s\in \mathcal{O}_{IS}$ we duplicate it to $k$ copies if $s$ is associated with $k$ intervals. Each copy is now associated with a single interval of $\mathcal{I}_o$. This creates a total of $\tilde{N}_{IS}=\sum_{o\in \mathcal{O}_{IS}}|\mathcal{I}(o)|$ objects.
Now we build a data structure $\mathcal{D}_{IC}$ to solve \Cref{def:DS-1} on this set of objects with preprocessing time $\OO(P(\tilde{N}_{IS}))$. For each query \textsc{IntervalSearch}$(q)$, we recursively issue queries to $\mathcal{D}_{IC}$. Specifically, we start with $I=[1, n]$. If $I$ is completely covered by the union of the intervals associated with objects in $\mathcal{O}_{IS}$ that intersect $q$ (which is checked by a query to $\mathcal{D}_{IC}$ with $q$ and $I$), we output $I$ and we are done. Otherwise, if $I$ is completely avoided, we output $\emptyset$ and we are also done. For the other case, we will recurse. We divide $I$ into two intervals of equal length, $I_1$ and $I_2$, and issue queries $(q, I_1)$ and $(q, I_2)$ with $\mathcal{D}_{IC}$. In the end, we will output the union of all the intervals that are fully covered by the intervals associated with objects in $\mathcal{O}_{IS}$ that intersect $q$. 

The running time for a query $q$ is dependent on the total number of queries issued to $\mathcal{D}_{IC}$ recursively. Notice that all query intervals are dyadic intervals.
In addition, recursion stops when an interval $I$ is completely covered by the union of intervals $\mathcal{S}(q)$ or completely avoided. Thus only the dyadic intervals whose parent partially overlaps with a query output interval will ever trigger a query. The total number of such intervals is in the order of $O(|\mathcal{I}_{out}(q)|\cdot \log n)$. Recall that each query to $\mathcal{D}_{IC}$ takes time  $Q(\tilde{N}_{IS})$. Summing up everything, we have the claim in the Lemma. 
\end{proof}

\subsection{Data Structure for Square Graphs}\label{subsec:ds-square}

We now solve the rainbow colored intersection searching problem  (\Cref{def:DS-2}) for a set of axis-parallel squares of possibly different size.
We use an approach that can be commonly found in previous work on colored range searching~\cite{gjs-frgis-1995,GuptaSURVEY}: for each color class, we build a set of new objects, so that colored range searching reduces to standard range searching on all the new objects.

\begin{figure}[ht!]
\centering
\includegraphics[scale=1]{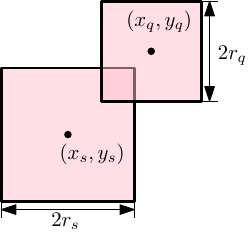}
\includegraphics[scale=1]{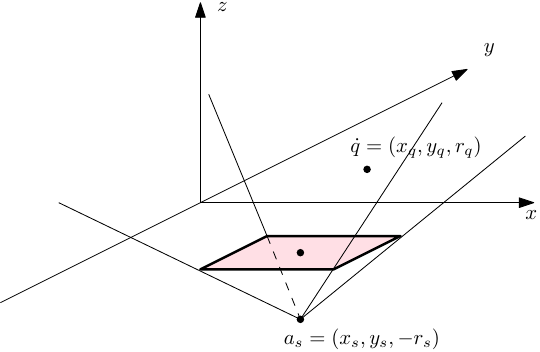}
\caption{Left: a square centered at $(x_s, y_s)$ with side length $2r_s$ and a query square centered at $(x_q, y_q)$ with side length $2r_q$. Right: the cone $C_s$ and point $\dot q=(x_q, y_q, r_q)$.}\label{fig:cone}
\end{figure}

Consider the input squares as being in the plane $z=0$ in 3D. For each square $s$ of center $(x,y,0)$ and side length $2r$ (or $\ell_\infty$-radius $r$) consider the point $a_s=(x,y,-r)\in \mathbb R^3$ and the cone $C_s$ with apex $a_s$ whose intersection with the plane $z=0$ is the square $s$. (If we imagine the $z$ axis pointing vertically up, then this cone opens upward.) See \Cref{fig:cone} for an example. For a collection $S$ of squares and the corresponding cones, consider a new square $q$ with center $(x_q,y_q,0)$ and side length $2r_q$. Notice that the normals of the planes bounding any cone $C_s$ is the intersection of four upper half-spaces, and the normals of these upper-half-spaces are $(1,0,1),(0,1,1),(0,-1,1)$ and $(-1,0,1)$.

\begin{observation}\label{obs:lift}
    The square $q$ intersects $s$ if and only if $\dot q \coloneqq (x_q,y_q,r_q)\in C_s$.
\end{observation}

\begin{proof}
    For a point $(x_q,y_q,0)$ outside $s$ we have that its $\ell_\infty$ distance to the square $\bd s$ is $\min(|x_s-x_q|,|y_s-y_q|)-r_s$. On the other hand, the vertical line through $(x_q,y_q,0)$ intersects the cone at exactly
    \[(x_q,y_q,\max(|x_s-x_q|,|y_s-y_q|)-r_q),\]
    that is, the signed vertical distance from $(x_q,y_q,0)$ to $\bd C_s$ is equal to the $\ell_\infty$ distance from $(x_q,y_q,0)$ to $s$. In particular, the square of $\ell_\infty$ radius $r_q$ centered at $(x_q,y_q,0)$ intersects $s$ if and only if $(x_q,y_q,r_q)$ is above $\bd C_s$, i.e., if and only if $(x_q,y_q,r_q)\in C_s$.
\end{proof}

As a consequence of the above observation, the square $q$ intersects some square among some set $S$ of squares if and only if $\dot q\in \bigcup_{s\in S} C_s$. In particular, if we have a convenient data structure to represent $\bigcup_{s\in S} C_s$, then we can quickly answer the query: given an axis-aligned square $q$, does it intersect at least one square from $S$?

\paragraph*{Detecting intersection with some square from $S$.}
We will now work on representing $U_S\coloneqq \bigcup_{s\in S} C_s$. Observe that $U_S$ is the union of translates of a fixed convex cone of constant complexity, thus it has linear union complexity. Indeed, each face $f$ of $\bd U_S$ is bounded from below, and the bottommost vertex (i.e., the vertex of minimum $z$-coordinate) on $f$ cannot be the intersection of a cone edge and a cone face nor the intersection of three cone faces, as a simple case distinction shows that all such vertices have an incident edge in $f$ where this vertex is strictly above the other endpoint. Thus the bottommost vertex of $f$ is the apex of some cone. On the other hand, each cone apex can be assigned to at most 4 faces (as there cannot be two faces $f,f'$ of $\bd U_S$ within the same cone face). We conclude that there are at most $4|S|$ faces in $\bd U_S$. By Euler's formula we have that $\bd U_S$ has complexity $O(|S|)$.

Consider the vertical projection $U_S^0$ of $\bd U_S$ into the plane $z=0$. Notice that this is exactly an additively weighted $\ell_\infty$ Voronoi diagram (where weights are the radii of the squares). Using standard techniques~\cite{Fortune87,Klein89} this diagram and $\bd U_S$ itself can be computed in $\OO(|S|)$ time.

We obtain a planar subdivision where edges are either axis-aligned or they are aligned with a 45 degree rotation of the axes.
We decompose this subdivision into $O(|S|)$ trapezoids with two vertical sides (or right-angle isoceles triangles with axis-aligned legs, as well as some unbounded polygons with at most two non-vertical sides) using the standard trapezoidation used for point location data structures~\cite{PointLoc} in $O(|S|\log |S|)$ time; let $T_S$ denote the resulting subdivision of size $O(|S|)$. More precisely, in order to get a \emph{partition} of the plane into faces, on boundary edges with normals $(0,1),(1,0),(1,1),(-1,1)$ we require weak inequalities, while we require strong inequalities for boundary edges with normals $(0,-1),(-1,0),(-1,-1),(1,-1)$.

We project $T_S$ vertically to get a $3$-dimensional subdivision $\cT$ of $U_S$ into convex vertical slabs: here each region is a vertical slab bounded by one face of $U_S$ from below and $\bd f \times \mathbb R$ on the sides, where $f$ is a face of $T_S$.

Note that the complexity of $\cT$ is $O(|S|)$ and it was computed in $O(|S|\log |S|)$ time. Moreover, each slab $T\in \cT$ is bounded by faces whose normals can have $12$ possible directions: there are $4$ possible normals for faces coming from $\bd U_S$, and $4\cdot 2$ for the vertical faces, as each of these are parallel to one of four directions in the plane $z=0$.

To check whether a point $\dot q$ lies in some region $T\in \cT$, we need to verify if it is contained in each half-space given by $\bd T$. Each such condition is of the form $\langle \dot q,\nu_j\rangle \leq c^j_T$ (or $<c^j_T$) where $\nu_j$ is one of $12$ possible normals and $c^j_T$ is a constant that depends only on $T$. (We define $c^j_T=\infty$ if $T$ does not have a face with normal direction $\nu_j$.) For a fixed region $T$ all of these linear conditions can be written as
\[\dot T \coloneqq (c^1_T,c^2_T,\dots,c^{12}_T)\in \mathrm{ort}_q\coloneqq ((-\infty, \langle \dot q,\nu_1\rangle]\times (-\infty, \langle \dot q,\nu_2\rangle] \times (-\infty, \langle \dot q,\nu_3\rangle) \dots \times (-\infty, \langle \dot q,\nu_{12}\rangle)).\]

Thus, our problem of deciding if $q$ intersects at least one square from $S$ is reduced to the following: given a query square $q$, we compute a $12$-dimensional orthogonal range 
that contains \emph{exactly one} point among $\{\dot T|T\in \cT\}$ if and only if $q$ intersects at least one square from $S$.
This problem can be solved with $12$-dimensional orthogonal range searching~\cite{CGAA}, which requires $\OO(|\cT|)=\OO(|S|)$ pre-processing time and space and $\OO(1)$ query time, to decide if the query range $\mathrm{ort}_q$ contains some point $\dot T$.

\paragraph*{Solving rainbow colored intersection searching.}

Suppose now that we are given a set of objects $\cO$, each associated with some color; let $\cS$ be the partition of $\cO$ into its color classes. For each color class $S\in \cS$ we set up the subdivision $\cT_S$ and compute the corresponding points $\{\dot T\mid T\in \cT_S\}$. Then we set up a standard orthogonal range counting data structure on the $12$-dimensional point set $T_\cS \coloneqq \bigcup_{S\in \cS} \{\dot T\mid T\in \cT_S\}$. This takes $\sum_{S\in \cS} \OO(|S|)=\OO(|\cO|)$
preprocessing time and space, and for any orthogonal query we can return the number of points in the range in $\OO(1)$ time.

Given a query square $\dot q$ we can compute the orthant query $\mathrm{ort}_q$ and observe that the number of points in $\mathrm{ort}_q$ is equal to the number of classes $S\in \cS$ such that $q$ intersects at least one square from $S$. Thus, $q$ intersects all color classes if and only if $\mathrm{ort}_q$ contains exactly $|\cS|$ points from $T_\cS$.
\newcommand{\Env}{{\cal E}}
\newcommand{\LE}{\textsc{LE}}
\newcommand{\UE}{\textsc{UE}}

\subsection{Data Structure for Unit Disks}
\label{SS:ds-unitdisk}

In this subsection, we directly solve the interval cover problem for unit disks restricted to a fixed modulo class, as needed in our diameter algorithm and distance oracle for unit-disk graphs.  (We do so without going through rainbow colored intersection searching, to get better time bounds.)
As noted in \Cref{sec:unit-disks}, this problem reduces to a corresponding interval cover problem about pseudolines:

\begin{problem}
\label{prob:pseudoline}
We are given an input set $S$ of $N$ pseudolines\footnote{We assume $O(1)$ time oracle access to deciding if a point is above/on/below a pseudoline, as well as to find the intersection of a pair of pseudolines (or determine that no itnersection exists).} in the plane, where each pseudoline $s\in S$ has an associated interval $I_s$.  We want to build a data structure to answer the following type of queries: given a query point $q$ and interval $I$, test whether $\displaystyle \bigcup_{\substack{s\in S\\ \text{$s$ below $q$}}} \! I_s$\, contains~$I$.
\footnote{
One way to interpret the problem is to think of each pseudoline $s$ as being ``active'' for a time window $I_s$; a query is to determine whether a given point $q$ stays above the upper envelope of the active pseudolines for the entire duration of the time window $I$.  We will not need this viewpoint for our algorithm.
}
\end{problem}

The rest of this subsection is dedicated to showing that \Cref{prob:pseudoline} can be constructed in $N \cdot 2^{O(\sqrt{\log N\log\alpha(N)})}$ preprocessing time and answering a query takes $2^{O(\sqrt{\log N\log\alpha(N)})}$ time, where $\alpha(\cdot)$ is the slow-growing inverse Ackermann function. This is sufficient to prove \Cref{L:ds-pseudolines}.  

To appreciate the difficulty of the problem, the reader may first consider the special case when $I=\R$, which is already nontrivial.
Our idea is to explicitly construct the region of all query points $q$ for which the answer is yes.  Interestingly, we are able to prove that this region has near-linear combinatorial complexity.  After constructing the region, answering queries in the case when $I=\R$ would become easy.

To prove this combinatorial fact and at the same time design a data structure for general $I$,
we will use a divide-and-conquer strategy.

\paragraph{Decomposing intervals into canonical intervals.}
Assume that the endpoints of all intervals $I_s$, as well as $I$, are integers bounded by $O(N)$ (by replacing numbers by their ranks).
Fix a parameter $b$.  
A \EMPH{canonical interval} refers to an interval of the form
$[j\cdot b^i,(j+1)\cdot b^i)$ for some $i$ and $j$.  Any interval can be expressed as a union of
$O(b\log_b N)$ canonical intervals.  
This is a well-known fact (e.g., in analyzing a $b$-ary range tree \cite{AgarwalE99,CGAA}).  For completeness,
we include a quick proof in the following paragraph:

Let $J=[x,y]$ be an original interval, and suppose that the largest canonical interval covered by $J$ has size $b^k\leq N$.
Remove the maximum number of such intervals. Notice that this operation removes some middle part $M$ of $J$ consisting of at most $b$ intervals of size $b^{k_1}$, and leaves an interval $J_1$ on the left of $M$ and $J_2$ to the right of $M$, both having size less than $b^k$ and one endpoint that is an integer power of $b^k$.
Now if $J_1=[x,\ell_x\cdot b^k]$, then we can shift it to the interval $J_1=[x-\ell_x\cdot b^k,0]$. If $(-z_k\dots z_1z_0)_b$ is the base-$b$ representation of the left endpoint of this interval, then it naturally decomposes this interval into $\sum_i z_i \leq 1+ b\log_b N$ intervals.
All of these intervals can be shifted by $\ell_y\cdot b^k$ to get a decomposition of $J_1$. Similarly, we can decompose $J_2=[\ell_y\cdot b^k,y]$ by considering the base-$b$ representation of $J'_2=[0,y-\ell_y\cdot b^k]$.
The resulting representation has at most $b+2+2b\log_b N=O(b\log_b N)$ intervals, and it can be found in $O(b\log_b N)$ time.

We replace each interval in the input and queries by canonical intervals.
If we do this procedure for all of our $N$ intervals then we end up with $O(N\cdot b\log_b N)$ canonical input intervals.
For each pseudoline $s$ whose original interval $J_s$ has been decomposed into $k_s$ canonical intervals, we will have $k_s$ copies of $s$ instead, each associated with one such canonical interval. Thus, we have $N'=O(Nb\log_b N)$ pseudolines, each associated with a single canonical interval. With slight abuse of notation, we will keep using $S$ for this set of pseudolines (where a single pseudoline may appear several times as long as their associated canonical intervals are different). 
Similarly, when a query interval is decomposed into canonical intervals, the query cost goes up by at most an $O(b\log_b N)$ factor.

\paragraph{Preprocessing.}
Let \EMPH{$\LE(X)$} and \EMPH{$\UE(X)$} denote the \EMPH{lower and upper envelope} of a set $X$ of $x$-monotone pseudolines, respectively. 

For each canonical interval $I$, let $\EMPH{$S_{\subseteq I}$} \coloneqq \set{s\in S : I_s\subseteq I}$
and $\EMPH{$S_I$} \coloneqq \set{s\in S: I_s=I}$.
Let \EMPH{$\Env_{\subseteq I}$} be the boundary of the region of all points $q\in\R^2$ such that 
\[ 
\bigcup_{\substack{s\in S_{\subseteq I}\\ \text{$s$ below $q$}}} I_s\ =\ I.
\]
Then $\Env_{\subseteq I}$ is an $x$-monotone chain in the arrangement of $S$---we can view this as a kind of ``generalized envelope''.
We will show that this generalized envelope
has near-linear combinatorial complexity and can be computed in near-linear time for a sufficiently large choice of $b$.

\begin{figure}[t]
\centering
\includegraphics[scale=0.8]{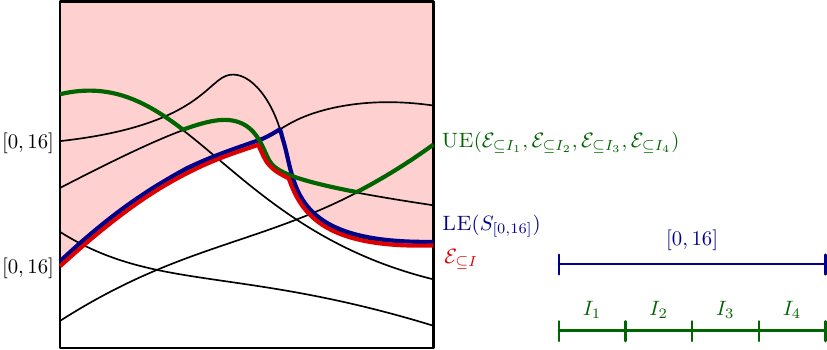}
\caption{The region $\Env_{\subseteq I}$ of points $q$ such that the intervals associated with the pseudolines under them covers $I=[0,16]$ (red shaded region). The boundary of this region is below all pseudolines associated with $[0,16]$ (blue envelope) and below the upper envelope of the regions associated with the canonical child intervals (green envelope).}\label{fig:general-envelope}
\end{figure}

Decompose $I$ into $b$ ``child'' canonical intervals $I_1,\ldots,I_b$. 
Note that $S_{\subseteq I} = S_I\cup S_{\subseteq I_1}\cup\cdots\cup S_{\subseteq I_b}$.
Then we have the
following recursive formula for the generalized envelope $\Env_{\subseteq I}$ (see Figure~\ref{fig:general-envelope}): 
\[  
\Env_{\subseteq I}\ =\ \LE\Paren{ \Set{\big. \LE(S_I),\ \UE( \set{\Env_{\subseteq I_1},\ldots,\Env_{\subseteq I_b}} ) }}. 
\]

Let \EMPH{$|\Env_{\subseteq I}|$} denote the combinatorial complexity (number of arcs) of $\Env_{\subseteq I}$.
The upper envelope of $f$ pseudo-segments is known to have combinatorial complexity $O(f\cdot \alpha(f))$, 
through Davenport-Schinzel sequences~\cite{AGARWAL20001,Pettie_DSseq_15}.
Thus, $\UE(\{\Env_{\subseteq I_1},\ldots,\Env_{\subseteq I_b}\})$ has combinatorial
complexity 
$O( (|\Env_{\subseteq I_1}| + \cdots+ |\Env_{\subseteq I_b}|) \cdot \alpha(N))$.
Now, $\LE(S_I)$ has combinatorial complexity $O(|S_I|)$.
Thus, $|\Env_{\subseteq I}| \le O( |S_I| +  (|\Env_{\subseteq I_1}| + \cdots+ |\Env_{\subseteq I_b}|) \cdot \alpha(N))$.
The maximum combinatorial complexity, \EMPH{$E(n)$}, of $\Env_{\subseteq I}$ among those with $|S_{\subseteq I}|=n$,
satisfies the recurrence
\[ 
E(n)\ \le  \max_{n_0,\ldots,n_b:\ n_0+\cdots+n_b=n} 
\Paren{ O(\alpha(N))\cdot \big(E(n_1)+\cdots+E(n_b)) + O(n_0) },
\]
which solves to $E(n) = n\cdot \alpha(N)^{O(\log_b N)}$.

For the data structure, we store $\Env_{\subseteq I}$ as well as $\LE(S_I)$ for each canonical interval $I$.  The preprocessing time satisfies the recurrence
\[ T(n)\ \le  \max_{n_0,\ldots,n_b:\ n_0+\cdots+n_b=n} \big(T(n_1)+\cdots + T(n_b) + 
\OO(E(n_1)+\cdots+E(n_b)+n_0)\big),
\]
which solves to $T(n) = \OO(n\cdot \alpha(N)^{O(\log_b N)})$.

\paragraph{Querying.}
Given a query point $q$ and a canonical interval $I$,
we check that $q$ is above $\Env_{\subseteq I}$ by binary search in the generalized envelope, or that $q$ is above $\LE(S_{I'})$ for some ``ancestor'' canonical interval $I'\supset I$
(there are $O(\log_b N)$ such intervals $I'$).  The query time
is $\OO(1)$.

\paragraph{Conclusion.}
After including the $O(b\log_b N)$ factor,
the overall preprocessing time is $\OO(bN\cdot \alpha(N)^{O(\log_b N)})$
and query time is $\OO(b)$.
Setting $b \coloneqq 2^{\sqrt{\log N\log\alpha(N)}}$, we get
$N2^{O(\sqrt{\log N\log\alpha(N)})}\le N^{1+o(1)}$ preprocessing time
and $2^{O(\sqrt{\log N\log\alpha(N)})}\le N^{o(1)}$ query time.
This concludes the proof.

\subsection{Data Structure for Unit Squares}\label{subsec:unitsq-DS}

\newcommand{\ZZ}{{\cal Z}}
\newcommand{\Stair}{\textsc{stair}}
\newcommand{\Uni}{\textsc{union}}
\newcommand{\Dtall}{\cD_{\mathrm{tall}}}
\newcommand{\cZ}{\mathcal Z}

In this subsection, 
we directly solve the interval cover problem for unit squares.
(Again, we do so without going through rainbow colored intersection searching, to get better time bounds.)
\begin{theorem}\label{thm:ds_unitsqure}
    There is a data structure $\cD_{\mathrm{square}}$ that solves the inteval cover problem (\Cref{def:DS-1}) for axis-aligned unit square objects, each associated with a single interval with $N\cdot 2^{O(\sqrt{\log N})}=N^{1+o(1)}$ preprocessing time and $2^{O(\sqrt{\log N})}=N^{o(1)}$ query time.
\end{theorem}

Observe that for the geometric intersection graph of unit side-length squares, we can replace each of the squares with squares of side-length $2$ with the same center such that a pair $s,t$ of original squares intersect if and only if the center of $s$ is contained in the scaled square~$t'$. As a result, the data structure problem is modified as follows: given a set $S$ of squares, where each $s\in S$ is associated with an interval $I_s$, we need a data structure to decide if the intervals of the squares containing the query \emph{point} $q$ will cover the query interval $I$.

\begin{figure}[ht!]
\centering
\includegraphics[scale=1]{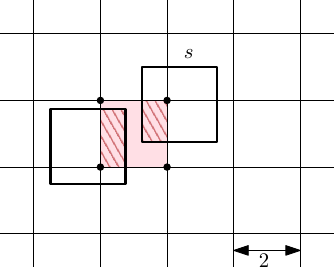}
\caption{A square $s$ of side length of $2$ intersecting cell $\square$ and the intersection (shaded) is an orthant containing exactly one vertex of $\square$.}\label{fig:orthant}
\end{figure}

Instead of the above variant, we overlay a grid of side length 2 (such that no grid line is collinear with any square of $S$); let $\Gamma$ denote the set of grid cells. Notice that if $q$ is in a given grid cell $\square\in \Gamma$, then for each square $s\in S$ we have that $s\cap \square$ appears as an \EMPH{orthant}, i.e., a rectangle\footnote{We use the term \emph{orthant} to distinguish these rectangles from other rectangles in the proof.} containing exactly one vertex of $\square$. Thus, in each cell $\square$ we have the following data structure problem. See \Cref{fig:orthant} for an example. 

\begin{problem}\label{prob:orthintervals}
We are given an input set $S$ of $N$ orthants in a cell $\square$ where each orthant covers exactly one vertex of $\square$ in $\R^2$, where each orthant $s\in S$ has an associated interval $I_s$. We want to build a data structure to answer the following type
of queries: given a query point $q$ and interval $I$, test whether $\displaystyle\bigcup_{s\in S:\ q\in s} I_s$ contains $I$.
\footnote{
Alternatively, if we think of intervals as living in a third dimension, the problem is equivalent to the following: given a set of axis-aligned boxes in $\R^3$ where the $xy$-projection of each box is an orthant, determine whether a query line segment parallel to the $z$-axis is completely contained in the union of the boxes.
We will not need this viewpoint for our algorithm (though this type of 3-dimensional data structure problem seems interesting in its own right).  As mentioned, unlike traditional range searching, this problem is not decomposable.
}
\end{problem}

The rest of this subsection is dedicated to showing that \Cref{prob:orthintervals} can be constructed in $N \cdot 2^{O(\sqrt{\log N})}$ preprocessing time and answering a query takes $2^{O(\sqrt{\log N})}$ time. This is sufficient to prove \Cref{thm:ds_unitsqure}, as we can use this data structure in each cell: the preprocessing time is $\sum_{\square\in \Gamma} N_\square \cdot 2^{O(\sqrt{\log N_\square})}=N2^{O(\sqrt{\log N})}$ where $N_\square$ is the number of orthants in cell $\square$ and $\sum_\square N_\square=4N$. To answer queries, we switch to the cell $\square_q$ containing $q$ in $O(1)$ time and answer the query using the data structure of $\square_q$ in $2^{O(\sqrt{\log N_\square})}\le 2^{O(\sqrt{\log N})}$ time.

We will use a divide-and-conquer strategy like in \Cref{SS:ds-unitdisk}, but the combinatorial complexity of the regions we want may no longer have near linear complexity (because we do not restrict orthants to a fixed type), so extra ideas are needed.

Let $\Uni(X)$ denote the union of a set $X$ of rectangles. As seen in \Cref{SS:ds-unitdisk}, we will later set some number $b$ and use canonical intervals  of the form
$[j\cdot b^i,(j+1)\cdot b^i)$ for some $i$ and $j$.
As seen in \Cref{SS:ds-unitdisk}, for each orthant $s$ whose interval $I_s$ has been decomposed to $k_s$ canonical intervals, we will have $k_s$ copies of $s$ instead, each associated with one such canonical interval. Thus, we have $N'=O(Nb\log_b N)$ objects, each associated with a single canonical interval. With slight abuse of notation, we will keep using $S$ for this set of objects. 
For an interval $I$ we again denote by $S_{\subseteq I}$ and $S_I$ the set of orthants whose intervals are subsets of $I$ or equal to $I$, respectively.

\paragraph{Preprocessing.}

\begin{quote}
Let $\ZZ_{\subseteq I}$ be the region of all points $q\in\R^2$ such that 
$\bigcup_{s\in S_{\subseteq I}:\ q\in s} I_s\ \neq\ I$.
\end{quote}
Unfortunately, the combinatorial complexity of $\ZZ_{\subseteq I}$ may be quadratic.
Instead, we will maintain a set of rectangles $Z_{\subseteq I}$ with
$\Uni(Z_{\subseteq I}) = \ZZ_{\subseteq I}$.  
In other words, instead of maintaining the region $\ZZ_{\subseteq I}$ explicitly, we implicitly represent $\ZZ_{\subseteq I}$ as a union of (possibly overlapping) rectangles.
We will show that a near linear 
number of rectangles suffices (for a sufficiently large $b$).

With this representation scheme, we can union two regions trivially.  However, intersection is a trickier operation.
In the lemma below, we show how to perform intersection with 
$\Uni(S)^c$ for a set $S$ of orthants, which is sufficient for our purposes.  Here, for a region $U$, we let \EMPH{$U^c$}$:=\square\setminus U$ denote the complement of $U$ in $\square$. 

\begin{lemma}
Given a set $S$ of orthants and a set $Z$ of rectangles in $\square$ we can construct a set $Z'$ of $O(|S|+|Z|)$ rectangles in $\OO(|S|+|Z|)$ time, such that
$\Uni(S)^c \cap \Uni(Z)\ =\ \Uni(Z')$.
\end{lemma}
\begin{proof}

A \EMPH{tallest-edge data structure} solves the following problem. We are given a set $Z$ of axis-aligned rectangles in the plane. Then, given a query segment $e$, we want to find the rectangle $z^*\in Z$ where the top side of $z^*$ has the maximal $y$ coordinate (i.e., $z^*$ is the tallest) among the rectangles $z\in Z$ covering $e$. We also allow queries in the other three axis directions, i.e., instead of the tallest reaching rectangle covering $e$, we also want to be able to find the leftmost, rightmost, or bottommost reaching rectangle covering $e$. 
Such queries can be answered using range trees in $\poly(\log |Z|)$ query time and $\OO(|Z|)$ preprocessing~\cite{CGAA}.
We start our construction by making a tallest-edge data structure $\Dtall$ for $Z$.

Let $S_1\cup S_2\cup S_3 \cup S_4$ be the partition of $S$ according to the vertex of $\square$ covered by the orthants. The \EMPH{staircase} $i$ for $i=1,2,3,4$ is the polygonal path 
$\left(\bdry \bigcup_{s\in S_i} s\right)\cap  \square$.

Set $Z_0\coloneqq Z$ and $\Dtall^0 \coloneqq \Dtall$. For each set $S_i$ we will do the following computation in the order of their indices ($i=1,\dots,4$). Suppose without loss of generality that $S_i$ covers the bottom left corner of $\square$; the other cases will be obtained from this via rotation.
Observe that $\Uni(S_i)^c$ is the region above a staircase.
The staircase has $O(|S_i|)$ edges.
Intersect the staircase with the boundaries of the rectangles of $Z$.
Subdivide the edges of the staircase at those intersection points. Note that the edge of the staircase intersected by a given edge of a rectangle $z$ can be found with a simple binary search.
The staircase now has $O(|S_i|+|Z_{i-1}|)$ edges, and it has been constructed in $\OO(|S_i|+|Z_{i-1}|)$ time.

For each edge $e$ of the staircase, we query $\Dtall^i$ to find the rectangle $z_e\in Z_{i-1}$ containing $e$ with
the highest top side in $\poly(\log |Z_{i-1}|)$ time.
Define $z_e'$ to be the rectangle with bottom side $e$ and
top side touching the top side of $z_e$.  Add $z_e$ to $Z_i$.

For each rectangle $z\in Z_{i-1}$, if the bottom side of $z$ intersects
the staircase at a point $p_z$, define $z'$ to be the part of $z$ to the right of $p_z$.
Add this rectangle $z'$ to $Z_i$.  If $z$ is completely above the staircase, add $z$ to~$Z_i$.

\begin{figure}[ht!]
\centering
\includegraphics[scale=1]{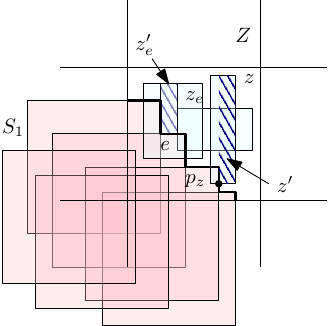}
\caption{$S_1$ is the set of squares that cover the bottom left vertex of $\square$. The staircase of $S_1$ is shown in a solid polygonal path. The figure shows the rectangles $z'_e$ and $z'$ added to $Z_i$.  }\label{fig:staircase}
\end{figure}

Then $Z_i$ has $O(|S_i|+|Z_{i-1}|)$ rectangles and satisfies the stated property. Finally, we set up a tallest-cover data structure for $Z_i$ in $\OO(|Z_i|)$ time. The total time for step $i$ is therefore $\OO(|S_i|+|Z_{i-1}|)$.

We can handle each of the sets $S_i$ one after another, and we set $Z'\coloneqq Z_4$.
The resulting number of rectangles is
\[|Z_4|=O(|S_4|+|Z_3|)=O(|S_4|+O(|S_3|+|Z_2|))=\dots = O(\sum_i|S_i| + |Z_0|) = O(|S|+|Z|).\]
The total running time is $\sum_i\OO(|S_i|+|Z_{i-1}|)=\OO(|S|+|Z|)$.
\end{proof}

Recall that the canonical interval $I$ can be decomposed into $b$ ``child'' canonical intervals $I_1,\ldots,I_b$. Suppose that there are $n_j$ orthants in $S_{\subseteq I_j}$ and $n_0$ orthants in $S_I$. We can compute $\cZ_{\subseteq I}$ using the following recursive formula:
\[  \ZZ_{\subseteq I}\ =\ \Uni(S_I)^c \,\cap\, (\ZZ_{\subseteq I_1}\cup\cdots\cup \ZZ_{\subseteq I_b}).
\]
We can apply the lemma to find a set $Z_{\subseteq I}$ of $O(|S_I|+|Z_{\subseteq I_1}\cup\cdots\cup Z_{\subseteq I_b}|)$ rectangles with $\Uni(Z_{\subseteq I}) = \ZZ_{\subseteq I}$.

The number of rectangles in $Z_{\subseteq I}$, assuming $|S_{\subseteq I}|=n$, satisfies the
recurrence
\[ E(n)\ \le  \max_{n_0,\ldots,n_b:\ n_0+\cdots+n_b=n} (O(1)\cdot \big(E(n_1)+\cdots+E(n_b)) + O(n_0)\big),
\]
which solves to $E(n) = O(n\cdot 2^{O(\log_b N)})$.

To construct the data structure $\cD_{\mathrm{square}}$, we store $Z_{\subseteq I}$ and $S_I$ in individual rectangle stabbing data structures~\cite{chazelle83,Shi2005-bk}, 
for each canonical interval $I$. The data structure for a given canonical interval $I$ can therefore be made in $\OO(E(n))$ time.

Consequently, the preprocessing time satisfies the recurrence
\[ T(n)\ \le  \max_{n_0,\ldots,n_b:\ n_0+\cdots+n_b=n} \big(T(n_1)+\cdots + T(n_b) + 
\OO(E(n_1)+\cdots+E(n_b)+n_0)\big),
\]
which solves to $T(n) = \OO(n\cdot 2^{O(\log_b N)})$.

\paragraph{Querying.}
Given a query point $q$ and a canonical interval $I$,
we check that $q$ is not stabbing any rectangle in $Z_{\subseteq I}$,
or that $q$ stabs some orthant in $S_{I'}$ for some ``ancestor'' canonical interval $I'\supset I$. Since there are $O(\log_b N)=O(\log N)$ ancestor canonical intervals, the query time is $\OO(1)$.

To answer the query about the original interval $J$, we make individual queries on each of the $O(b\log_b N)$ canonical intervals in its decomposition, and answer ``yes'' if and only if each canonical interval was covered.

\paragraph{Conclusion.}
After including the $O(b\log_b N)$ factor,
the overall preprocessing time is $\OO(bN\cdot 2^{O(\log_b N)})$
and the query time is $\OO(b)$.
Setting $b=2^{\sqrt{\log N}}$, we get
$N2^{O(\sqrt{\log N})}\le N^{1+o(1)}$ preprocessing time
and $2^{O(\sqrt{\log N})}\le N^{o(1)}$ query time, and conclude the proof of \Cref{thm:ds_unitsqure}.

\section{Switching Interval Representation between Different Stabbing Paths}
\label{SS:reordering}

We are given a set system $(X,\cS)$ with at most $n = |X|$ elements and $m = |\cS|$ sets with dual shatter dimension of $(X,\cS)$ is $d$.
Throughout the rest of the section we assume the existence of an \EMPH{element reporting oracle} that, given $S\in\cS$, can enumerate all elements of $S$ in \EMPH{$T_0(n)$} time, where $T_0(n) \ge n$. 

Let $\spath$ be an ordering of $X$.
We say that a set $S$ \EMPH{crosses} a pair $(x,y)$ if $x\in S$ and $y\not\in S$, or vice versa.  The number of consecutive pairs in $\spath$ crossed by $S$ is at most twice the size $|\Rep_\spath(S)|$.
For any collection $\cR$, define the \EMPH{equivalence relation $\equiv_\cR$} over $X$, where $x\equiv_\cR y$ if and only if 
no set in $\cR$ crosses $(x,y)$. 
(In other words, 
$\{S\in\cR: x\in S\}=\{S\in\cR: y\in S\}$.) 
Then $\equiv_\cR$ has $O(|\cR|^d)$ equivalence classes since the dual shatter dimension is at most $d$.
For every $x$ and $y$ in $X$, the \EMPH{crossing number $c_\cS(x,y)$} is the number of sets in $\cS$ crossing $(x,y)$.
(Notice that $c_\cS(\cdot, \cdot)$ forms a pseudometric.)

For the purpose of the remaining section, we will fix a \EMPH{$\rho$-sampling $\cR$} of $\cS$, where each set in $\cS$ chosen with probability $\rho/m$. 
(Later on we will restrict $\cR$ to subcollection $\cS'$ of $\cS$ and obtain $\cR'$; we can still think of $\cR'$ as obtained from $\cS'$ by sampling each element with probability $\rho/m$, even though we do not explicit sample from $\cS'$. Notice that the parameter $m$ does \emph{not} change even if $\cS'$ gets smaller.)

\medskip
Our first goal is to prove that any $\rho$-sampling of $\cS$ has low crossing number and thus can be used to construct a stabbing path for $(X,\cS)$.

\begin{lemma}
\label{lem:net}
Let $\cR$ be $\rho$-sampling of $\cS$.
Then for every $x,y\in X$ with $x\equiv_\cR y$, crossing number $c_\cS(x,y)$ is at most $O((m/\rho)\log n)$ with high probability.
\end{lemma}

\begin{proof} 
This follows by a standard hitting set argument. Consider any two elements $x, y \in X$ with crossing number $c_\cS(x,y)$. By standard Chernoff bounds, if $c_\cS(x,y) = \Omega((m/\rho)\log n)$, we would have sampled one of the sets in $\cR$ that cross $(x, y)$ with high probability, i.e., probability $1-1/n^c$ for a large constant $c$, but $x\equiv_\cR y$ which is a contradiction. 
The conclusion follows after taking a union bound over the $n^2$ pairs of elements.
\end{proof}

Let $\cS'$ be an arbitrary subcollection of $\cS$.
Denote the restriction of the fixed $\rho$-sampling $\cR$ of $\cS$ in $\cS'$ as \EMPH{$\cR'$}; in notation, $\cR' \coloneqq \cS' \cap \cR$. 
Notice that $\cR'$ is also a $\rho$-sampling.
Given any set system $(X,\cS)$,
a stabbing path $\spath$ of $(X,\cS)$ is \EMPH{$\cR'$-respecting} if 
each equivalence class of $\equiv_{\cR'}$ appears contiguously in $\spath$ for the restriction $\cR'$.  (The equivalent classes of $\equiv_{\cR'}$ is with respect to the restriction $\cR'$, not $\cR$.)
The above proof can be adapted so that the resulting stabbing path is $\cR$-respecting (by choosing $\cS'=\cS$):

\respect*

\begin{proof}
Let $\cR$ be a $\rho$-sampling of $\cS$; then
$|\cR| = \OO(\rho)$ with high probability.
We first enumerate the elements in all $R\in\cR$ in $\OO(T_0(n)\cdot \rho)$ time, and compute the $O(\rho^d)$ equivalence classes of $\equiv_R$ in $\OO(n\rho)$ time.
(Each class will appear contiguously in the stabbing path $\spath$ to be constructed.)
Within each equivalence class $C_i$, we order its elements $x_1^{(C_i)},\ldots, x_{|C_i|}^{(C_i)}$ arbitrarily.
We recursively compute an ordering of \smash{$\Set{\big. x_1^{(C_i)} : i \in [1:\rho^d]}$} by invoking the main statement of the lemma itself (which is inductively $\cR$-respecting), with run time $\OO(T_0(n)\cdot (\rho^d)^{1/d})=\OO(T_0(n) \cdot \rho)$.
We then order the classes $C_i$ (as intervals of elements) according to the order of $\Set{\big. x_1^{(C_i)} : i \in [1:\rho^d]}$.

Since $2\cdot |\Rep_\spath(S)|$ is equal to the number of consecutive pairs in $\spath$ crossed by $S$, with high probability
\begin{eqnarray*}
\sum_{S\in\cS} 2\cdot|\Rep_\spath(S)| &=& \sum_{C_i} \left(c_\cS(x_1^{(C_i)},x_2^{(C_i)})+c_\cS(x_2^{(C_i)},x_3^{(C_i)}) + \cdots + c_\cS(x_{|C_i|-1}^{(C_i)},x_{|C_i|}^{(C_i)}) + c_\cS(x_{|C_i|}^{(C_i)}, x_1^{(C_{i+1})}) \right)\\
& \le & 
\sum_{C_i} \left(2c_\cS(x_1^{(C_i)},x_2^{(C_i)})+2c_\cS(x_2^{(C_i)},x_3^{(C_i)}) + \cdots + 2c_\cS(x_{|C_i|-1}^{(C_i)},x_{|C_i|}^{(C_i)}) + c_\cS(x_{1}^{(C_i)}, x_1^{(C_{i+1})})\right)\\
&\le & \OO(mn/\rho) + \sum_{C_i} c_\cS(x_{1}^{(C_i)}, x_1^{(C_{i+1})}),
\end{eqnarray*}
where the first inequality follows from applying the triangle inequality of $c_\cS(\cdot,\cdot)$ (because $c_\cS(\cdot, \cdot)$ forms a pseudometric) on the last term $c_\cS(x_{|C_i|}^{(C_i)}, x_1^{(C_{i+1})})$, and the second inequality is from Lemma~\ref{lem:net}.
Since we recurse on the first element of every equivalence class, by recursion we have
\[
\sum_{S\in\cS} |\Rep_\spath(S)| ~\le~ \OO(mn/\rho + m(\rho^d)^{1-1/d}) = \OO(mn/\rho + m\rho^{d-1}).
\]
The ordering is clearly $\cR$-respecting. 
The total running time is $\OO(T_0(n) \cdot \rho)$.
\end{proof}

\begin{lemma}
\label{lem:convert-order}
Consider a fixed $\rho$-sampling $\cR$ of $\cS$.
We are given two $\cR$-respecting stabbing paths $\spath$ and $\spath'$ of $(X,\cS)$ (along with the equivalence classes of $\equiv_\cR$).
Let $\cT$ be an arbitrary subcollection of $\cS$.
Given $\Rep_\spath(S)$ for all $S\in\cT$, we can compute $\Rep_{\spath'}(S)$ for all $S\in\cT$ in $\OO(mn/\rho + m\rho^d)$ total time with high probability.
\end{lemma}

\begin{proof}
Given $S\in\cT$ and an equivalent class $C$ of $\equiv_\cR$, we compute the part of $\Rep_{\spath'}(S)$ within $C$ as follows.
Fix one representative element $x_C\in C$.
\begin{itemize}
\item Case 1: $x_C\not\in S$.  We enumerate all $x\in C$ in $S$, 
by examining the union of intervals of $\Rep_\spath(S)$.
We then concatenate $\langle x\rangle$ (singletons) over all such $x$ in the order determined by $\spath'$.
\item Case 2: $x_C\in S$.  We enumerate all $x\in C$ not in $S$,
by examining the complement of the union of intervals of $\Rep_\spath(S)$.
We then concatenate $\langle x\rangle$ (singletons) over all such $x$ in the order determined by $\spath'$, and take
the complement of the resulting union of intervals.
\end{itemize}
In both cases, the run time is linear in the number of $x\in C$ such that $S$ crosses $(x_C,x)$.
So, the total run time over all $S\in\cT$ is upper-bounded by $\sum_C\sum_{x\in C} c_\cS(x_C,x) = \OO(mn/\rho)$ with high probability by Lemma~\ref{lem:net}.

Finally, we concatenate the different parts of $\Rep_{\spath'}(S)$ over all the classes, in the order determined by $\spath'$.  This takes additional $O(m\rho^d)$ total time.
\end{proof}

\shrinkexpand*

\begin{proof}
For (1), notice that the equivalence classes for $\equiv_{\cR}$ are refinements of the equivalence classes for $\equiv_{\cR'}$.
Let $\spath''$ be an ordering obtained by taking $\spath$, and re-ordering the classes for $\equiv_{\cR}$ so that classes inside a common class of $\equiv_{\cR'}$ appear contiguously, which takes $O(m\rho^d)$ time.
This way, $\spath''$ is both $\cR$-respecting  and $\cR'$-respecting.
Now, we can apply Lemma~\ref{lem:convert-order} twice, to convert from $\Rep_\spath(S)$ to $\Rep_{\spath''}(S)$ and from $\Rep_{\spath''}(S)$ to $\Rep_{\spath'}(S)$ for all $S \in \cT$. 
This takes $\OO(mn/\rho + m\rho^d)$ time.

Similarly for (2), 
Let $\spath''$ be an ordering obtained by taking $\spath$, and re-ordering the classes for $\equiv_{\cR}$ so that classes inside a common class of $\equiv_{\cR'}$ appear contiguously.
This way, $\spath''$ is both $\cR$-respecting  and $\cR'$-respecting.
Now, we can apply Lemma~\ref{lem:convert-order} twice, to convert from $\Rep_{\spath'}(S)$ to $\Rep_{\spath''}(S)$ and from $\Rep_{\spath''}(S)$ to $\Rep_{\spath}(S)$.
\end{proof}

\section{Handling Small Pieces}

\subsection{Patterns}
Let $P$ be a piece in some LDD of $G$ with diameter $\Delta$. 
Recall that the set of boundary vertices of $P$ is denoted by $\bdry P$. Fix an arbitrary sequence of vertices $\sigma_P = \langle s_1, s_2, \dots, s_{|\bdry P|}\rangle$. For each vertex $v \in V(G)$, let $d(v, P)$ denote the distance between $v$ and any vertex of $P$. 
We denote a \EMPH{pattern} of $v$ with respect to the ordering $\sigma_P$, denoted by $\patt_v$ to be the following $|\bdry P|$ dimensional vector:
\[ \patt_v[i] = d(v, s_i) - d(v, P) \qquad \text{for every } 1 \le i \le |\bdry P|.\]
We remark that instead of subtracting by an offset of $d(v, P)$, we could have subtracted by any other offset. For example, \cite{le2023vc} instead use the offset of $d(v, s_1)$. 

Le and Wulff-Nilsen \cite{le2023vc} showed a bound on the total number of patterns with respect to $\sigma_P$ if the distance encoding VC-dimension is bounded. The proof also works for generalized distance VC-dimension.

\begin{lemma}
\label{lem:pd_vcdim}
Let $P$ be a piece in a graph $G$ with general distance VC-dimension $d$ and $\sigma_P$ an arbitrary ordering on $\bdry P$. Let $\Patt = \{\patt_v \mid v\in V(G)\}$ be the set of patterns with respect to $\sigma_P$. Then $|\Patt| = O(|\bdry P|^d \Delta^d)$.
\end{lemma}

\begin{proof}
Consider the set system $(V_G\times\Z, \GB)$ of generalized neighborhood balls, and the set system where we restrict the ground set $(\bdry P \times [\Delta], \GB)$. This restriction of the ground set does not increase the VC-dimension of the set system. 
There is a clear bijection between $\patt_v\in \Patt$ and the generalized neighborhood ball:
$\tilde N^{d(v,P)}[v] \cap (\bdry P \times [\Delta])  = \{(u, r): u\in \bdry P, r\in [\Delta], d(u,v)\le d(u,P) + r\}$.
So the number of patterns is bounded by the number of unique sets of $(\bdry P \times [\Delta], \GB)$. By the Sauer-Shelah Lemma (see \Cref{lem:vcfacts}), this is at most $O(|\bdry P|^d|\Delta|^d)$.
\end{proof}

\subsection{Diameter and Eccentricities using Patterns}
\label{SS:diameter-smallpiece}

The following algorithm computes the eccentricities of all vertices in a piece $P$ of the graph $G$.
\begin{enumerate}
    \item Compute the pairwise distance between pairs of vertices in $P$.  Let $d_v'$ denote the distance to the farthest vertex from $v$ that is within $P$.
    \item Compute all patterns $\Patt$ for $P$, and for each pattern $\patt\in \Patt$ find the farthest vertex $u\in V(G)$ that attains that pattern. Let $d_\patt$ be $d(u, s_0)$, the base distance for the pattern.
    \item For each pattern $\patt\in \Patt$, compute the distance $d(\patt, v)$ from the pattern to each $v\in P$ by doing a \EMPH{boundary weighted BFS}, i.e., a BFS where the boundary vertex distances are initialized to the values of the pattern $\patt$, and for each vertex $v\in P$ compute $d_v = \max_{\patt \in \Patt} d(\patt, v) + d_\patt$.
    \item Return $\max\{d_v, d_v'\}$.
\end{enumerate}

Step 1 can be implemented by running a BFS within $P$ from each vertex.
By \Cref{lem:pd_vcdim} the number of patterns computed in step 2 is at most $O(|\bdry P|^d \Delta^d)$, and it takes $O(n|\bdry P|)$ time to consider all distances to compute the pattern.
Running a BFS for each pattern in step 3 takes time $T(P)$ per pattern where $T(P)$ is the time it takes to run a boundary weighted BFS in $P$. 
\diamsmallpiece*

\subsection{Boundary Weighted BFS in Geometric Intersection Graphs}
\label{SS:weighted-bfs}

One approach to compute a shortest path tree in a unit disk graph of $n$ disks uses a semi-dynamic data structure, developed in~\cite{Efrat2001-hm}, that in $O(\log n)$ amortized time finds a disk containing a query point and deletes it from the set. Thus one can repeatedly apply the data structure 
to find the disks at the $(i+1)$-hop frontier in a BFS tree from the $i$th hop frontier-- for each disk at $i$-hop away from the root, repeated query the center of the disk to look for disks that intersect with it until such disks are exhausted. This gives a running time of $O(n\log n)$ to compute a BFS tree, since each disk is only deleted once.
The semi-dynamic data structure uses a grid of side length $1/2$. For each cell $Q$ of the grid, maintain the set of disks whose center lies in $Q$. Furthermore, maintain the upper envelope $S_1$ of the disks that intersect $Q$ with centers below the line through the lower boundary of $Q$, and similarly maintain the envelopes $S_2, S_3, S_4$ for the other three boundaries. Therefore, if a query point $q$ lies in a cell $Q$, all the disks that are centered inside $Q$ would contain $q$ and can be returned. Further, query $q$ against the upper envelope $S_1$ (check if $q$ is below $S_1$) to look for additional candidates. And repeat the same procedure for the other three envelopes. The upper envelope is maintained by a binary tree similar to a segment tree.

The boundary weighted BFS problem in a unit disk graph can be solved by a slight modification of this procedure: vertices on the boundary appear as query points when the shortest path tree has reached a sufficient depth. Therefore we have the following observation. 
\begin{observation}
\label{lm:bdr-BFS-unitdisk}
The boundary weighted BFS problem in a unit disk graph can be implemented in $O(|P|\log |P|)$ time.
\end{observation}

For boundary weighted BFS in the intersection graph of axis aligned squares (of varying sizes), we can use the same idea above. We need the following semi-dynamic data structure for a set of axis-parallel squares:
given a query square $q$ return a square $r$ that intersects $q$, and then delete $r$. If $q$ and $r$ intersect, either some corners of $q$ is inside $r$ or some corners of $r$ are inside $q$. Thus, the above query can be implemented by running an orthogonal range query of $q$ on the set of corner points of current set of squares, as well as a point enclosure query~\cite{chazelle83} (also called a rectangle stabbing query~\cite{Shi2005-bk}) of each of the corners of $q$ against the set of current squares. 
These queries can be answered by 2D orthogonal range trees or 2D segment trees. 
By using dynamic fractional cascading with deletion only, both query and deletion can be handled in $O(\log s)$ amortized time if we have $s$ squares~\cite{Chazelle1986-at, Chazelle1986-wk}. Therefore, we have the following lemma.

\begin{lemma}\label{lm:bdr-BFS-square}
The boundary weighted BFS problem in the intersection graph of axis-aligned squares can be implemented in $O(|P|\log |P|)$ time.
\end{lemma}

\subsection{Exact Distance Oracles}
\label{app:oracle-smallsize}
The lemmas in this section are implicit in the distance oracles of \cite{le2023vc}, but we present their proofs in full to keep our exposition self-contained.

\begin{lemma}[Section 3.2.3 of \cite{le2023vc}]
\label{lm:do-sparse-exact-constant}
    Let $G$ be a graph on $n$ vertices with bounded generalized distance VC-dimension $d$ and $P$ be a piece in $G$ with boundary $\bdry P$ and diameter $\Delta$. 
    There exists an exact distance oracle for queries in which at least one end point lies within $P$ with 
    $O(n + |\bdry P|^d \Delta^d|P| + |P|^2)$ space and $O(1)$ query time.
    
    Furthermore if distances from $\bdry P$ to all vertices of $G$ are known, the distance oracle can be computed in $O(n|\bdry P|+(|\bdry P|^d \Delta^d + |P|) \cdot T(P))$ precomputation time, where $T(P)$ is the time it takes to run vertex weighted BFS on $P$ with weights at most $\Delta$.
\end{lemma}

\begin{proof}
For each vertex $v\in P$, store the distances to all other vertices in $P$.
Every other vertex of the graph $u\in G\setminus P$ stores a pointer to their respective pattern $\patt_u$, and the distance $d(u, P)$.
Also store the distance $d(\patt, v)$ for each pattern $\patt \in \Patt$ to each $v\in P$.

To handle a query between two vertices of $P$, we can look up the distance between the vertices in constant time. For one vertex $v\in P$, and another vertex $u\in G\setminus P$, we know that:
\[ d(u, v) = d(u, P) + d(\patt_u, v).\]
and we can look up $d(u,P)$, $\patt_u$, and $d(\patt_u, v)$ in constant time.

The total space needed for the oracle is $O(|P|^2)$ for the distances between pairs of vertices in $P$, $O(n)$ for the pointers from vertices $u\in G\setminus P$ to their respective patterns, and $O(|\bdry P|^d\Delta^d|P|)$ to store the pattern to $P$ distances.

The precomputation time is the same as in \Cref{lm:diameter-smallpiece} for eccentricities.
\end{proof}

\lmdoexact*
%

\begin{proof}
Store the distances between pairs of vertices in $P$.
For every other vertex $u\in G\setminus P$, consider the sequence of balls $B(u, r_1), \dots, B(u, r_k)$ such that $B(u, r_1)$ is the smallest ball that contains at least one vertex of $\bdry P$, and $B(u, r_i)$ is the smallest ball containing at least one vertex of $\bdry P \setminus B(u, r_i)$ (note that $k\le |\bdry P|)$. Store a pointer to each of these balls, and the set of vertices $Y_i = B(u,r_i) \cap P$ (and $Y_0 = \varnothing$)  in a data structure that allows for $O(1)$ time membership lookup.
For each relevant $Y_i$, store the distance $d(Y_i, v):= \min_{s\in Y_i} d(s, v)$.

If two vertices $u$ and $v$ are within $P$, we can look up their distance in $O(1)$ time. Otherwise, if $v\in P$ and $u\in G\setminus P$, then we can binary search over $Y_0, Y_1 ..., Y_k$ to find the first $Y_i$ where $v\not\in Y_i$ and $v\in Y_{i+1}$ in $O(\log k) = O(\log |\bdry P|)$ time. Then, we can look up the distance $d(Y_i, v)$ in constant time and return the distance:
\[d(u, v) = d(u, r_i) + d(Y_i, v).\]

The space required to store distances between pairs of vertices in $P$ is at most $O(|P|^2)$.
The space required is $O(n|\bdry P|)$ to store the pointers between $u\in G\setminus P$ and their respective $Y_0, Y_1, \dots, Y_k$ since $k\le |\bdry P|$. The total number of balls is at most $O(|P|^d)$ by \Cref{lem:vcfacts} (Sauer's lemma). 

To compute this distance oracle, we need to compute $Y_1, \dots, Y_k$ for each vertex $u\in G \setminus P$. To do so, we can cluster these vertices into vertices with the same pattern $\patt_u$, and consider $Y_1, \dots, Y_k$ with respect to each pattern. This can be done as the BFS to compute $d(\patt_u, v)$ for every vertex $v\in P$ also implicitly computes the balls $Y_1, \dots Y_k$, as well as the distances $d(Y_i, v)$. To compute a pointer from $u$ to $Y_i$, we can look up the balls we computed by storing all $Y_i$s in a data structure that supports $O(1)$ lookup for sets (e.g. a hashing based data structure).
The precomputation time analysis is the same as in \Cref{lm:diameter-smallpiece} for eccentricities.
\end{proof}

\end{document}